\documentclass[submission, Phys]{SciPost}
\usepackage{amsmath,amssymb,mathtools,xspace, mathrsfs}
\usepackage{booktabs,multirow,graphicx,tabularx,slashed}
\usepackage{hyperref}
\usepackage{color,xcolor}
\usepackage[normalem]{ulem}
\usepackage{enumitem}
\usepackage{braket}
\usepackage{stackrel}
\usepackage{tikz}
\usepackage{float}
\usepackage{bm}
\usepackage[force]{feynmp-auto}
\usepackage{dcolumn,booktabs}
\usepackage{gensymb}

\graphicspath{{./plots/}}

\makeatletter
\def\BState{\State\hskip-\ALG@thistlm}
\makeatother

\makeatletter
\@ifundefined{pdfoutput}{}{\DeclareGraphicsRule{*}{mps}{*}{}}
\makeatother

\makeatletter
\DeclareRobustCommand*{\bfseries}{%
   \not@math@alphabet\bfseries\mathbf
   \fontseries\bfdefault\selectfont
   \boldmath
}
\makeatother

\setitemize{itemsep=2pt,topsep=2pt,parsep=0pt,partopsep=0pt,leftmargin=*}
\setenumerate{itemsep=0pt,topsep=2pt,parsep=0pt,partopsep=0pt,labelindent=3pt,leftmargin=*}
\newcolumntype{d}[1]{D{.}{.}{#1}}

\definecolor{Gcolor}{HTML}{3b528b}
\definecolor{Dcolor}{HTML}{e41a1c}

\tikzstyle{generator} = [rectangle, rounded corners, minimum width=3cm, minimum height=1cm,text centered, draw=Gcolor]
\tikzstyle{discriminator} = [rectangle, rounded corners, minimum width=3cm, minimum height=1cm,text centered, draw=Dcolor]
\tikzstyle{io} = [circle, trapezium left angle=70, trapezium right angle=110, minimum width=1cm, minimum height=1cm, text centered, draw=black]

\tikzstyle{process} = [rectangle, minimum width=1cm, minimum height=1cm, text centered, draw=black]
\tikzstyle{decision} = [rectangle, minimum width=1cm, minimum height=1cm, text centered, draw=black]

\tikzstyle{arrow} = [thick,->,>=stealth]
\usepackage{xcolor}

\newcommand{\D}{\mathrm{d}}

\newcommand{\Langle}{\big\langle}
\newcommand{\Rangle}{\big\rangle}

\newcommand{\XXLangle}{\Bigg\langle}
\newcommand{\XXRangle}{\Bigg\rangle}

\newcommand\one{\leavevmode\hbox{\small1\normalsize\kern-.33em1}}

\newcommand{\lag}{\mathscr{L}}
\newcommand{\loss}{\mathcal{L}}

\newcommand{\qqquad}{\qquad \qquad}

\newcommand{\br}{\text{BR}}

\def\slashchar#1{\setbox0=\hbox{$#1$}           
   \dimen0=\wd0                                 
   \setbox1=\hbox{/} \dimen1=\wd1               
   \ifdim\dimen0>\dimen1                        
      \rlap{\hbox to \dimen0{\hfil/\hfil}}      
      #1                                        
   \else                                        
      \rlap{\hbox to \dimen1{\hfil$#1$\hfil}}   
      /                                         
   \fi}

\newcommand{\ie}{\textsl{i.e.}\;}

\newcommand{\pytorch}{\textsc{PyTorch}\xspace}

\newcommand{\adam}{\textsc{Adam}\xspace}

\begin{document}
\begin{fmffile}{feynman}

\begin{center}{\Large \textbf{
      Two Invertible Networks for the Matrix Element Method
}}\end{center}

\begin{center}
Anja Butter\textsuperscript{1}, 
Theo Heimel\textsuperscript{1}, 
Till Martini\textsuperscript{2}\footnote{Work on this article was conducted while employed at Humboldt-Universit\"at zu Berlin, Institut f\"ur Physik, Berlin, Germany},
Sascha Peitzsch\textsuperscript{3*}, and 
Tilman Plehn\textsuperscript{1}
\end{center}

\begin{center}
{\bf 1} Institut f\"ur Theoretische Physik, Universit\"at Heidelberg, Germany \\
{\bf 2} Fraunhofer Zentrum SIRIOS, Fraunhofer Institute for High-Speed Dynamics EMI, Berlin, Germany \\
{\bf 3} Fraunhofer Zentrum SIRIOS, Fraunhofer Institute for Open Communication Systems FOKUS, Berlin, Germany
\end{center}

\begin{center}
\today
\end{center}

\section*{Abstract}
         {\bf The matrix element method is widely considered the
           ultimate LHC inference tool for small event numbers.
           We show how a combination of two
           conditional generative neural networks encodes the QCD
           radiation and detector effects without any simplifying
           assumptions, while keeping the computation of
           likelihoods for individual events numerically efficient.
           We illustrate our
           approach for the CP-violating phase of the top Yukawa
           coupling in associated Higgs and single-top
           production. Currently, the limiting factor for the
           precision of our approach is jet combinatorics.}

\vspace{10pt}
\noindent\rule{\textwidth}{1pt}
\tableofcontents\thispagestyle{fancy}
\noindent\rule{\textwidth}{1pt}
\vspace{10pt}

\newpage
\section{Introduction}
\label{sec:intro}

In the search for optimal analysis methods at colliders, the matrix
element method (MEM) has been playing a key role since it was
developed for the Tevatron~\cite{Kondo:1988yd,Kondo:1991dw}. If offers
an especially simple and interpretable link between theory predictions
and hypothesis tests. Its optimality is derived directly from the
Neyman-Pearson theorem, which means it includes all available
information encoded in phase space configurations and evaluates it
using an optimal hypothesis test. The MEM is based on the observation
that we can compute the likelihood of individual events, given a
theory hypothesis, as the scattering amplitude from first-principle
quantum field theory. Two different theory hypotheses or parameter
points then define a likelihood ratio for a given event. The
log-likelihood ratio of an event sample follows from adding individual
events' log-likelihood ratios, but unlike essentially all other
inference method the combination of a number of events into a
distribution is not necessary.

The first
application of the MEM was the Tevatron measurement of the top mass
based on a limited number of statistically defined top quark
events~\cite{D0:1998eiz,Abazov:2004cs,CDF:2006nne,Fiedler:2010sg}. Also
at the Tevatron, it was used to discover the single-top production
process~\cite{Giammanco:2017xyn}. At the LHC, several studies~\cite{Andersen:2012kn,Artoisenet:2013vfa,Englert:2015dlp,FerreiradeLima:2017iwx} and
applications~\cite{CMS:2015enw,ATLAS:2015jmq,CMS:2015cal,Gritsan:2016hjl,FerreiradeLima:2017iwx} of the MEM
exist. The challenge in applying the MEM is that we have to integrate over
the scattering amplitudes at the parton level for each measured event.
This makes MEM applications extremely CPU-expensive.  For instance
Madgraph can already compute parton-level amplitudes for a given event
automatically at leading order~\cite{Artoisenet:2010cn}. To really use
the power of the method we need to base it on precision predictions,
including QCD jet radiation~\cite{Alwall:2010cq} and NLO QCD
corrections, as shown for color-neutral particles in the final
state~\cite{Campbell:2012cz,Campbell:2013hz} and for jet
production~\cite{Williams:2013kfb}.  A consistent treatment of the
MEM at NLO has been developed for electron-positron collisions with
final-state radiation~\cite{Martini:2015fsa} and for hadronic
collisions with modified jet algorithms~\cite{Martini:2017ydu}, and
standard jet algorithms~\cite{Kraus:2019qoq,Baumeister:2016maz}.

We will show how modern machine learning (ML) can enhance the
MEM. Fast and invertible LHC simulations benefit from generative
networks~\cite{Butter:2020tvl,Butter:2022rso,ml_notes} like generative
adversarial networks (GANs)~\cite{Butter:2020qhk}, variational
autoencoders (VAEs), normalizing flows, and their invertible network
(INN) variant~\cite{inn}. Within the established simulation chain,
such networks can be applied to loop
integrals~\cite{Winterhalder:2021ngy}, phase space
integration~\cite{maxim,Chen:2020nfb}, phase space
sampling~\cite{Bothmann:2020ywa,Gao:2020vdv,Gao:2020zvv,Danziger:2021eeg},
event subtraction~\cite{Butter:2019eyo}, event
unweighting~\cite{Verheyen:2020bjw,Backes:2020vka}, parton
showering~\cite{shower,locationGAN,juniprshower,Dohi:2020eda},
super-resolution enhancement~\cite{DiBello:2020bas,Baldi:2020hjm},
or detector simulations~\cite{Paganini:2017dwg,Erdmann:2018jxd,Buhmann:2020pmy,Krause:2021ilc,Krause:2021wez}.
Once we control the forward direction with NN-based event
generators~\cite{dutch,gan_datasets,DijetGAN2,Butter:2019cae,Alanazi:2020klf,Butter:2021csz},
conditional GANs and INNs also allow us to invert the simulation
chain, to unfold detector
effects~\cite{Datta:2018mwd,fcgan,Andreassen:2019cjw} or to extract
the hard scattering process in a statistically consistent
manner~\cite{cinn,Bellagente:2020piv}. The fully calibrated inverted
simulation uses the same conditional INN (cINN) as simulation-based
inference~\cite{Bieringer:2020tnw,Bister:2021arb} or kinematic
reconstruction~\cite{Leigh:2022lpn}. Obviously, any application of
(generative) networks to LHC physics requires an uncertainty
treatment~\cite{Bellagente:2021yyh,Butter:2021csz}. A related
ML-approach to likelihood extraction is based on simulated versus
observed event samples~\cite{Brehmer:2019xox}.
A connection between the MEM and modern ML-methods, mentioned in
Ref.~\cite{HEPSoftwareFoundation:2017ggl}, was demonstrated in
Ref.~\cite{Bury:2020ewi}, specifically using a deep regression network
to evaluate the MEM integral.

In this paper we show how a combination of two cINNs allows for a
better modeling of QCD and detector effects while keeping the MEM
numerically efficient. First, a Transfer-cINN learns the effects of
the parton shower, detector resolution, and reconstruction on
simulated events. Second, an Unfolding-cINN provides a phase space
mapping for the integration over the hard-scattering phase space at
parton level. For both networks we use a Bayesian network version of
the INN~\cite{Bellagente:2021yyh,Butter:2021csz} to track their
reliability. The Bayesian Transfer-cINN also provides an uncertainty
estimate for the extracted likelihood.  Our toy example is the search
for CP violation in the top Yukawa coupling, based on the kinematics
of single top and Higgs production. In Sec.~\ref{sec:proc} we
introduce the physics process and the effect of a CP-phase on the
event kinematics. In Sec.~\ref{sec:mem} we introduce our dual-cINN
architecture, which we benchmark on simulated events with a leptonic
and hadronic top decay in Sec.~\ref{sec:perf}. While we do not (yet)
include a full NLO calculation of the event likelihood, we do combine
different jet numbers through initial state radiation as a first step
in this direction in Sec.~\ref{sec:perf_hadr}.  We discuss some
remaining challenges for the network precision due to combinatorics
once we include many jets. Once those issues can be overcome, our
method naturally extends to NLO likelihood predictions.

\section{LHC process}
\label{sec:proc}

\begin{figure}[b!]
\centering
\fmfset{arrow_len}{2mm}
\begin{fmfgraph*}(100,80)
    \fmfstraight
    \fmfleft{i1b,i1,i1c,i2}
    \fmfright{o1b,o1,o1c,o2}
    \fmf{fermion,width=0.6,tension=1,label=$b$}{i1,v1}
    \fmf{fermion,width=0.6,tension=2}{v1,v1b}
    \fmf{fermion,width=0.6,tension=2,label=$t$,lab.side=right}{v1b,o1}
    \fmf{fermion,width=0.6,tension=1,label=$u$}{i2,v2}
    \fmf{fermion,width=0.6,tension=1,label=$d$}{v2,o2}
    \fmffreeze
    \fmf{boson,width=0.6,tension=1,label=$W$}{v1,v2}
    \fmffreeze
    \fmf{dashes,width=0.6,tension=1,label=$H$,lab.side=left}{v1b,o1c}
\end{fmfgraph*}
\hspace{1cm}
\begin{fmfgraph*}(100,80)
    \fmfstraight
    \fmfleft{i1b,i1,i1c,i2}
    \fmfright{o1b,o1,o1c,o2}
    \fmf{fermion,width=0.6,tension=2,label=$b$}{i1,v1b}
    \fmf{fermion,width=0.6,tension=2}{v1b,v1}
    \fmf{fermion,width=0.6,tension=1,label=$t$}{v1,o1}
    \fmf{fermion,width=0.6,tension=1,label=$u$}{i2,v2}
    \fmf{fermion,width=0.6,tension=1,label=$d$}{v2,o2}
    \fmffreeze
    \fmf{boson,width=0.6,tension=1,label=$W$}{v1,v2}
    \fmffreeze
    \fmf{dashes,width=0.6,tension=1,label=$H$}{v1b,o1b}
\end{fmfgraph*}
\hspace{1cm}
\begin{fmfgraph*}(100,80)
    \fmfstraight
    \fmfleft{i1b,i1,i1c,i2}
    \fmfright{o1b,o1,o1c,o2}
    \fmf{fermion,width=0.6,tension=1,label=$b$}{i1,v1}
    \fmf{fermion,width=0.6,tension=1,label=$t$}{v1,o1}
    \fmf{fermion,width=0.6,tension=1,label=$u$, label.angle=-60}{i2,v2}
    \fmf{fermion,width=0.6,tension=1,label=$d$}{v2,o2}
    \fmffreeze    
    \fmf{boson,width=0.6,tension=1,label=$W$}{v1,v3,v2}
    \fmffreeze
    \fmf{dashes,width=0.6,tension=1,label=$H$}{v3,o1c}    
\end{fmfgraph*}
\caption{Leading-order Feynman diagrams for the hard process $pp
  \to tHj$. We neglect the second diagram in the limit of a massless
  bottom quark. The diagrams also appear with an inverted
  light-quark line.}
\label{fig:feyn}
\end{figure}
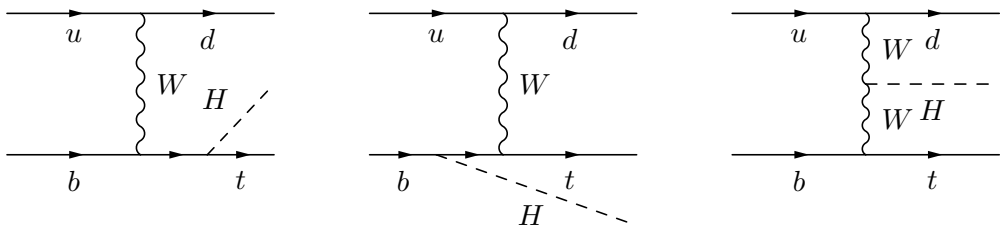

\begin{table}[t]
    \centering
    \begin{small}
    \begin{tabular}{l l r r}
        \toprule
        Dataset & cut & rate [ab] & fraction \\
        \midrule
        leptonic & $\sigma$ & $43.6 \cdot 10^3$ & \\
        & $\sigma \times \br$ & 7.38 & \\
        & $\geq 2$ photons with $p_T>20$~GeV and $\eta < 2.4$ & 3.58 & 0.485 \\
        & $\geq 1$ muon with $p_T>20$~GeV and $\eta < 2.4$ & 2.29 & 0.310 \\
        & $\geq 2$ jets & 1.69 & 0.230 \\
        & 1 $b$-jet with $p_T>25$~GeV and $\eta < 2.4$ & 1.00 & 0.136 \\
        & $\geq 1$ jets with $p_T>25$~GeV and $\eta < 2.4$ & 0.41 & 0.055 \\
        \midrule
        hadronic, no ISR & $\sigma$ & $43.6 \cdot 10^3$ & \\
        & $\sigma \times \br$ & 44.28 & \\
        & $\geq 2$ photons with $p_T>20$~GeV and $\eta < 2.4$ & 19.56 & 0.442 \\
        & $\geq 4$ jets & 7.09 & 0.160 \\
        & 1 $b$-jet with $p_T>25$~GeV and $\eta < 2.4$ & 3.93 & 0.089 \\
        & $\geq 3$ jets with $p_T>25$~GeV and $\eta < 2.4$ & 1.23 & 0.028 \\
        \midrule
        hadronic, with ISR & $\sigma$ & $43.6 \cdot 10^3$ & \\
        & $\sigma \times \br$ & 44.26 & \\
        & $\geq 2$ photons with $p_T>20$~GeV and $\eta < 2.4$ & 18.37 & 0.415 \\
        & $\geq 4$ jets & 12.67 & 0.286 \\
        & 1 $b$-jet with $p_T>25$~GeV and $\eta < 2.4$ & 6.44 & 0.146 \\
        & $\geq 3$ jets with $p_T>25$~GeV and $\eta < 2.4$ & 3.06 & 0.069 \\
        \bottomrule
    \end{tabular}
    \end{small}
    \caption{Cut flow for $pp \to tHj$ with $H\to \gamma \gamma$ and
      for SM events ($\alpha=0^\circ$).  We assume $m_b=0$ and
      intermediate on-shell particles.}
    \label{tab:partonxs}
\end{table}

To illustrate how we can use generative networks for measurements
using the MEM, we choose associated single-top and Higgs production
\begin{align}
    p p \to t H j \; .
\end{align}
This process will allow us to study a CP-phase in the top Yukawa
coupling at future LHC runs~\cite{Buckley:2015vsa,Ren:2019xhp,Bortolato:2020zcg,Bahl:2020wee,Martini:2021uey,Goncalves:2021dcu,Barman:2021yfh,Bahl:2021dnc,Kraus:2019myc}, unfortunately with limited expected event
numbers. This limitation means that we need an optimal analysis
framework for this measurement, specifically the matrix element method
based on likelihood ratios. We choose the decay $H \to \gamma \gamma$
to illustrate our point with a focus on the signal process. Our
methods can be generalized to other decay processes, for which we
would also need to include continuum backgrounds.

To extract the likelihoods corresponding to different theory
hypotheses for a given phase space configuration, we will use a
combination of two neural networks.  The crucial ingredient for our NN
training are paired events at the hard-scattering level and after
parton shower and detector simulation. The usual Monte Carlo
simulation starts from the hard scattering matrix element and
successively adds parton showers and detector effects.  For the
simulation of our signal process we use Madgraph5, v3.1.0, with
LO-NNPDF and $\alpha_s= 0.119$~\cite{DelDebbio:2004xtd}. We produce
the heavy top and Higgs on their respective mass shells and decay them
in a second step. Our simulation includes the standard
Pythia~\cite{pythia8} parton shower, Delphes~\cite{delphes} as a fast
detector simulation, and Fastjet~\cite{fastjet} to reconstruct
anti-$k_T$ jets~\cite{anti-kt} of size 0.4.  As illustrated in
Fig.~\ref{fig:feyn} we consider massless incoming bottoms, while the
jet in the final state always comes from a light quark.  In the
Standard Model, the dominant contribution stems from the first diagram
where the Higgs couples to the top.  Throughout our analysis we
neglect the second diagram because of the small bottom Yukawa.

We generate three different datasets. First, the $W$ decays
leptonically, into a muon and a neutrino,
\begin{align}
  p p \to t H j \to (b \mu^+ \nu_\mu) \; (\gamma \gamma) \; j \; .
  \label{eq:proc_lept}
\end{align}
Second, the $W$ decays hadronically, resulting in two jets
\begin{align}
  p p \to t H j \to (b jj) \; (\gamma \gamma) \; j \; .
  \label{eq:proc_had}
\end{align}
In both cases, we do not generate initial state radiation (ISR) and
multi-parton interactions by disabling the corresponding settings
in Pythia. For the third dataset, we again consider
hadronic decays, but this time including ISR jets,
\begin{align}
  p p \to t H j \to (b jj) \; (\gamma \gamma) \; j + \text{QCD jets}\; .
  \label{eq:proc_isr}
\end{align}
We allow for up to four additional jets in our datasets. They can come
from final state radiation, or, in the case of the third dataset, from
initial state radiation.  The total proton-proton cross section for
$tHj$ production is 43.6~fb, where we always combine top and anti-top
production. Table~\ref{tab:partonxs} provides an overview of the cross
sections and the detector-level cuts. We do not apply cuts in $p_T$ or $\eta$
at the hard-scattering level. To illustrate the ML-based
numerical approach to the MEM we limit ourselves to the more
challenging signal, with the narrow Higgs mass peak, and ignore all
continuum backgrounds.

\subsubsection*{CP-violating Yukawa coupling}

\begin{table}[t]
    \centering
    \begin{small}
    \begin{tabular}{lrrrrr}
    \toprule
     Dataset           &               $A$ [fb] &               $B$ [fb] &                $C$ [fb] &               $D$ [fb] &                $E$ [fb] \\
    \midrule
     leptonic           & $4.07 \cdot 10^{-4}$ & $2.37 \cdot 10^{-3}$ & $-1.22 \cdot 10^{-3}$ & $1.86 \cdot 10^{-6}$ & $-2.90 \cdot 10^{-7}$ \\
     hadronic, no ISR   & $1.23 \cdot 10^{-3}$ & $7.59 \cdot 10^{-3}$ & $-3.78 \cdot 10^{-3}$ & $1.24 \cdot 10^{-5}$ & $-7.96 \cdot 10^{-6}$ \\
     hadronic, with ISR & $3.06 \cdot 10^{-3}$ & $2.05 \cdot 10^{-2}$ & $-9.50 \cdot 10^{-3}$ & $1.90 \cdot 10^{-5}$ & $-5.77 \cdot 10^{-6}$ \\
    \bottomrule
    \end{tabular}
    \end{small}
    \caption{Fit parameters for the fiducial cross sections, for the
      formula given in Eq.\eqref{eq:fiteq}.}
    \label{tab:fiducial_fit}
\end{table}

To study the top Yukawa coupling independently of the top mass, we
allow for a mixture of CP-even and CP-odd
interactions~\cite{Artoisenet:2013puc}. The top-Higgs interaction is
parameterized by
\begin{align}
  \lag_{t\bar{t}H} = - \frac{y_t}{\sqrt{2}}
  \Big[ a\cos \alpha \; \bar{t}t + 
  ib\sin \alpha \; \bar{t}\gamma_5 t \Big] H \; .
\label{eq:LttH}
\end{align}
with $a = 1$ and $b=2/3$~\cite{Demartin:2015uha}, so the total $gg\to
H$ cross section remains constant when we vary $\alpha$.  This model
has only one free parameter, the CP-angle $\alpha$, interpolating
between a CP-even $(\alpha=0^\circ)$ and a CP-odd
$(\alpha=180^\circ)$ Yukawa coupling.  Because of this coupling structure,
all observables $O$ for the $tHj$ process obtained by integrating over
the hard-scattering phase space take the functional form
\begin{align}
  O(\alpha) =
  A + B(1-\cos \alpha ) + \sin \alpha \left(C\sin \alpha + D + E\cos \alpha \right) \;,
\label{eq:fiteq}
\end{align}
as long as we only consider higher-order QCD corrections. A fit of the
fiducial cross section to the angles $\alpha = -180^\circ, -90^\circ,
-45^\circ, 0^\circ, 22.5^\circ, 45^\circ, 90^\circ, 135^\circ,
180^\circ$ gives the parameters quoted in
Tab.~\ref{tab:fiducial_fit}. With $D,E \ll A,B,C$ we see that there is
an approximate degeneracy in the sign of the CP-phase.

\begin{figure}[b!]
 \includegraphics[width=0.325\textwidth]{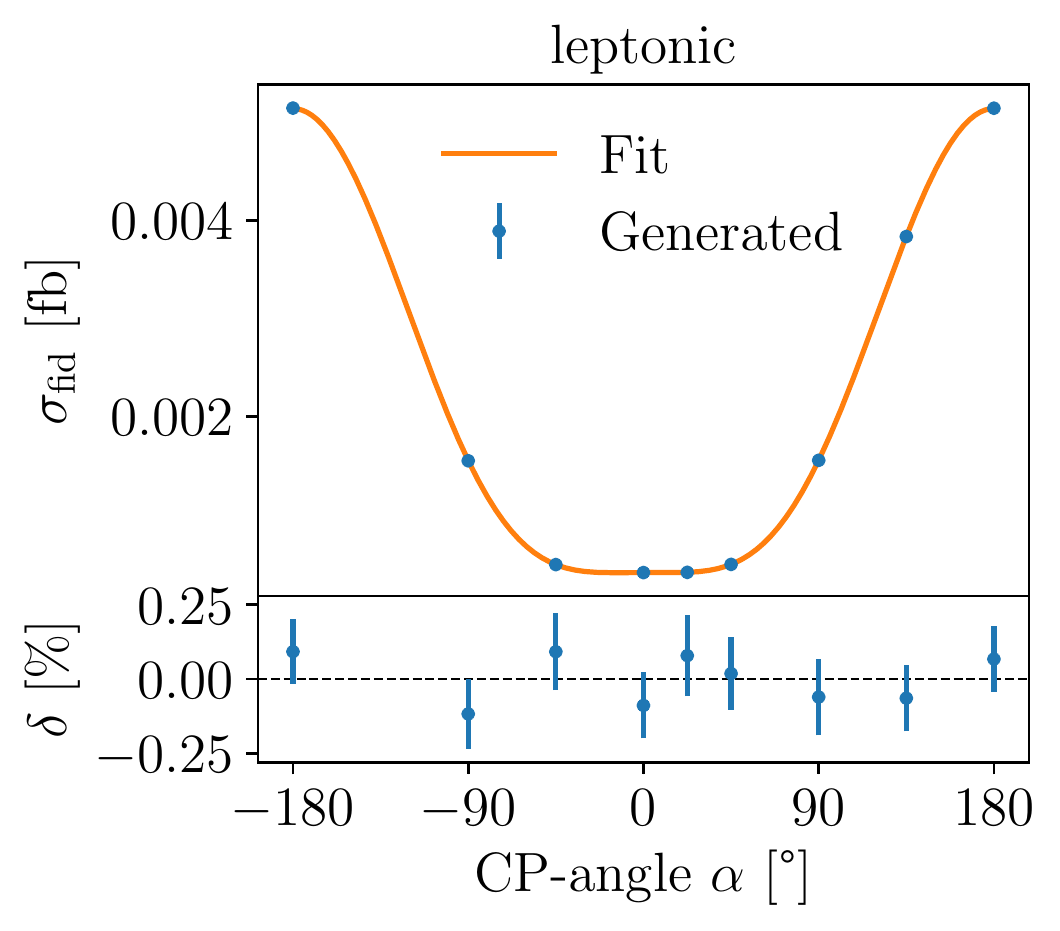}
 \includegraphics[width=0.325\textwidth]{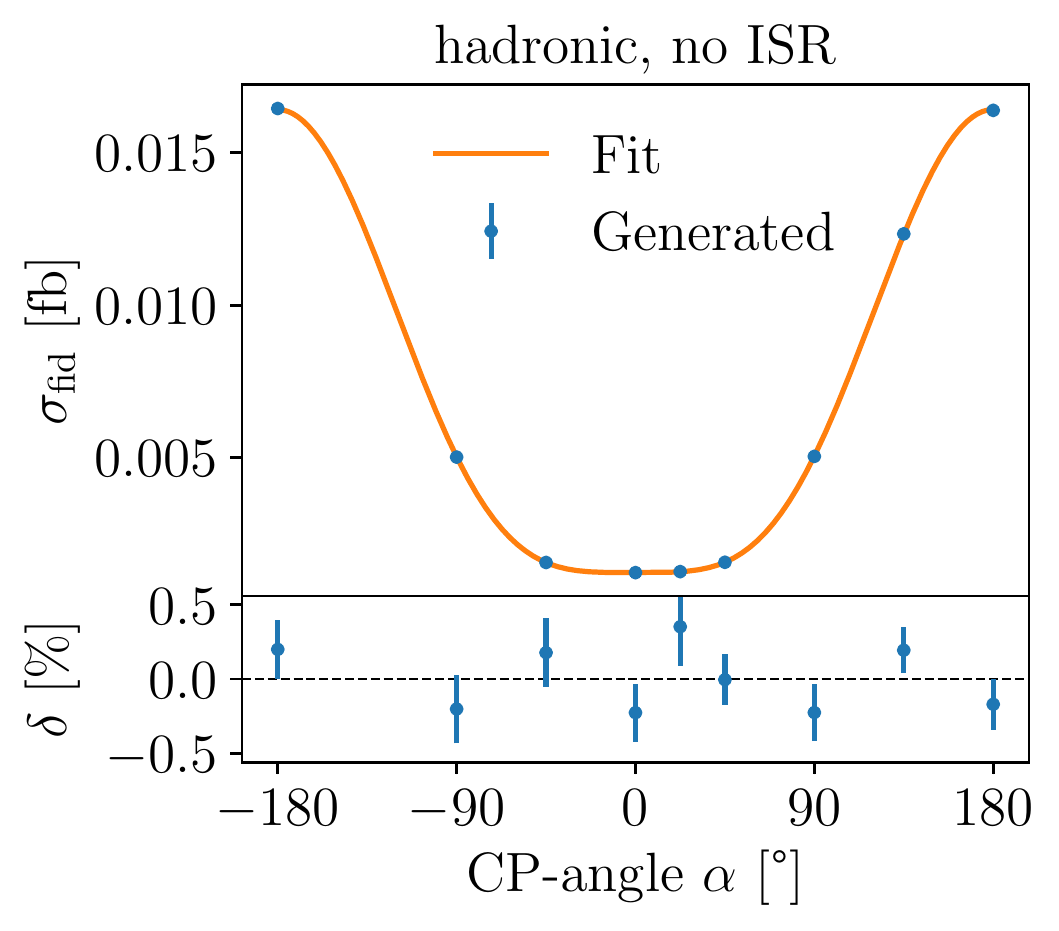}
 \includegraphics[width=0.325\textwidth]{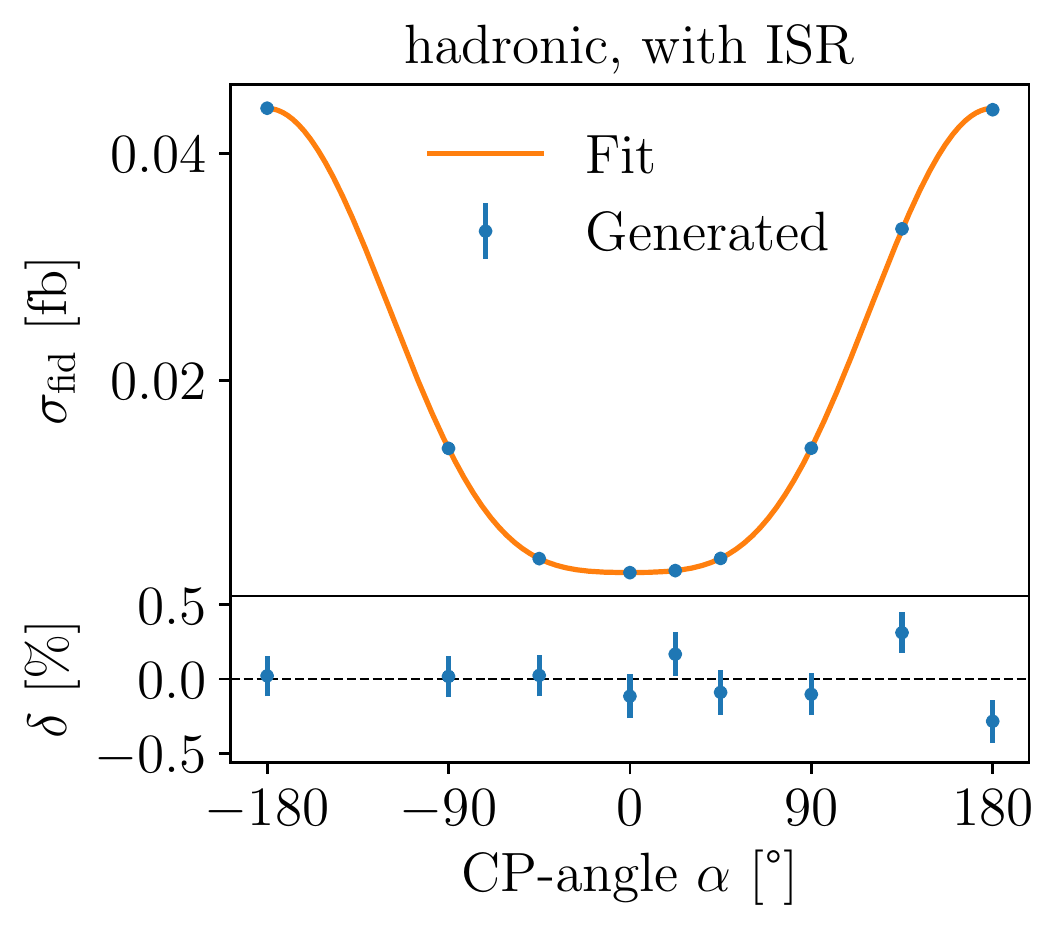}
\caption{Fiducial cross section including decays and after cuts as a
  function of the CP-angle $\alpha$. The lower panels illustrate the
  agreement between the generated data and the fitted continuous
  function defined in Tab.~\ref{tab:fiducial_fit}.}
\label{fig:xsec}
\end{figure}

In Fig.~\ref{fig:xsec} we show the fiducial $tHj$ cross section
including decays after cuts as a function of $\alpha$. Typical rates
especially around the Standard Model ($\alpha = 0^\circ$) are below
0.01~fb, which means that in an actual analysis we need to extract as
much information as possible from a small number of events and their
kinematic features.  The rate increase with $\alpha$ is driven by
the interplay of the leading top Yukawa contribution, shown to the
left in Fig.~\ref{fig:feyn}, and the sub-leading gauge coupling to the
right. The angle $\alpha$ defines a relative phase between the two
diagrams, which leads to a destructive interference in the Standard
Model.  The change in the total rate reflects the shift from
destructive to constructive interference. From Fig.~\ref{fig:xsec} we
expect that small values $\alpha \lesssim 40^\circ$ will hardly be
measured from the total rate, especially once we include experimental
and theoretical uncertainties. This means we have to complement the
rate information with kinematic features.

\begin{figure}[t]
  \includegraphics[width=0.325\textwidth,page=4]{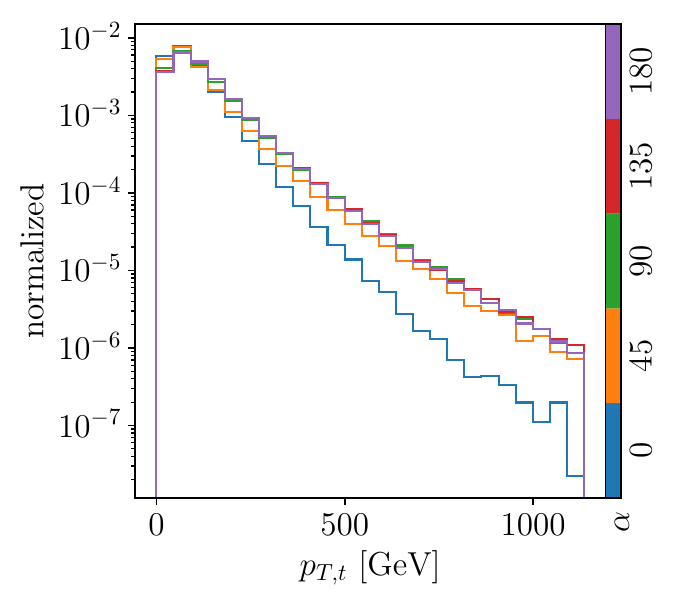}
  \includegraphics[width=0.325\textwidth,page=10]{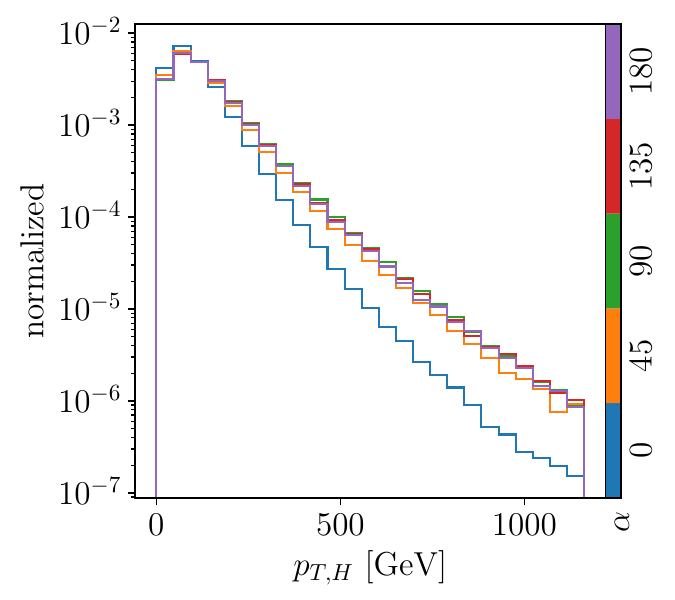}
  \includegraphics[width=0.325\textwidth,page=16]{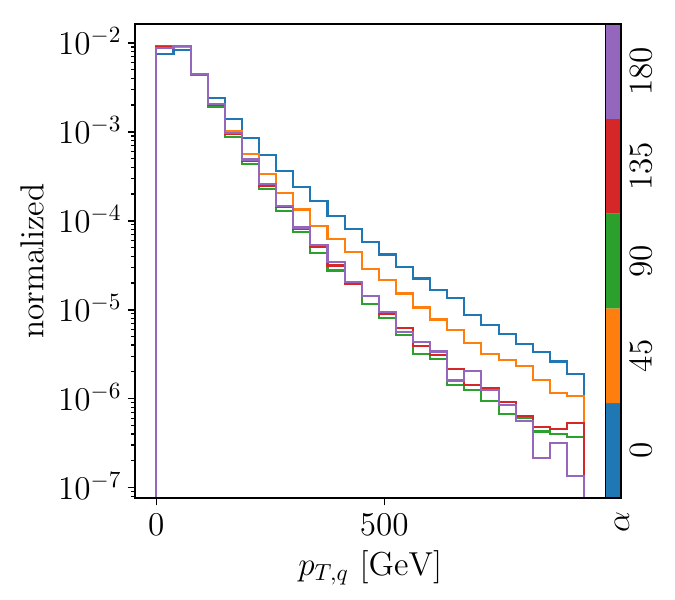}\\
  \includegraphics[width=0.325\textwidth,page=6]{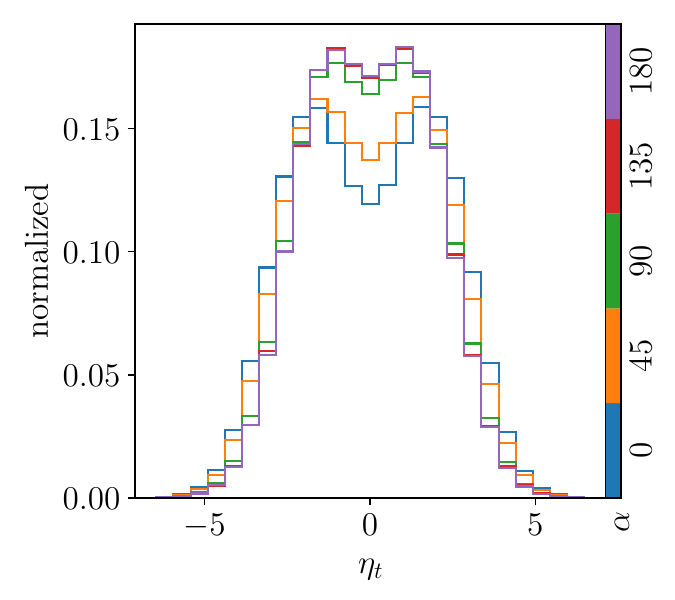}
  \includegraphics[width=0.325\textwidth,page=12]{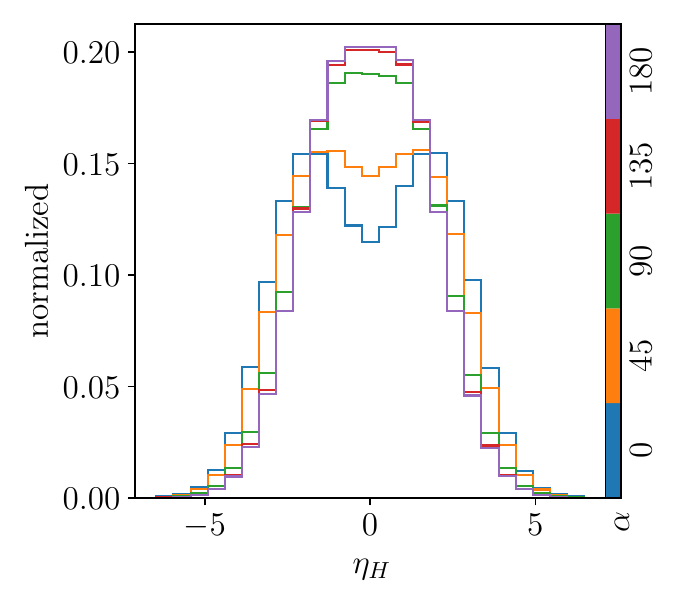}
  \includegraphics[width=0.325\textwidth,page=18]{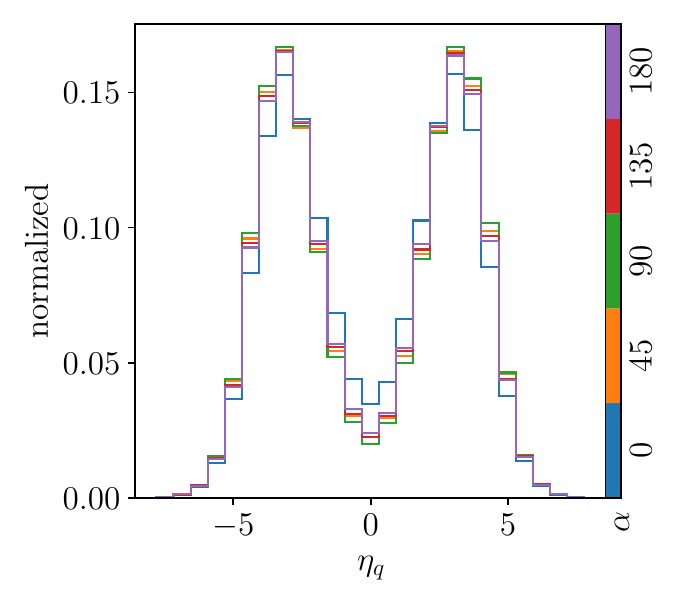}\\
  \includegraphics[width=0.325\textwidth,page=21]{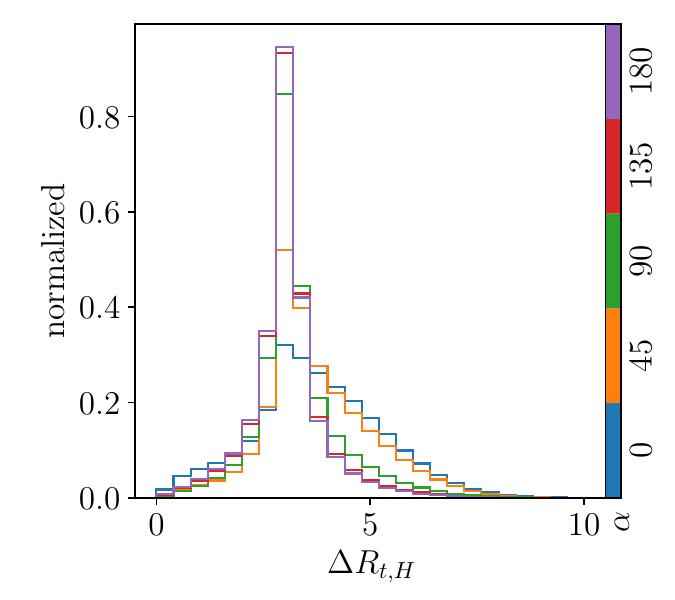}
  \includegraphics[width=0.325\textwidth,page=24]{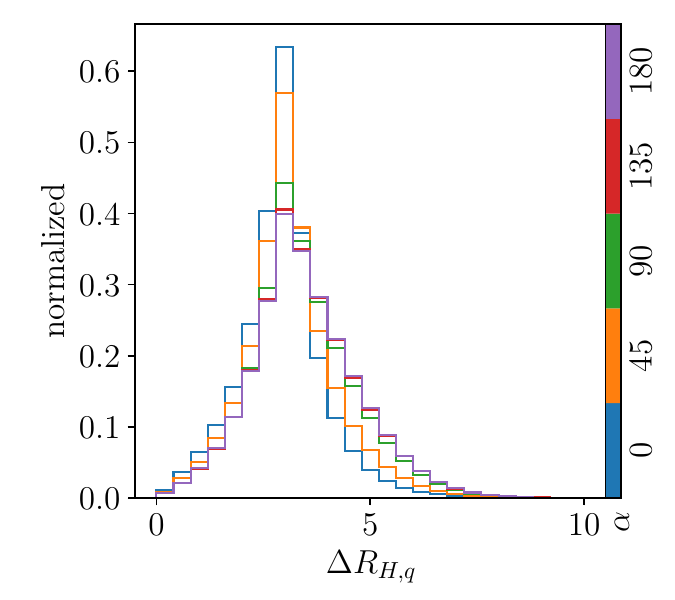}
  \includegraphics[width=0.325\textwidth,page=27]{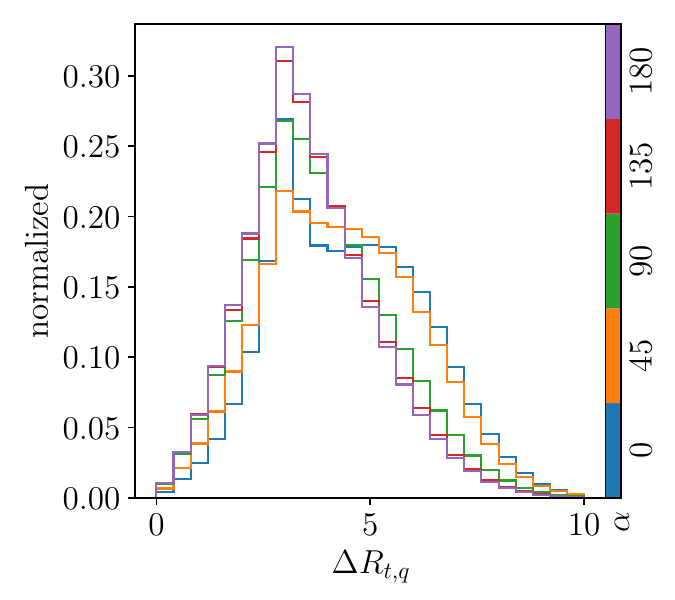}
\caption{Kinematic distributions for the hard-scattering $tHj$ final
    state for different values of the CP-angle.}
\label{fig:distris}
\end{figure}

We show the distributions for the hard-scattering $tHj$ kinematics in
Fig.~\ref{fig:distris}.  Again, the change in the kinematics is
driven by the interference between the leading top Yukawa contribution
and the subleading gauge contribution. For $p_{T,t}$ and $p_{T,H}$,
large phases lead to a harder transverse recoil of the heavy particles
and a less central scattering process in rapidity. In contrast, in the
Standard Model the two leading Feynman diagrams cancel in the central
region. In the angular separation $\Delta R_{tj}$ a second maximum
with a large angular separation vanishes when we switch from
destructive to constructive interference. Unlike for the total rate we
see that changing $\alpha$ from zero to $45^\circ$ leads to visible
kinematic shifts.

\subsubsection*{Data samples}

As mentioned above, we will work with three different datasets,
consisting of paired events at the hard-scattering level and after
parton shower, detector simulation, and reconstruction. All samples
share the same format for the hard scattering, including the CP-angle
$\alpha$ and the 4-momenta $\{ p_t, p_H, p_q \}$. For the leptonic top
decay, Eq.\eqref{eq:proc_lept}, the reco-level events are described by
the 4-momenta $\{ p_{\gamma,1}, p_{\gamma,2}, p_b, p_\mu, p_{j,1} \}$.
While we could use established methods to reconstruct the neutrino
momentum from the missing transverse energy, we do not include it as an
additional input to the network. Since it is a deterministic function
of the other momenta, it would not increase the number of degrees of
freedom and instead, only make the training more challenging.
Additional light jets can come from final state radiation, and the
photon and jet momenta are ordered in $p_T$. We do not allow for
initial state radiation, to simplify the problem as much as possible.
We train our networks on 1.3M paired events with $\alpha$ drawn from a
uniform distribution. The test dataset consists of 260k events for
each of the angles $\alpha \in \{0^\circ, 45^\circ, 90^\circ,
135^\circ, 180^\circ\}$.

The other two datasets assume a hadronic top decay, without ISR
(Eq.\eqref{eq:proc_had}) and with ISR (Eq.\eqref{eq:proc_isr}). The
reco-level event format includes $\{ p_{\gamma,1}, p_{\gamma,2}, p_b,
p_{j,1}, p_{j,2}, p_{j,3} \}$, plus potential additional light jet
momenta. For the hadronic decays, the network has to learn which of
the jets comes from the hard scattering. This problem gets more
challenging when we include ISR, because the additional jets can
lead to faulty or incomplete reconstruction. Hence, we
lose the clear correspondence between parton-level and reconstructed
jets in the high-multiplicity events. We extract the hard-scattering
momenta before we add ISR in our simulation chain, so our method does
not require a boost into the hard-scattering rest frame. In the ISR case,
we increase the number of training events to 3.4M.

\section{ML-MEM}
\label{sec:mem}

For our signal-only toy example, the matrix element method tracks the
dependence of the hard scattering cross section on one continuous
parameter of interest, the CP-phase $\alpha$ appearing in the
Lagrangian of Eq.\eqref{eq:LttH}. We denote the hard-scattering
momenta at parton level as $x_\text{hard}$ and split the differential
cross section into a total cross section and a probability density,
\begin{align}
  \frac{\D\sigma(\alpha)}{\D x_\text{hard}} = \sigma(\alpha) \; p(x_\text{hard}|\alpha) \; .
  \label{eq:dcsparton}
\end{align}
The likelihood for a single hard-scattering event $x_\text{hard}$ to
correspond to a given value for $\alpha$ is then
\begin{align}
 p(x_\text{hard}|\alpha) = \frac{1}{\sigma(\alpha)} \;
 \frac{\D\sigma(\alpha)}{\D x_\text{hard}} \; .
\end{align}
Next, we consider the effects of parton shower, hadronization,
detector, and reconstruction.  The corresponding forward simulation
maps $x_\text{hard}$ to a reco-level configuration $x_\text{reco}$,
provided it passes the cuts. In the forward direction
$p(x_\text{reco}|x_\text{hard})$ is the conditional probability for a
reco-level event $x_\text{reco}$, given $x_\text{hard}$ at parton
level.  In general, this conditional probability depends on our
parameter of interest, $p(x_\text{reco}|x_\text{hard},\alpha)$, so we
can use it to write the likelihood linking a single reco-level event
$x_\text{reco}$ to the parameter $\alpha$ as
\begin{align}
  p(x_\text{reco}|\alpha)
  &= \int \D x_\text{hard}\; p(x_\text{hard}|\alpha) \; p(x_\text{reco}|x_\text{hard},\alpha) \notag \\
  &= \frac{1}{\sigma(\alpha)} \int \D x_\text{hard}\; \frac{\D\sigma(\alpha)}{\D x_\text{hard}} \; p(x_\text{reco}|x_\text{hard},\alpha) \; .
  \label{eq:likeli_eff} 
\end{align}
The conditional probability $p(x_\text{reco}|x_\text{hard},\alpha)$
corresponds to the usual transfer function, which can sometimes be
approximated by a Gaussian. In general, it is only defined implicitly
through a complex forward simulation.
Using the single-event likelihoods at the reco-level we can compute
the likelihood for an event sample as a function of the parameter
of interest,
\begin{align}
  L(\alpha)
  & \approx \prod_{\text{events} \; j} p(x_{\text{reco}, j}|\alpha) \notag \\
  &= \prod_{\text{events} \; j}  \frac{1}{\sigma(\alpha)} \int \D x_\text{hard}\; \frac{\D\sigma(\alpha)}{\D x_\text{hard}} \; p(x_{\text{reco},j}|x_\text{hard},\alpha)   \; ,
\label{eq:likeli_set}
\end{align}
where we omit any prefactors related to the observed number of events~\cite{ml_notes}.

\subsubsection*{Transfer-cINN}

Because of the phase space cuts, the conditional probability in
Eq.\eqref{eq:likeli_eff} is not normalized to one. Instead, we can define
the acceptance rate $a(x_\text{hard},\alpha)$ to obtain
\begin{align}
    \int \D x_\text{reco}\; p(x_\text{reco}|x_\text{hard},\alpha) = a(x_\text{hard},\alpha) \; .
\end{align}
Alternatively, we can account for this efficiency by replacing the
full volume with the fiducial volume at the hard-scattering level.
Here we assume that there is a hard cut-off at the parton level
$x_\text{hard}$, even though we define our cuts at the detector-level.
This means we replace $a(x_\text{hard},\alpha) \ne
1$ by shifting $\sigma(\alpha) \to \sigma_\text{fid}(\alpha)$ in
Eq.\eqref{eq:likeli_eff},
\begin{align}
  p(x_\text{reco}|\alpha)
  &= \int_\text{fid} \D x_\text{hard}\; p(x_\text{hard}|\alpha) \; p(x_\text{reco}|x_\text{hard},\alpha) \notag \\
  &= \frac{1}{\sigma_\text{fid}(\alpha)} \int_\text{fid} \D x_\text{hard}\; \frac{\D\sigma(\alpha)}{\D x_\text{hard}}  \; p(x_\text{reco}|x_\text{hard},\alpha) \; .
  \label{eq:likeli_formula}
\end{align}
In this integral, the differential cross section is available
numerically and $p(x_\text{reco}|x_\text{hard},\alpha)$ can be encoded
in a neural network. Normally, this would be a regression task, but in
our case we do not have the explicit training data to train a
regression network. Instead, we train a normalizing flow, specifically
a conditional cINN, to reproduce the reco-level kinematics for a given
hard-scattering event from Gaussian random numbers
\begin{align}
  r \sim p(r) \quad \longleftrightarrow \quad x_\text{reco} \sim p(x_\text{reco}|x_\text{hard},\alpha) 
  \qqquad \text{Transfer-cINN.}
  \label{eq:transfer_cinn}
\end{align}
In the inverse direction, this network estimates and encodes the
conditional density over the reco-level phase space, and in the
forward direction, it is nothing but a fast detector simulation
generating reco-level events.  We will see that for our purpose and
implementation we can ignore the $\alpha$-dependence of
$p(x_\text{reco}|x_\text{hard},\alpha)$. The best way to train the
network is to use data with variable $\alpha$ and ignore this
input. Such a training improves the phase space coverage even for
extreme values of $\alpha$ and provides an averaging over any
remaining $\alpha$-dependence.

\subsubsection*{Unfolding-cINN}

\begin{figure}[b!]
\centering
\usetikzlibrary{arrows, shapes.geometric, arrows.meta, shapes, decorations.pathreplacing, fit}

\definecolor{Rcolor}{HTML}{E99595}
\definecolor{Gcolor}{HTML}{C5E0B4}
\definecolor{Bcolor}{HTML}{9DC3E6}
\definecolor{Ycolor}{HTML}{FFE699}

\tikzstyle{expr} = [circle, minimum width=1.8cm, minimum height=1.8cm, text centered, align=center, inner sep=0, fill=Ycolor, draw]
\tikzstyle{txt_huge} = [align=center, font=\Huge, scale=2]
\tikzstyle{txt} = [align=center, font=\LARGE]
\tikzstyle{cinn} = [double arrow, double arrow head extend=0cm, double arrow tip angle=130, shape border rotate=90, inner sep=0, align=center, minimum width=2.1cm, minimum height=2.3cm, fill=Gcolor, draw]
\tikzstyle{cinn_black} = [cinn, minimum height=2.5cm, fill=black]
\tikzstyle{arrow} = [thick,-{Latex[scale=1.0]}, line width=0.2mm, color=black]
\tikzstyle{line} = [thick, line width=0.2mm, color=black]

\begin{tikzpicture}[node distance=2cm, scale=0.65, every node/.style={transform shape}]

\node (likeli) [txt] {$p(x_\text{reco}|\alpha) = $};
\node (sigfid) [expr, right of=likeli, xshift = 0.7cm] {\LARGE $\frac{1}{\sigma_\text{fid}}$};
\node (langle) [txt_huge, right of=sigfid, xshift = -1.25cm] {$\langle$};
    \node (jac) [expr, right of=langle, xshift = -0.6cm] {\LARGE $\frac{\partial x_\text{hard}}{p(r)\, \partial r}$};
\node (dcs) [expr, right of=jac] {\LARGE $\frac{\mathrm{d} \sigma}{\mathrm{d} x_\text{hard}}$};
\node (dens) [expr, right of=dcs] {$p(x_\text{reco}|$\\$x_\text{hard})$};
\node (langle) [txt_huge, right of=dens, xshift = -1.25cm] {$\rangle$};

\node (unfcinn_b) [cinn_black, above of=jac, yshift=2.6cm, xshift=0.2cm] {};
\node (unfcinn) [cinn, above of=jac, yshift=2.6cm, xshift=0.2cm, fill=Bcolor] {Unfolding\\cINN};
\node (tracinn_b) [cinn_black, above of=dens, yshift=1cm] {};
\node (tracinn) [cinn, above of=dens, yshift=1cm] {Transfer\\cINN};

\node (alpha) [txt, above of=likeli, yshift=2.4cm] {$\alpha$};
\node (xr) [txt, above of=alpha, yshift=1.4cm] {$x_\text{reco}$};
\node (random) [txt, above of=unfcinn, yshift=0.3cm] {\normalsize$\{r\}_{r \sim p(r)}$};

\draw [arrow, color=black] (tracinn_b.south) -- (dens.north);
\draw [arrow, color=black] ([xshift=-0.2cm, yshift=0.1cm]unfcinn_b.south) -- (jac.north);
    \draw [arrow, color=black] ([xshift=0.2cm, yshift=0.1cm]unfcinn_b.south) -- ([xshift=0.2cm]unfcinn_b.south |- tracinn_b.west) -- (tracinn_b.west) node[midway,above,xshift=0.2cm]{$\{x_\text{hard}\}$};
\draw [arrow, color=black] (random.south) -- (unfcinn_b.north);
\draw [arrow, color=black] (alpha.east) -- ([yshift=-0.2cm]unfcinn_b.west);
\draw [arrow, color=black] (xr.east) -- (tracinn_b.north |- xr.east) -- (tracinn_b.north);
\draw [arrow, color=black] (alpha.east -| sigfid.north) -- (sigfid.north);
\draw [arrow, color=black] (xr.east -| sigfid.north) -- ([yshift=0.2cm]unfcinn_b.west -| sigfid.north) -- ([yshift=0.2cm]unfcinn_b.west);
\draw [arrow, color=black] ([xshift=0.2cm]dcs.north|- tracinn_b.west) -- ([xshift=0.2cm]dcs.north);
\draw [line, color=black] ([yshift=-1.0cm]sigfid.north|- tracinn_b.west) -- ([xshift=-0.35cm, yshift=-1.0cm]unfcinn_b.south|- tracinn_b.west);
\draw [arrow, color=black] ([xshift=-0.05cm, yshift=-1.0cm]unfcinn_b.south|- tracinn_b.west) -- ([xshift=-0.2cm, yshift=-1.0cm]dcs.north|- tracinn_b.west) -- ([xshift=-0.2cm]dcs.north);

\end{tikzpicture}
\vspace{-2em}
\caption{Dual-cINN setup of the MEM integrator evaluating
  Eq.\eqref{eq:likeli_integral} through sampling $r$. The
  Unfolding-cINN is conditioned on the CP-angle $\alpha$ and
  the reco-level event $x_\text{reco}$. The Transfer-cINN is conditioned
  on the hard-scattering event $x_\text{hard}$.}
\label{fig:mem_setup}
\end{figure}
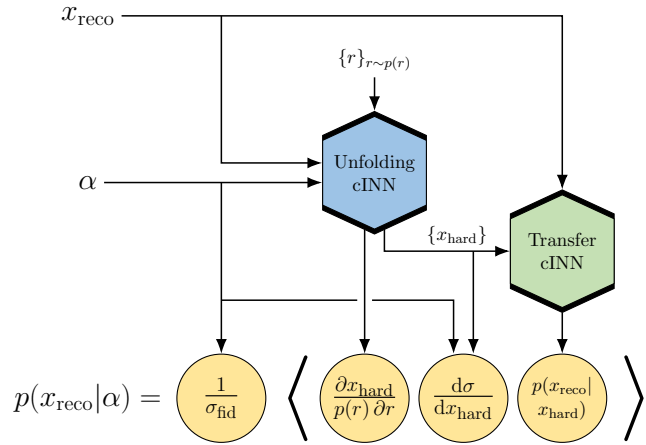

Even with a fast surrogate integrand, the integral in
Eq.\eqref{eq:likeli_formula} is numerically challenging, because the
squared matrix element spans several orders of magnitude and
$p(x_\text{reco}|x_\text{hard},\alpha)$ drops rapidly away from the
trivial mapping of the intermediate on-shell particles and hard
partons turning into single jets.  We can define an appropriate
mapping $x_\text{hard} \to q(x_\text{hard})$, or sampling of
$x_\text{hard} \sim q(x_\text{hard})$, such that
Eq.\eqref{eq:likeli_formula} becomes
\begin{align}
  p(x_\text{reco}|\alpha)
  &= \int_\text{fid} \D x_\text{hard} \; p(x_\text{hard}|\alpha) \; p(x_\text{reco}|x_\text{hard},\alpha)  \notag \\
  &= \XXLangle \frac{1}{q(x_\text{hard})} \; p(x_\text{hard}|\alpha) \; p(x_\text{reco}|x_\text{hard},\alpha)
     \XXRangle_{x_\text{hard} \sim q(x_\text{hard})} \\
  &= \Langle p(x_\text{reco}|\alpha) \Rangle_{x_\text{hard} \sim q(x_\text{hard})}
     \qquad \Leftrightarrow \qquad
     x_\text{hard}\sim q(x_\text{hard}) = p(x_\text{hard}|x_\text{reco},\alpha) \; .
     \notag
\end{align}
The last line uses Bayes' theorem and means our
$x_\text{hard}$-integration would become trivial if we could sample
$x_\text{hard}$ from the distribution
$p(x_\text{hard}|x_\text{reco},\alpha)$.

To sample the hard-scattering momenta $x_\text{hard}$ following such a
distribution we again train a conditional INN, mapping random numbers
with a latent distribution $p(r)$ to the target distribution in
momentum space,
\begin{align}
  r \sim p(r) \quad \longleftrightarrow \quad x_\text{hard}(r) \sim p(x_\text{hard}|x_\text{reco},\alpha)
  \qqquad \text{Unfolding-cINN.} 
  \label{eq:unfolding_cinn}
\end{align}
It turns out that sampling $x_\text{hard}$ from
$p(x_\text{hard}|x_\text{reco},\alpha)$ defines the standard cINN used
for unfolding or Bayesian inference in
Refs.~\cite{cinn,Bellagente:2020piv,Bieringer:2020tnw}. A better modeling
of the distribution of $x_\text{hard}$ will lead to a more efficient
integration. In Eq.\eqref{eq:likeli_formula} the Unfolding-cINN transforms
the $x_\text{hard}$-integration into an $r$-integration. In the
corresponding Jacobian we have to account for the full conditional
dependence of $x_\text{hard}(r; x_\text{reco},\alpha)$,
\begin{align}
  p(x_\text{reco}|\alpha)
  &= \frac{1}{\sigma_\text{fid}(\alpha)}
  \int \D r\; 
  \frac{\partial x_\text{hard}(r; x_\text{reco},\alpha)}{\partial r} \; 
  \left[ \frac{\D\sigma(\alpha)}{\D x_\text{hard}} p(x_\text{reco}|x_\text{hard},\alpha) \right]_{x_\text{hard}(r; x_\text{reco},\alpha)} \notag \\
  &\equiv \frac{1}{\sigma_\text{fid}(\alpha)}
  \XXLangle
  \frac{1}{p(r)} \;
  \frac{\partial x_\text{hard}(r; x_\text{reco},\alpha)}{\partial r} \; 
  \left[ \frac{\D\sigma(\alpha)}{\D x_\text{hard}} p(x_\text{reco}|x_\text{hard},\alpha) \right]_{x_\text{hard}(r; x_\text{reco},\alpha)}
  \XXRangle_{r \sim p(r)} \; .
  \label{eq:likeli_integral}
\end{align}
The dual-network architecture of our MEM integrator is illustrated in
Fig.~\ref{fig:mem_setup}.

\subsubsection*{Network architecture}

Both cINNs are built as a sequence of rational quadratic spline
coupling blocks~\cite{durkan2019neural}, each followed by a random
rotation matrix.  The spline coupling blocks implement a mapping
between hypercubes. To make them compatible with a Gaussian latent
space and the rotation matrices, we set their bounds to
$[-10,10]$. This ensures that the points passed through the network
are sufficiently far from the spline boundaries, after applying a
standard scaling to the training data. For a cINN that maps a batch of
$B$ data points $x_i$ to points $r_i$ in a Gaussian latent space, given
a condition $c_i$, the loss function is~\cite{cinn,ml_notes}
\begin{align}
    \loss_\text{cINN} = \sum_{i=1}^B \left( \frac{r_i(x_i;c_i)^2}{2}
    - \log \left| \frac{\partial r_i(x_i;c_i)}{\partial x_i} \right| \right) \; .
\end{align}
For the two networks we identify
\begin{alignat}{8}
  x &= x_\text{hard} &\qqquad
  c &= (x_\text{reco}, \alpha) &\qqquad
  \text{Unfolding-cINN,} \notag \\
  x &= x_\text{reco} &\qqquad
  c &= x_\text{hard} &\qqquad 
  \text{Transfer-cINN.} \notag 
\end{alignat}
The networks are implemented in \pytorch~\cite{pytorch}.  We use the
\adam~\cite{adam} optimizer with a one-cycle learning rate
scheduling~\cite{one-cycle-lr}.  After tuning the hyper-parameters of
the Unfolding-cINN, we found that the same setup and hyper-parameters
also yield good results for the Transfer-cINN.  The network
hyper-parameters are given in Tab.~\ref{tab:cinn}.

\begin{table}[b!]
\centering
\begin{small} \begin{tabular}{l|c}
\toprule
    Parameter & cINN \\
\midrule
Blocks & 20 \\
Block type & Rational quadratic \\
Layers per block & 5 \\
Units per layer & 256 \\
Spline bins & 30 \\
Epochs (Bayesian) & 100 (200) \\
Learning rate scheduling & One-cycle \\
Initial learning rate & $1 \cdot 10^{-4}$ \\
Maximum learning rate & $3 \cdot 10^{-4}$ \\
Batch size & 1024 \\
Training events & 1.3M \\
\bottomrule
\end{tabular} \end{small}
\caption{Identical setup and hyper-parameters for the Transfer-cINN
  and the Unfolding-cINN.}
\label{tab:cinn}
\end{table}

\subsubsection*{Uncertainty-aware cINN}

Bayesian neural networks allow us to test the training stability and
to estimate uncertainties on the network output.  They take the
architecture of standard regression, classification, or generative
networks and allow the individual network weights to fluctuate. The
uncertainty on the network output is then estimated by sampling from
the weight
distributions~\cite{bnn_early,bnn_early2,bnn_early3,deep_errors,Bollweg:2019skg,Kasieczka:2020vlh}. For
generative networks this concept has been applied to normalizing flows
or INNs~\cite{Bellagente:2021yyh,Butter:2021csz}. Here the network
encodes a density over phase space and the uncertainty on this density
over the same phase space. For more details on the Bayesian cINN we
refer to the original papers~\cite{Bellagente:2021yyh,Butter:2021csz}
and the lecture notes of Ref.~\cite{ml_notes}. By construction,
Bayesian networks include an L2-regularization, so with limited extra
numerical effort Bayesian networks deliver the same level of
performance as their deterministic counterparts.

Because the Unfolding-cINN is only used to improve the importance
sampling for the numerical integration, its uncertainty is irrelevant
for the actual integral, so we do not generalize it to a Bayesian
version for our final application. However, we will use a Bayesian
Unfolding-cINN to estimate its uncertainties and confirm its
reliability.

In contrast, the Transfer-cINN encodes the reco-level phase space
density, which means we can use a Bayesian cINN to estimate the
uncertainty of this learned density. Whenever we show results for this
density, we also indicate the corresponding uncertainty from the
network training. For tricky applications like the MEM this additional
check on the network performance is crucial.

\section{Performance}
\label{sec:perf}

To illustrate the advantages and the remaining challenges of a
ML-realization of the matrix element method, we show results for the
associated Higgs and single-top production. We only consider signal
events, because of the especially challenging Higgs mass pole. We test
the two cINNs independently, including an uncertainty analysis through
a Bayesian network setup.

\subsection{Leptonic top decay}
\label{sec:perf_lept}

\begin{figure}[t]
\includegraphics[width=0.49\textwidth,page=4]{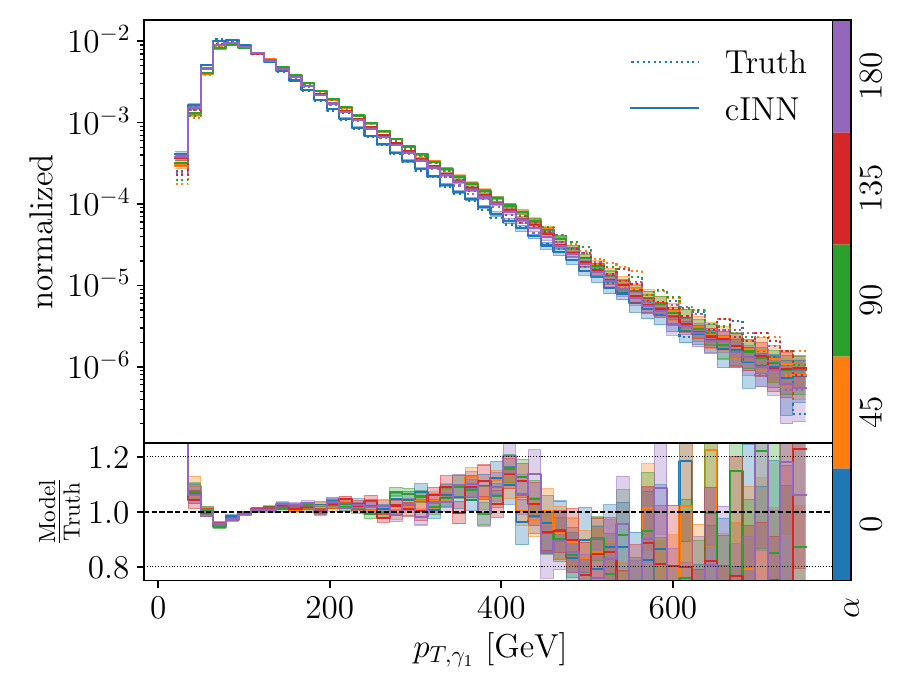}
\includegraphics[width=0.49\textwidth,page=6]{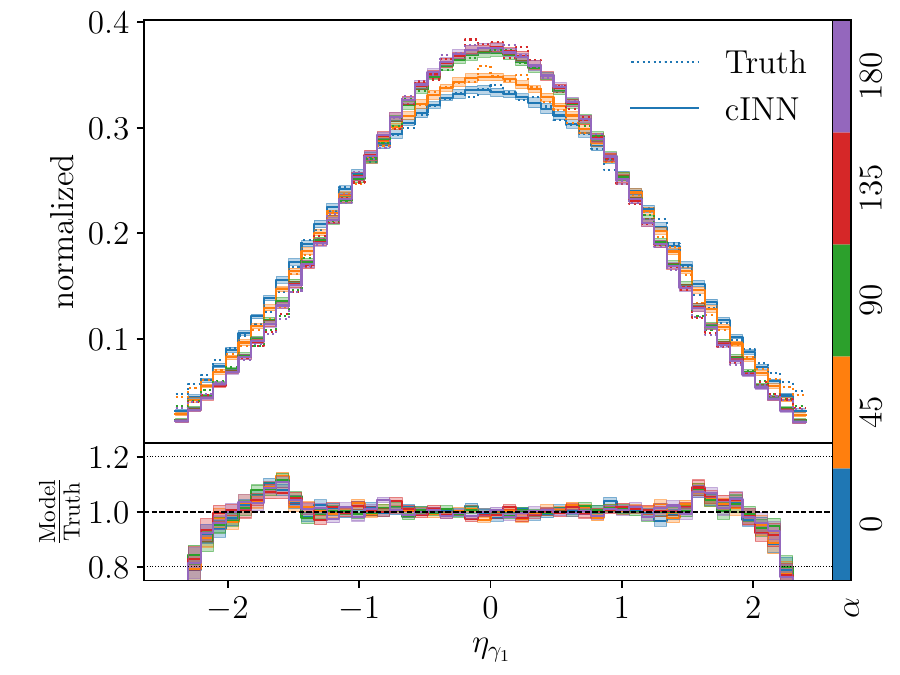}\\
\includegraphics[width=0.49\textwidth,page=22]{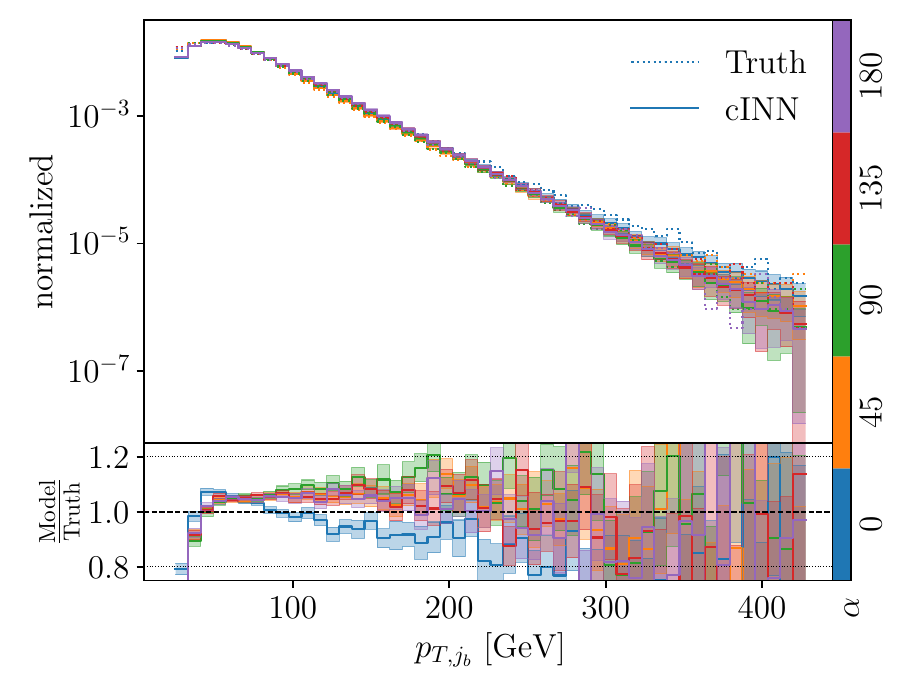}
\includegraphics[width=0.49\textwidth,page=28]{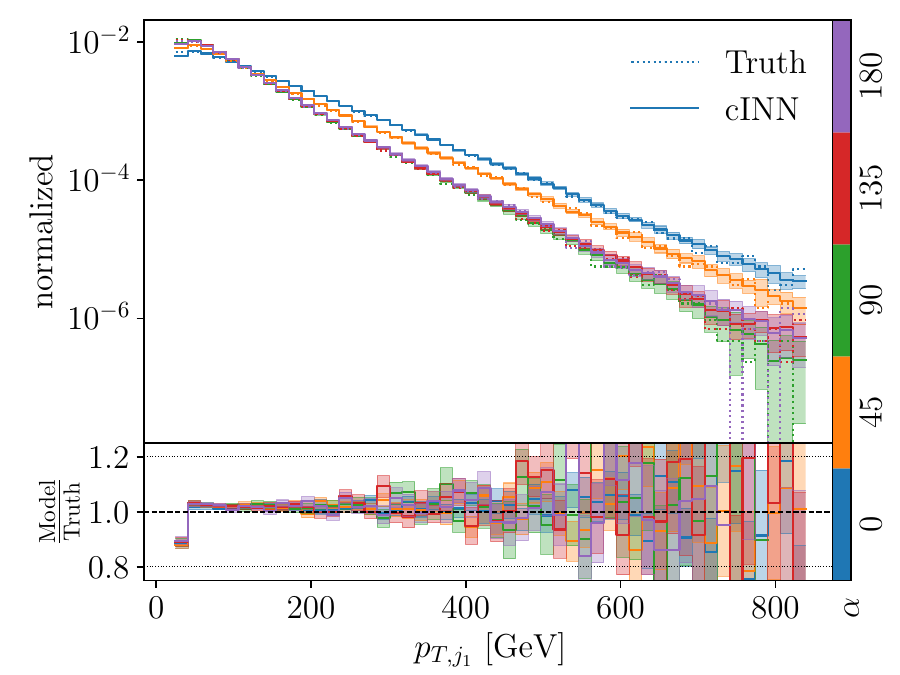}
\caption{Forward-simulated kinematic distributions for the leptonic
  top decay, assuming five different CP-angles and including
  uncertainties from the Bayesian cINN. These distributions test the
  Transfer-cINN.}
\label{fig:histograms_forward_leptonic_bayesian}
\end{figure}

The first results we discuss in detail are for the leptonic top decay,
as defined in Eq.\eqref{eq:proc_lept}.  We start with a test of the
Transfer-cINN from Eq.\eqref{eq:transfer_cinn} in the forward
direction.  As mentioned above, the network generates 4-momenta of
five final state particles at reco-level, including one light
jet. Assuming these particles to be produced on their mass shell we
remove the photon and muon energies from the network's degrees of
freedom, leaving us with a phase space dimensionality of $5 \cdot 4 -
3 = 17$. The forward generation is conditioned on the corresponding
set of three hard-scattering momenta, all of them assumed to be
on-shell.  In all cases we implement a standard scaling, including a
subtraction of the mean values and a normalization to standard
deviation one. All hyper-parameters are shown in Tab.~\ref{tab:cinn}.

The kinematic distributions from the network evaluated in the forward
direction are shown in
Fig.~\ref{fig:histograms_forward_leptonic_bayesian}, compared to the
true reco-level distributions from the training dataset. One
assumption we can test is that the Transfer-cINN does not have to be
conditioned on $\alpha$, which means that this detail of the
underlying model is numerically irrelevant. To make such a statement
we need the uncertainties of the network prediction as a reference
measure. The reliability of the network is best seen in the lower
panels, where we see that deviations from the true distributions
appear in the tails of the distributions.  The uncertainty estimate is
reliable in the bulk of all distributions, reflects the increased
uncertainty in the $p_T$-tails, and covers the rapidly dropping
rapidity distributions less well. Looking at this uncertainty we
confirm that the distributions differ for varying $\alpha$, but this
variation is explained entirely by the effects on the hard-scattering
distributions, combined with an $\alpha$-independent Transfer-cINN.

\begin{figure}[t]
\includegraphics[width=0.49\textwidth,page=4]{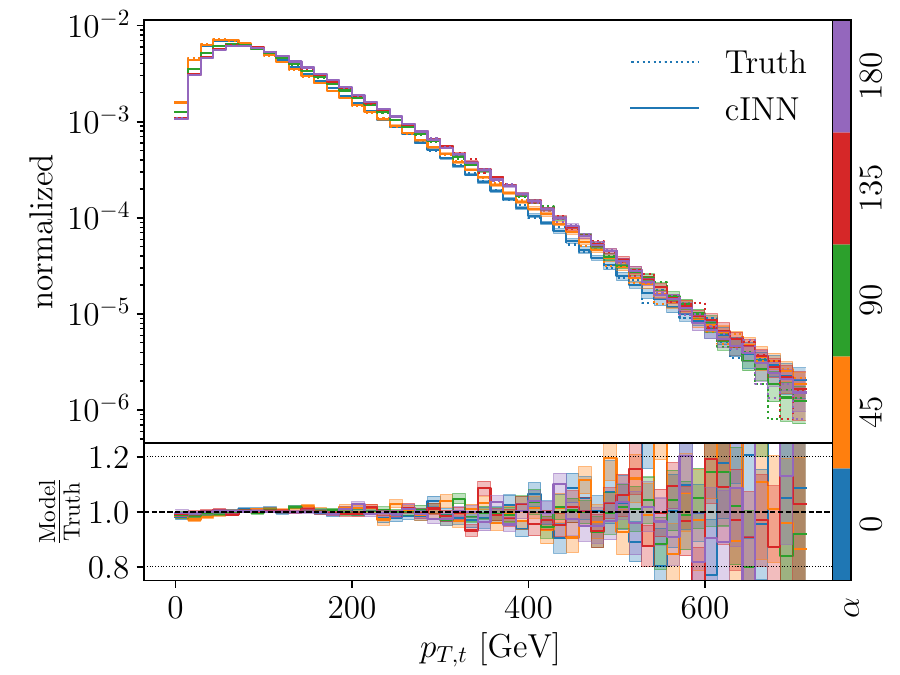}
\includegraphics[width=0.49\textwidth,page=6]{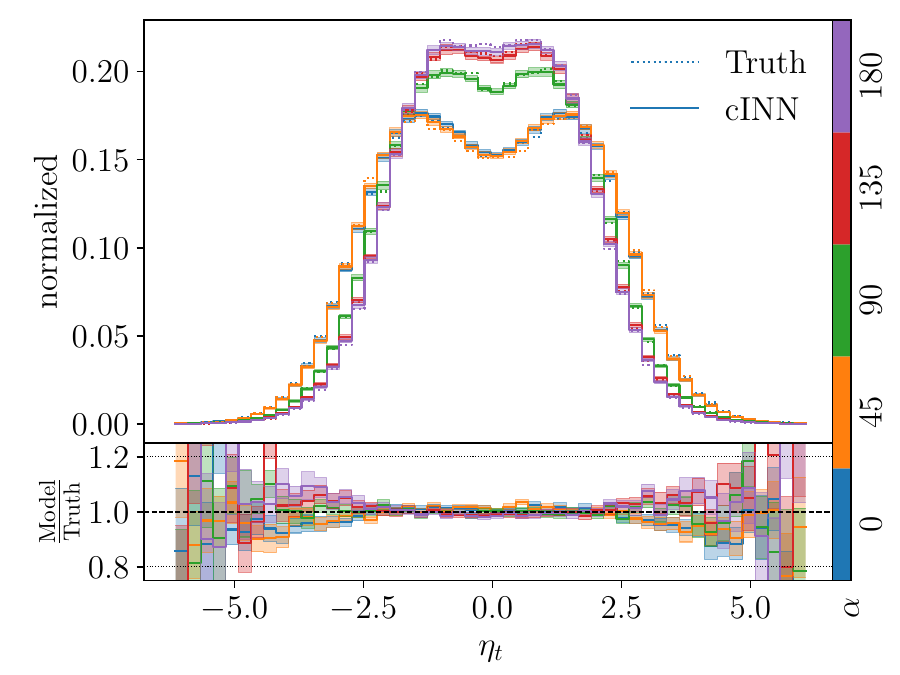}\\
\includegraphics[width=0.49\textwidth,page=12]{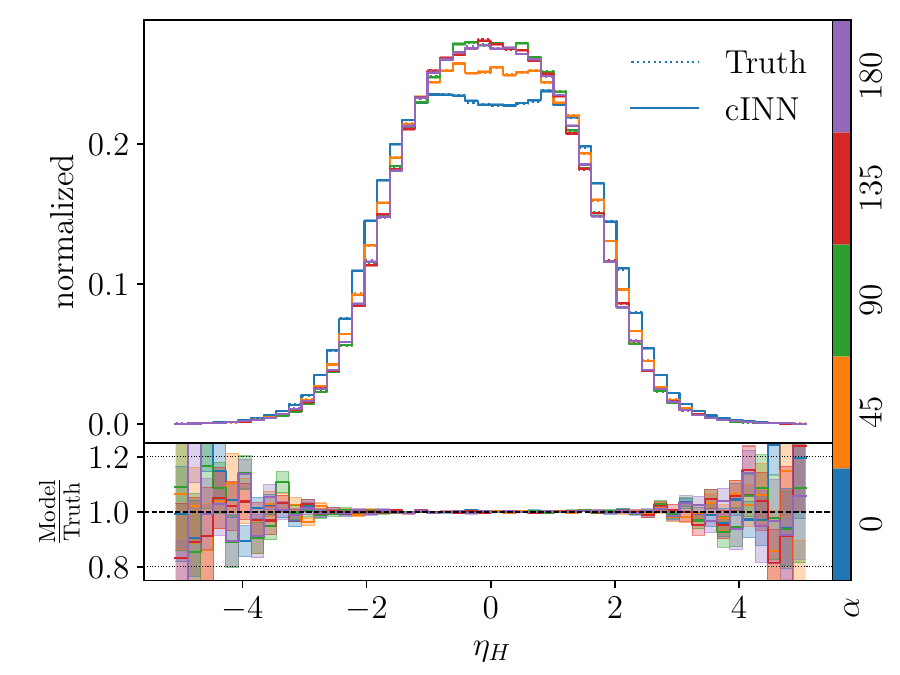}
\includegraphics[width=0.49\textwidth,page=18]{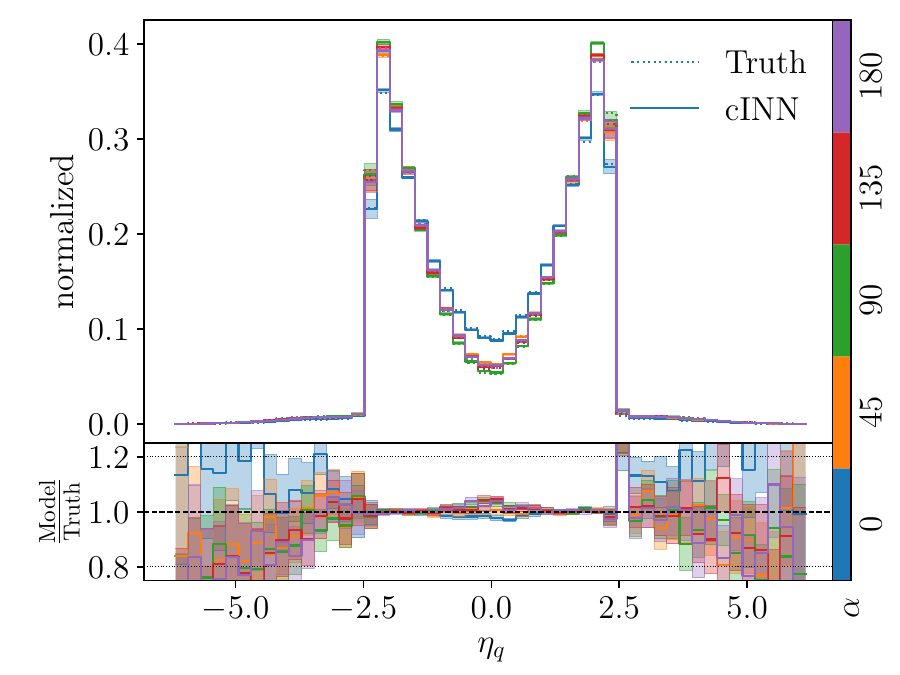}
\caption{Unfolded kinematic distributions for the leptonic top decay,
  assuming five different CP-angles and including uncertainties
  from the Bayesian cINN. These distributions test the
  Unfolding-cINN.}
\label{fig:histograms_inverse_leptonic}
\end{figure}

Next, we test the Unfolding-cINN, which we will use to improve the
numerical integration. The three generated momenta are defined at the
parton level, all particles are on-shell, and we assume momentum
conservation in the azimuthal plane. The corresponding 7-dimensional
phase space is spanned by the coordinates $(\vec{p}_t, \vec{p}_h,
p_q^z)$. The conditional input is the reco-level phase space, where we
allow for up to four additional jets, \ie altogether up to nine
4-momenta zero-padded. In addition, we condition on the angle
$\alpha$. As for the Transfer-cINN we implement a standard scaling for
all data. The results for the unfolding, with uncertainties, are shown
in Fig.~\ref{fig:histograms_inverse_leptonic}. Again, we see that the
network reproduces all features, including the $\alpha$-dependence,
and remaining differences between the cINN-unfolded and truth events
are covered by the network uncertainties.

\begin{figure}[t]
\includegraphics[width=0.32\textwidth]{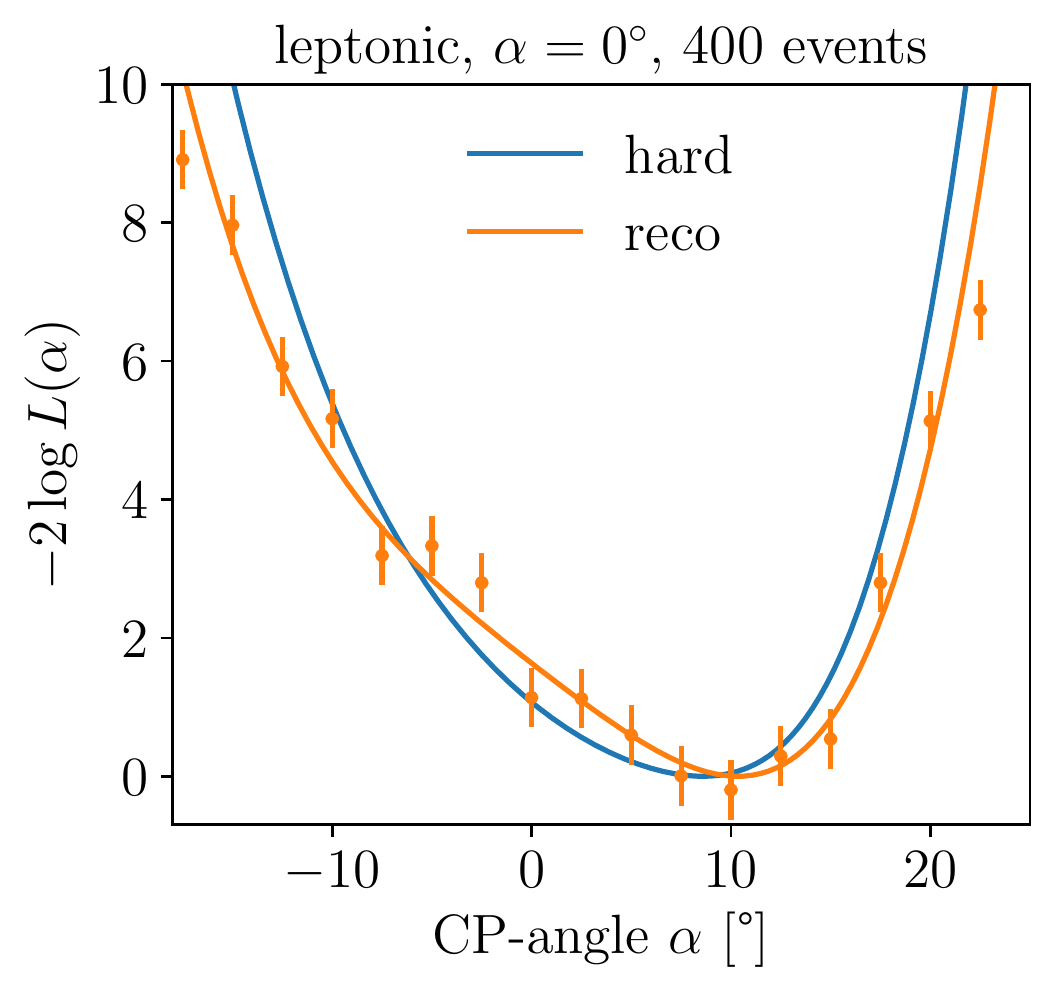}
\includegraphics[width=0.32\textwidth]{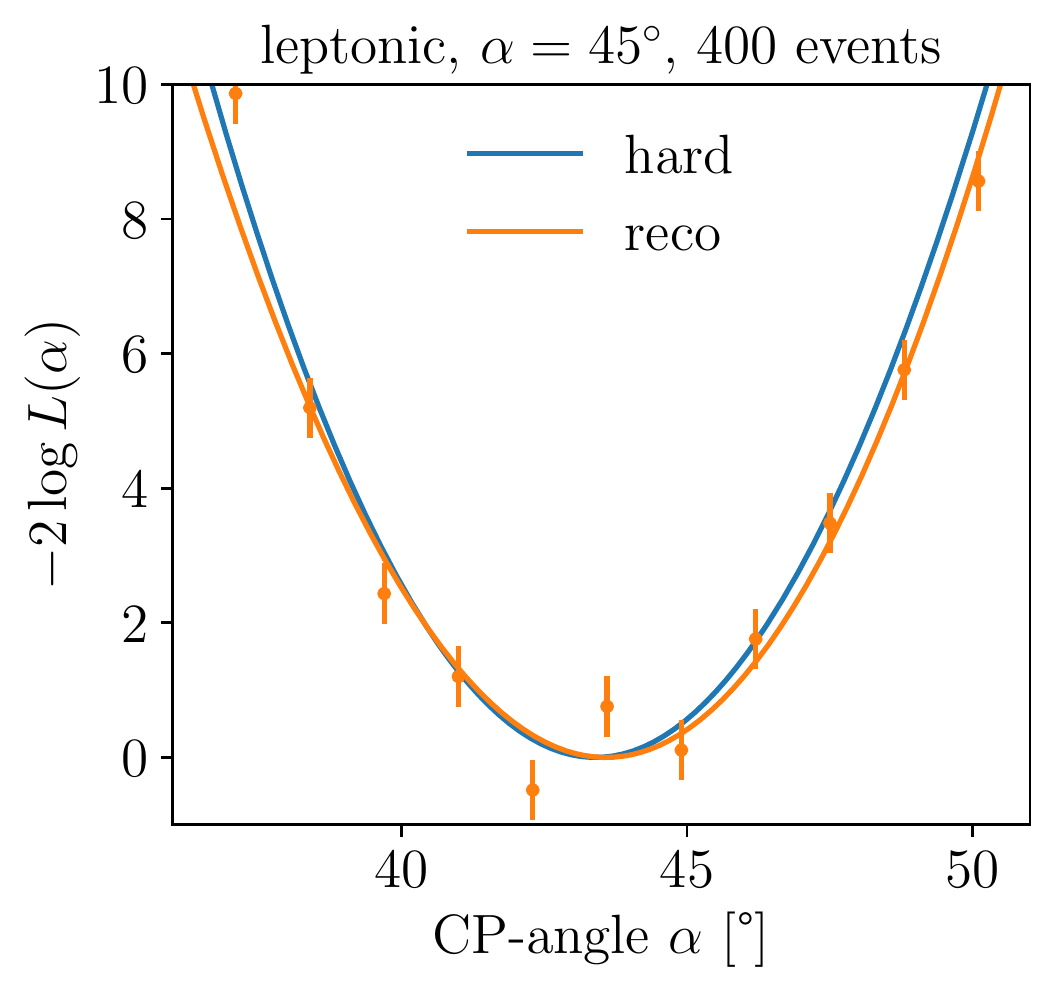}
\includegraphics[width=0.32\textwidth]{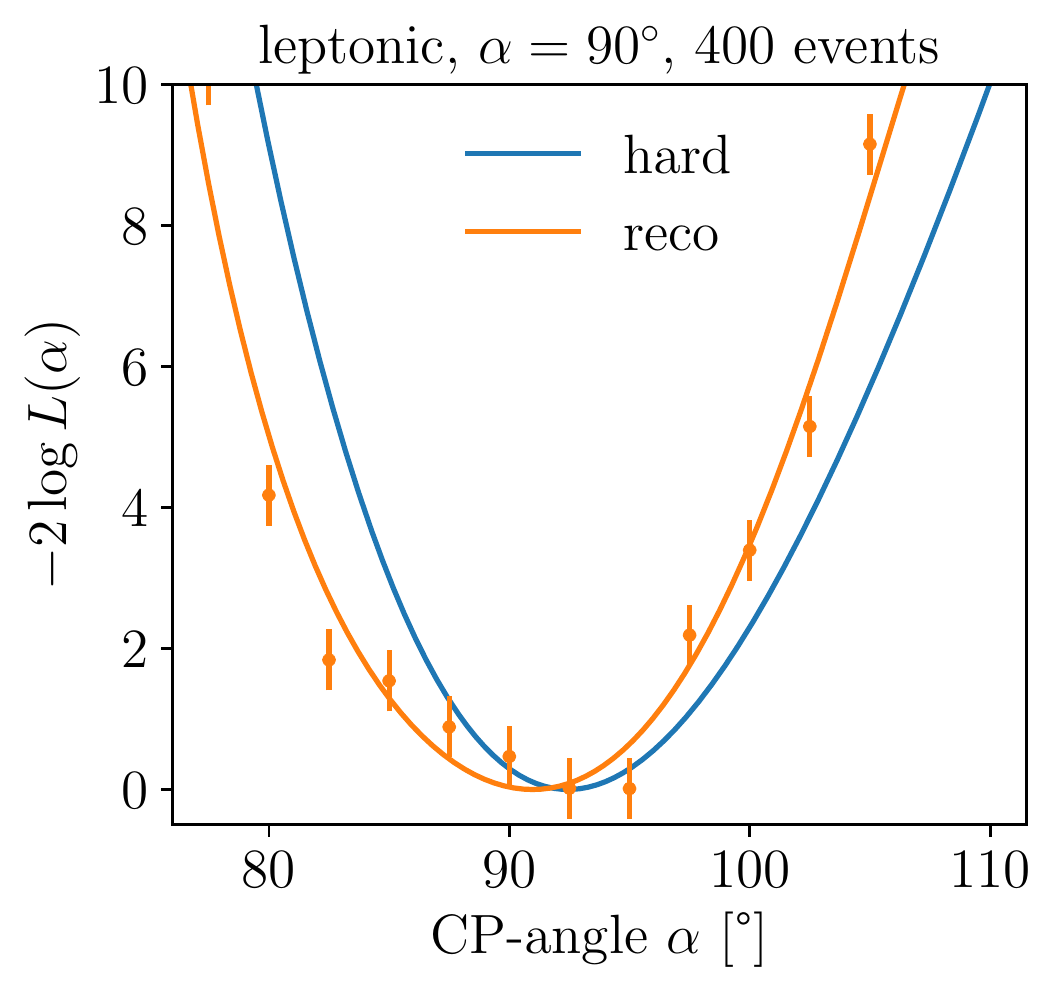} \\
\includegraphics[width=0.32\textwidth]{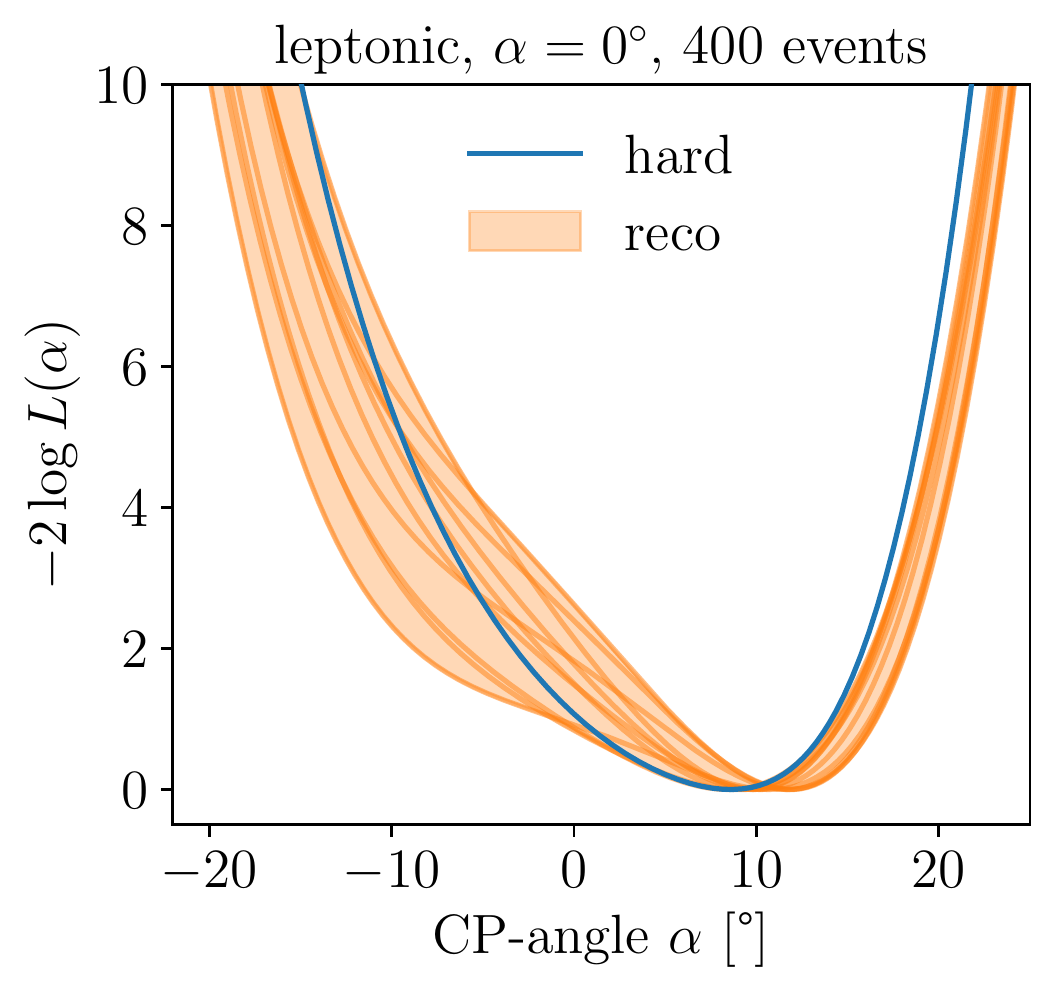}
\includegraphics[width=0.32\textwidth]{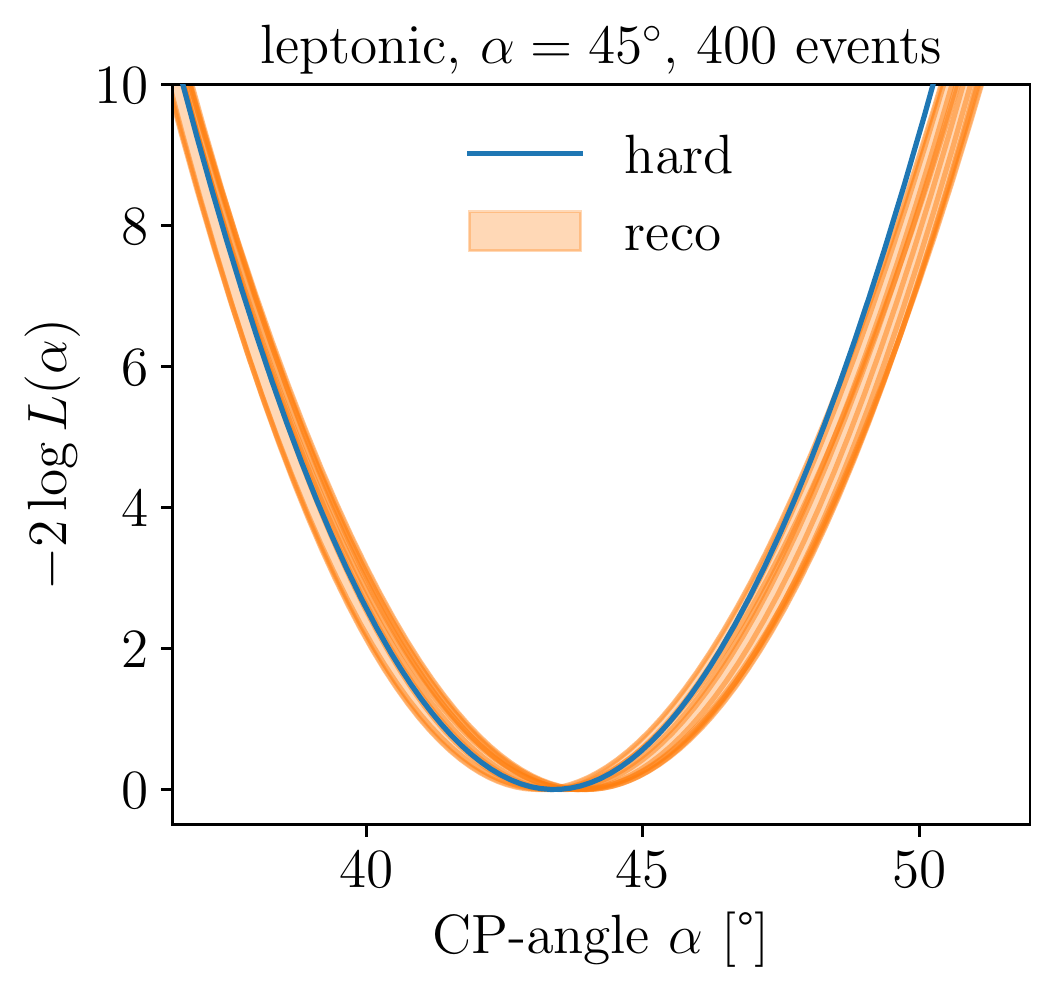}
\includegraphics[width=0.32\textwidth]{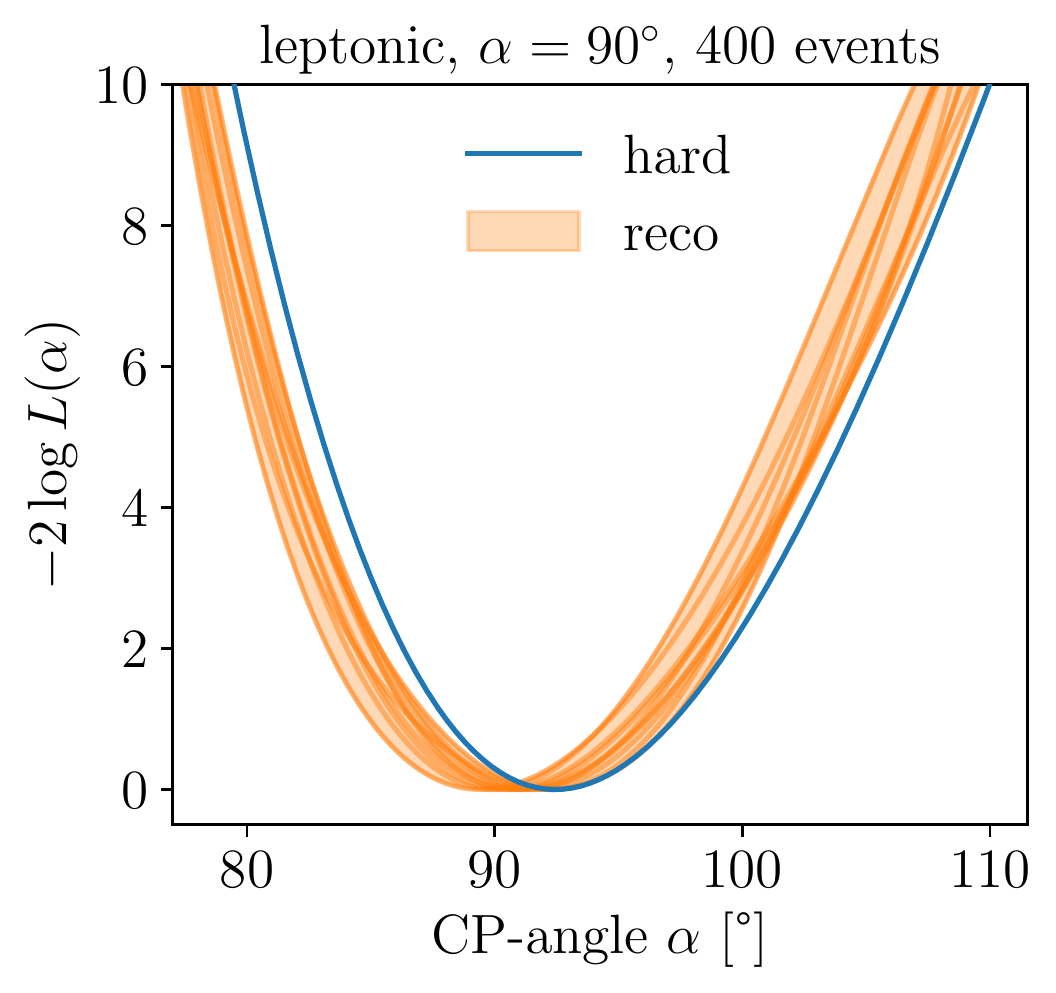}
\caption{Likelihoods for the leptonic top decay as a function of the
  CP-angle $\alpha$, extracted from 400 events for three assumed truth
  angles. For the Bayesian uncertainties we show the integrated
  likelihoods from 10 sampled networks.}
\label{fig:likeli400_leptonic}
\end{figure}

\begin{figure}[b!]
\includegraphics[width=0.32\textwidth]{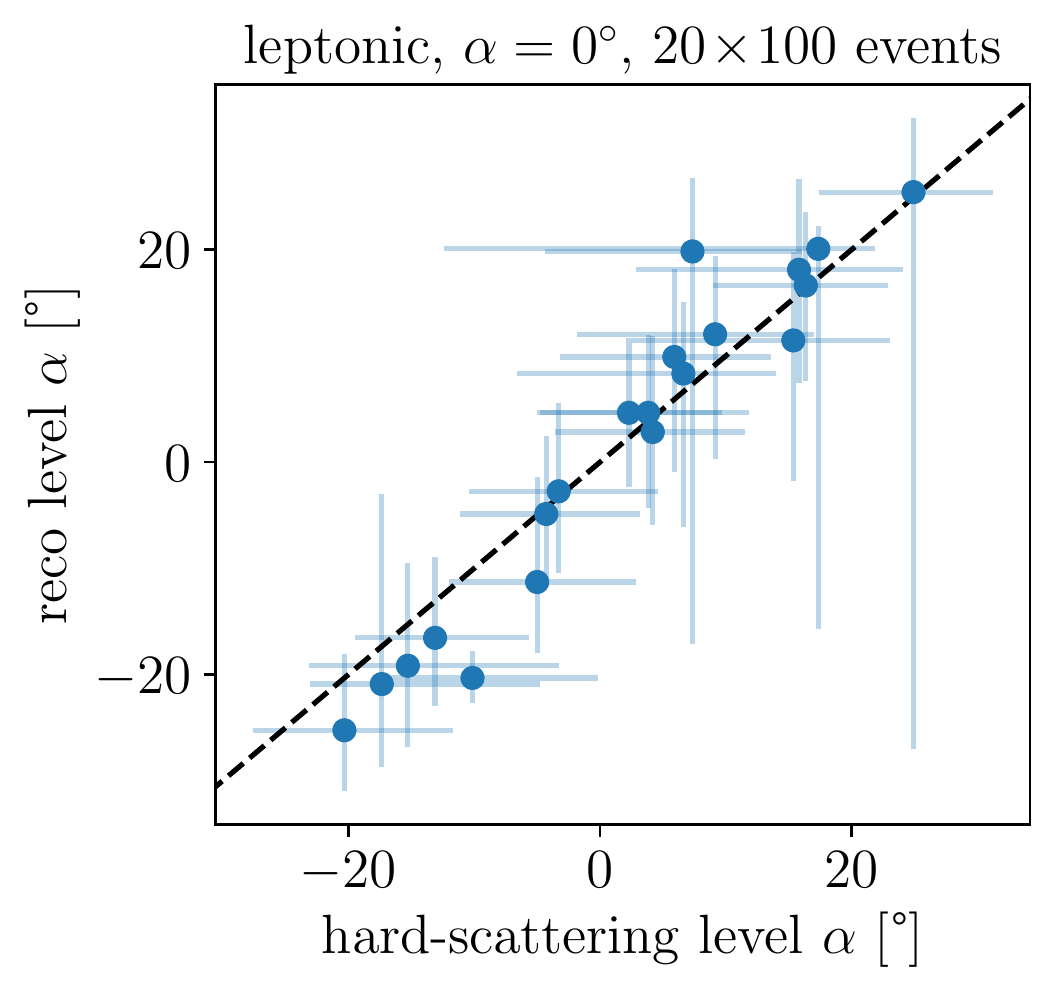}
\includegraphics[width=0.32\textwidth]{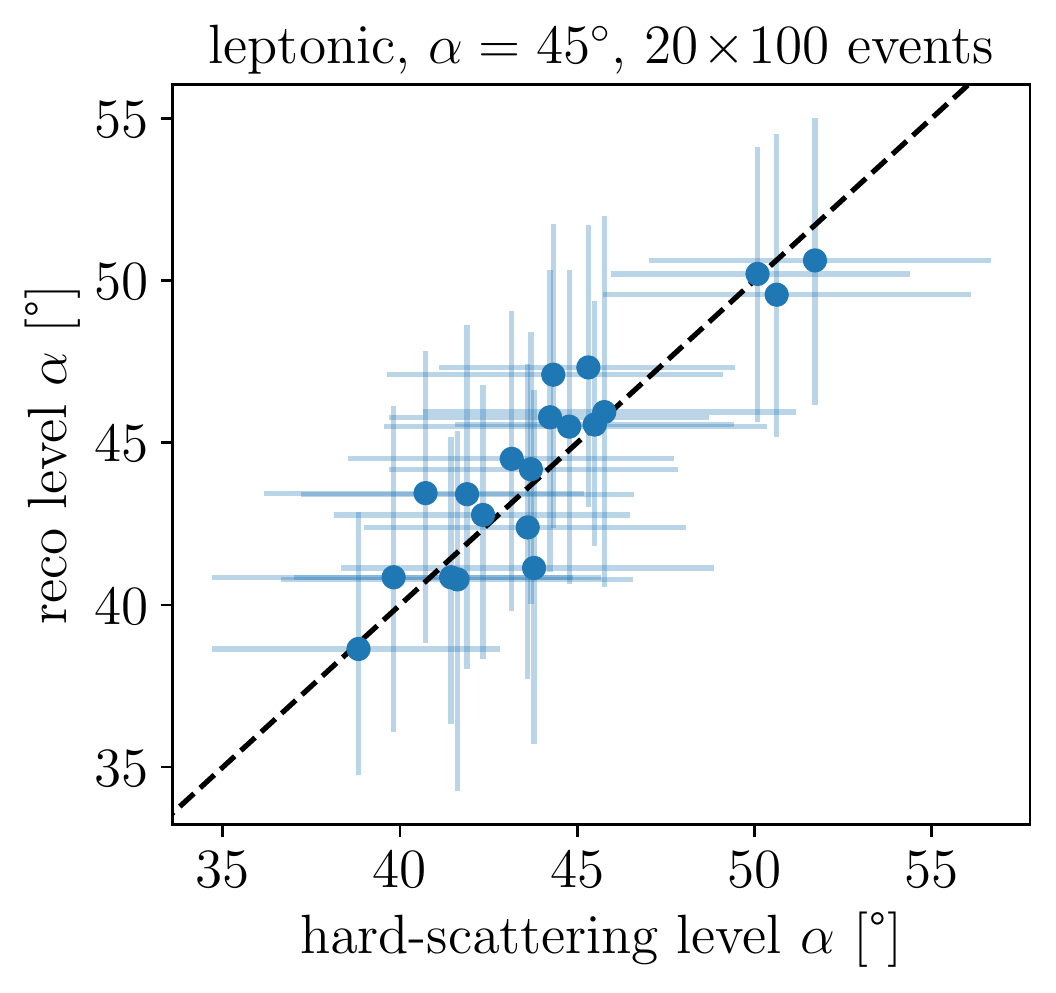}
\includegraphics[width=0.32\textwidth]{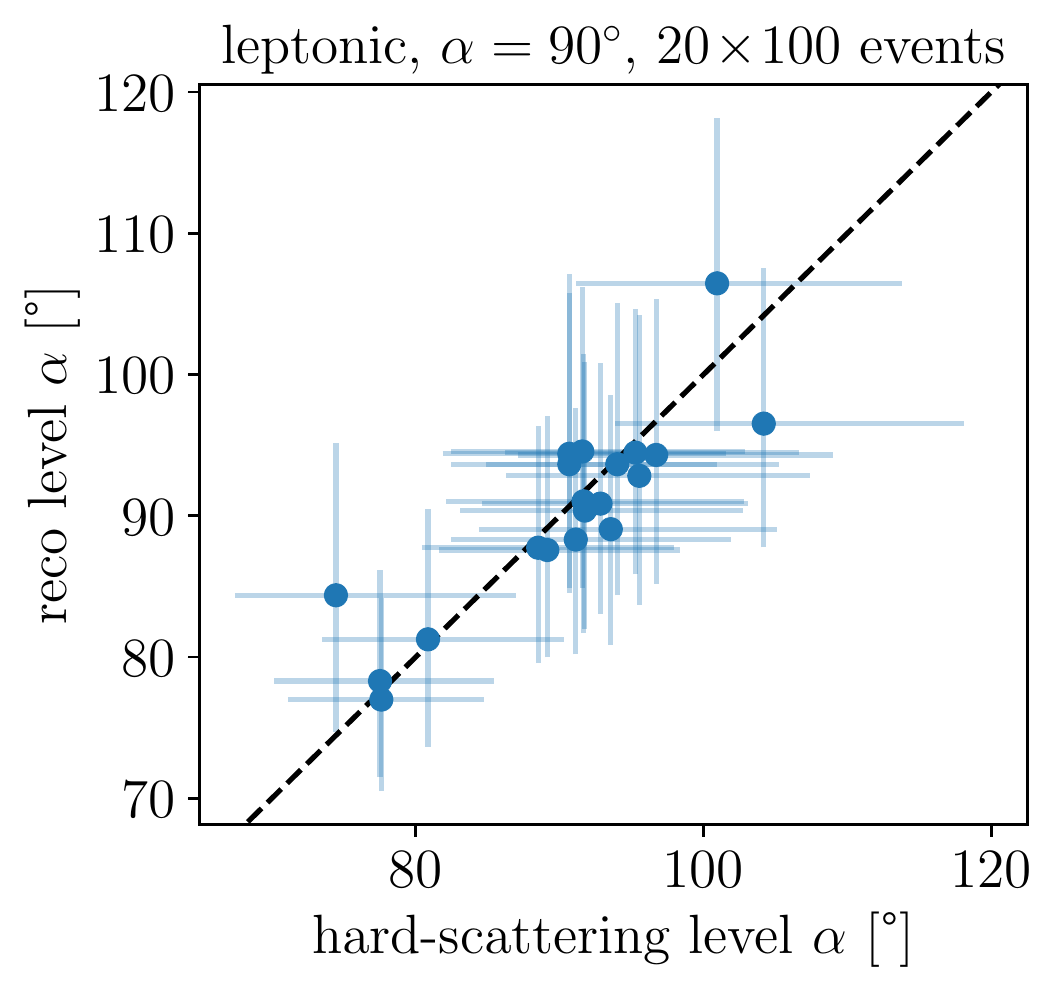}
\caption{Calibration of the $\alpha$-measurement from leptonic top
  decays, in terms of mean values and 68\% confidence intervals
  extracted from 20 sets of 100 events at parton level and measured.}
\label{fig:par_vs_int_leptonic}
\end{figure}

After testing both networks individually, we can use their combination
to extract likelihoods as a function of the CP-angle $\alpha$ for a
given set of reco-level events. While our method allows us to compute
these likelihood for individual events, we only show combinations of
400 events, to see if the corresponding distributions are reliable. In
the center panels of Fig.~\ref{fig:likeli400_leptonic} we show
likelihood distributions for an assumed true value $\alpha =
45^\circ$. According to Fig.~\ref{fig:distris} we expect the event
kinematics to be comparably sensitive to CP-angles around this value.

We compute the negative log-likelihood $- 2 \log
L_i(\alpha)$ for a given event $i$ from the integral given
in Eq.~\ref{eq:likeli_integral}, evaluated for 100k sampling points.
To improve the numerical stability we use trimmed means and standard
deviations for the integration, which means we leave out $1\%$ of
random numbers in the lower and upper tails when computing the
integral. Also in the integration we remove rare unphysical configurations,
for instance when the unfolding network generates events with momentum
fractions $x > 1$ or the leading-order differential cross section
turns negative.  The log-likelihoods for individual events is then
added to give smooth log-likelihood distributions for small event
samples.  In the upper central panel of
Fig.~\ref{fig:likeli400_leptonic} we show the likelihood values for
400 events as a function of $\alpha$.  We show the actual data points
as well as a polynomial fit to those points. The uncertainties on the
log-likelihood are computed using Gaussian error propagation of the
Monte Carlo integration error. Because we are interested in likelihood
ratios, we always show the difference in the log-likelihoods to the
minimum of the fitted curve. In the lower panel we show the results
from the Bayesian Transfer-cINN, where we visualize the training
uncertainty by repeating the likelihood calculation for 10 networks
sampled from the distribution over their trainable weights.  Comparing
the two panels we see that the variation of points around the
polynomial fit are what we expect from the network uncertainties. The
deviation from the hard-scattering truth distribution shows a small,
insignificant shift, which might come from the reconstruction of the
longitudinal neutrino momentum.

In the outer panels of Fig.~\ref{fig:likeli400_leptonic} we show the
same results for assumed CP-angles of $0^\circ$ and $90^\circ$. The
general pattern is the same as for $45^\circ$, but we see that the
quality of the measurement decreases for larger angles and becomes a
challenge towards the SM-value. The reason can be seen in
Fig.~\ref{fig:distris}, namely that the effect of small shifts in the
angle on the event kinematics is smaller than for $\alpha =
45^\circ$. An additional complication for the SM-value $\alpha =
0^\circ$ is that the total rate is symmetric under a sign flip of the
CP-angle, and Eq.\eqref{eq:fiteq} shows that this symmetry is
approximately also true for the kinematic distributions.

Finally, we check the calibration of the extracted CP-angle for 20
sets of 100 events defined at different angles. For each of those sets
we extract a mean and a 68\% confidence interval on the
hard-scattering level truth and extracted angles. We compute the
confidence intervals by assuming an approximately Gaussian likelihood
distribution with different lower and upper tails, such that the
likelihood values at the two limits are the same. This can lead to
asymmetric error bars. Instead of testing our method on a large number
of different CP phases, we use a small number of values for $\alpha$
and show the correlation between the MEM result at the reco- and
hard-scattering level for several sets of events to obtain approximate
calibration curves. In Fig.~\ref{fig:par_vs_int_leptonic} we show such
curves for the three assumed true $\alpha$-values. For
the best measurement at $\alpha = 45^\circ$ the true and extracted
values of the angle are nicely correlated. A slight bias towards
overestimating the angle can be removed through a proper
calibration. For $\alpha= 90^\circ$ the situation is similar, but, if
anything, the bias tends to underestimate the true angle. Finally, for
the challenging SM-value $\alpha = 0^\circ$ the range of the
correlation and the error bars increase, but the calibration is
perfectly fine.

\subsection{Hadronic top decay}
\label{sec:perf_hadr}

Moving on to the more challenging hadronic top decay, we use the same
neural network setup as before and see how it deals with the challenge
moving from the neutrino reconstruction to increasingly complex jet
combinatorics.

\subsubsection*{Without ISR}

\begin{figure}[t]
\includegraphics[width=0.32\textwidth]{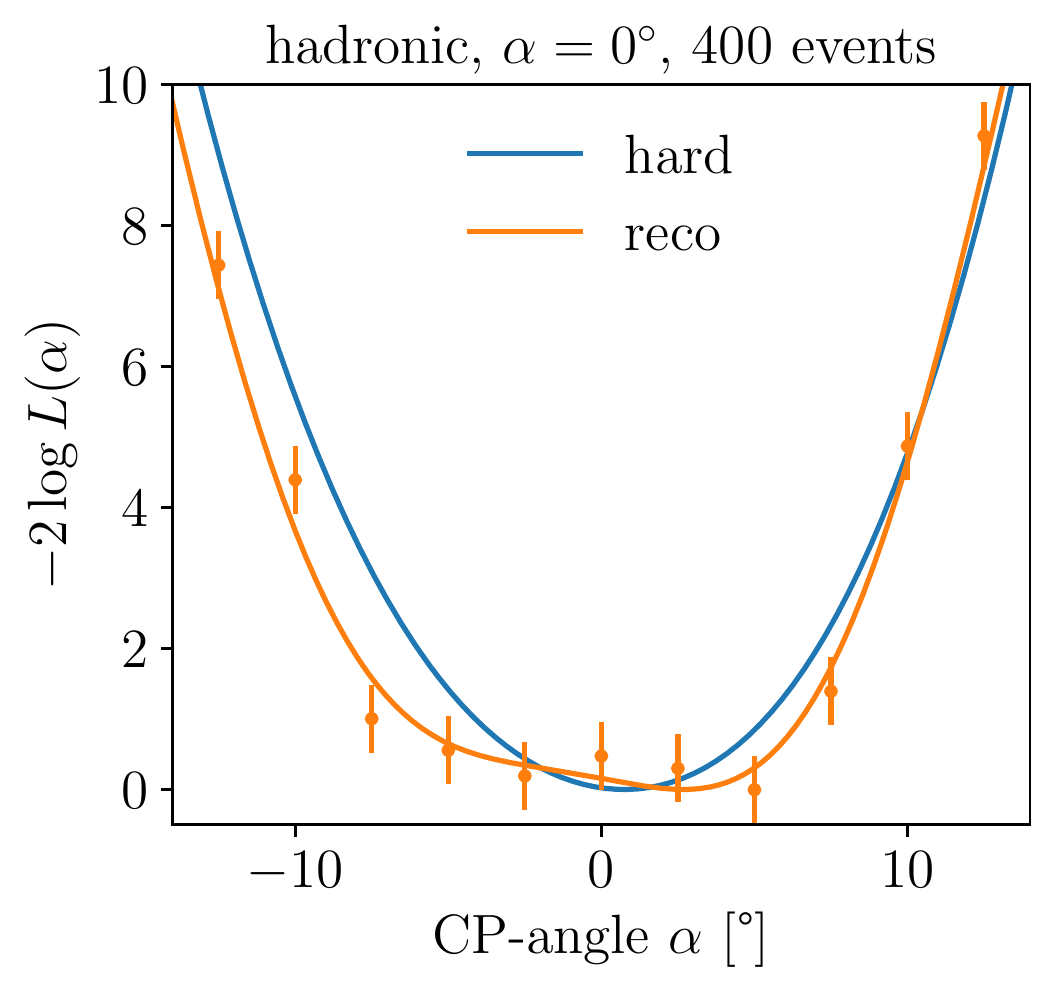}
\includegraphics[width=0.32\textwidth]{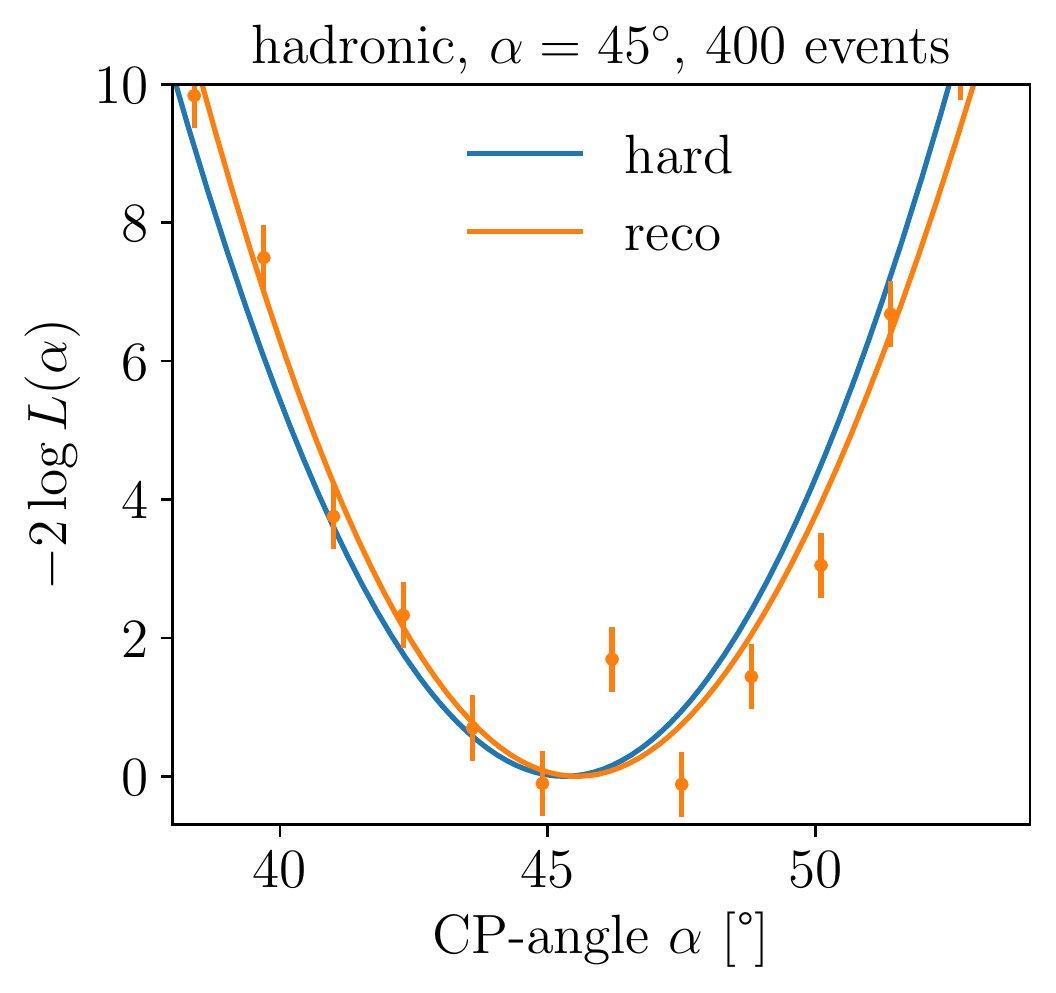}
\includegraphics[width=0.32\textwidth]{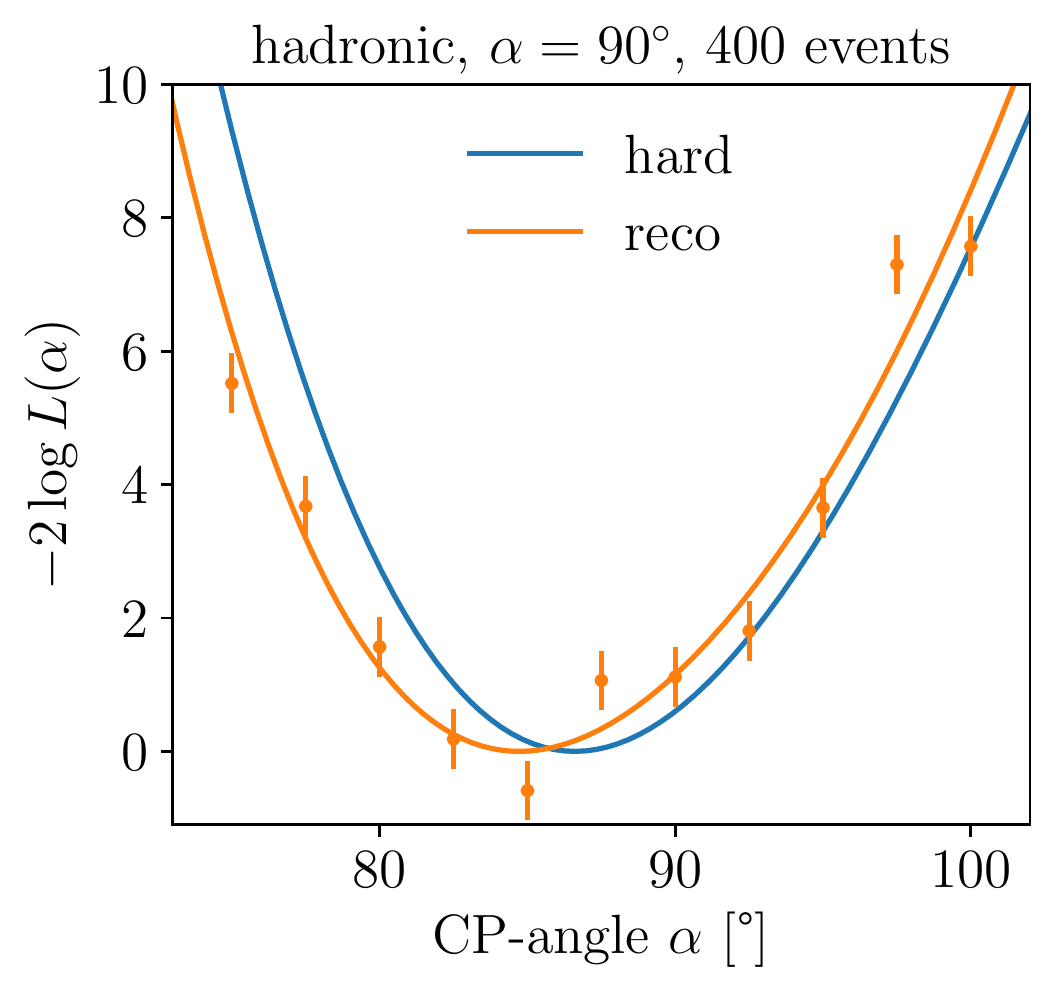} \\
\includegraphics[width=0.32\textwidth]{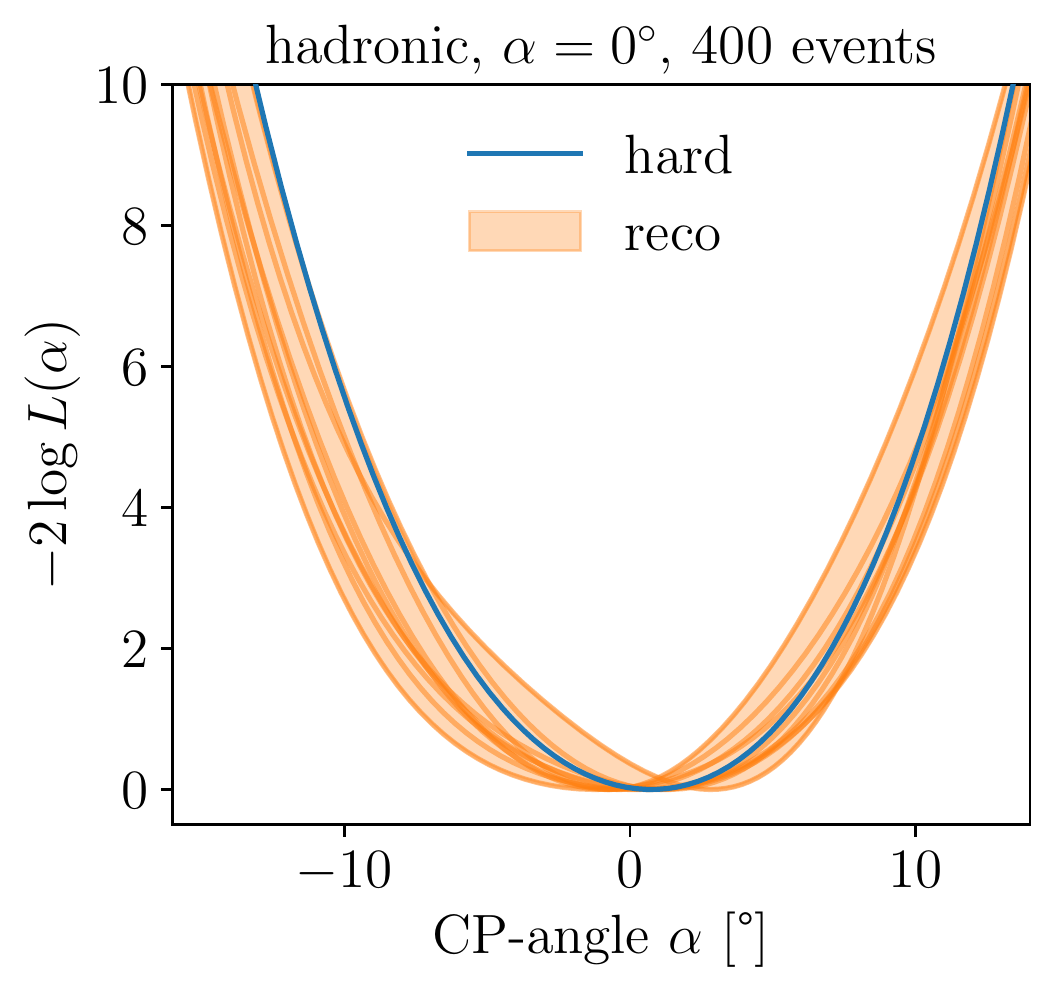}
\includegraphics[width=0.32\textwidth]{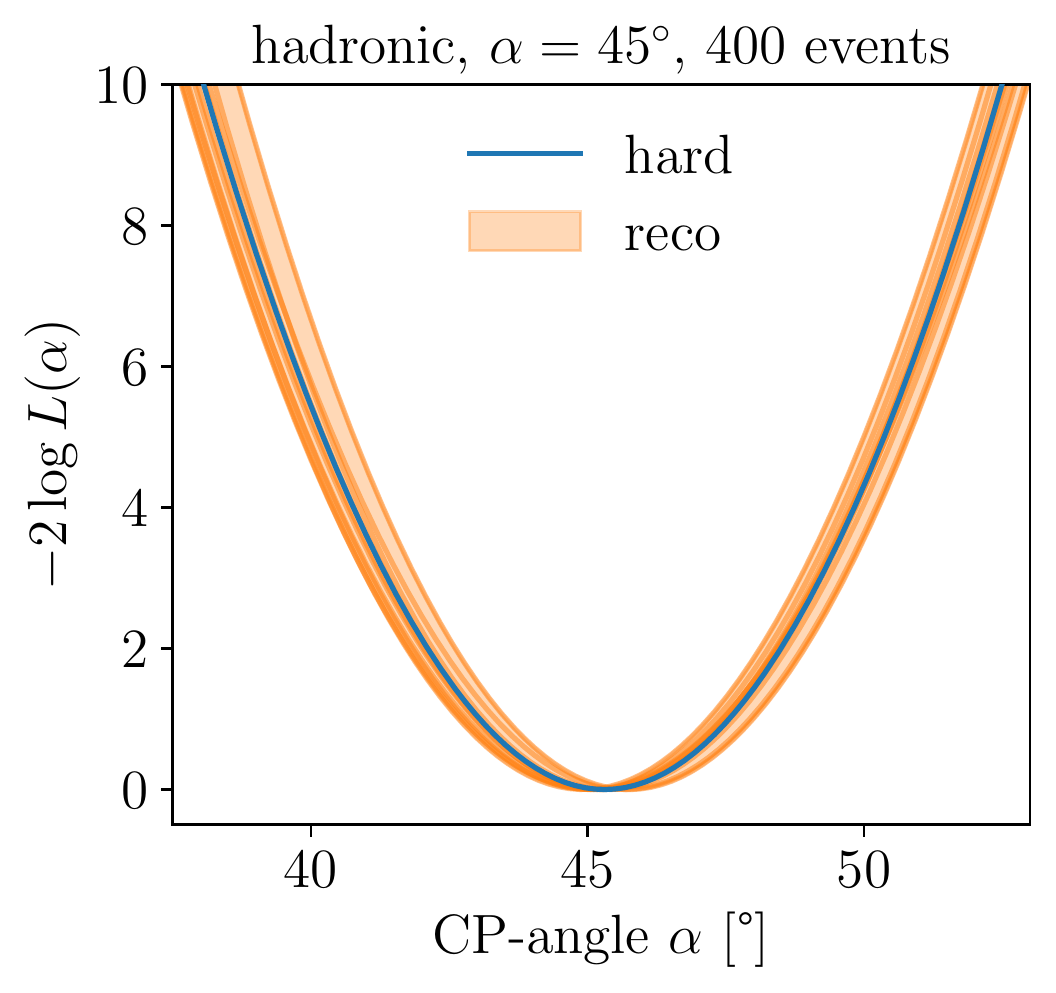}
\includegraphics[width=0.32\textwidth]{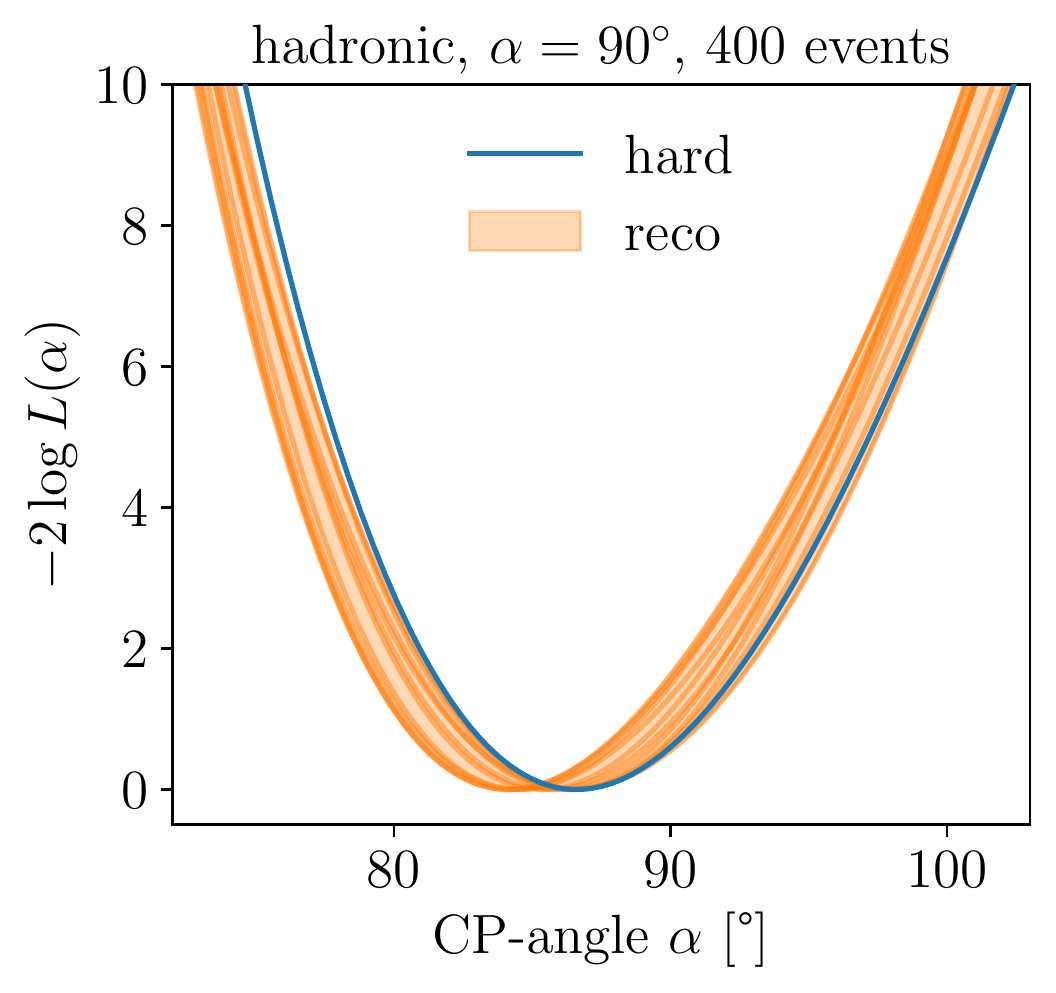}
\caption{Likelihoods for the hadronic top decay as a function of the
  CP-angle $\alpha$, extracted from 400 events for three assumed truth
  angles. For the Bayesian uncertainties we show the integrated
  likelihoods from 10 sampled networks.}
\label{fig:likeli400_noisr}
\end{figure}

\begin{figure}[b!]
\includegraphics[width=0.32\textwidth]{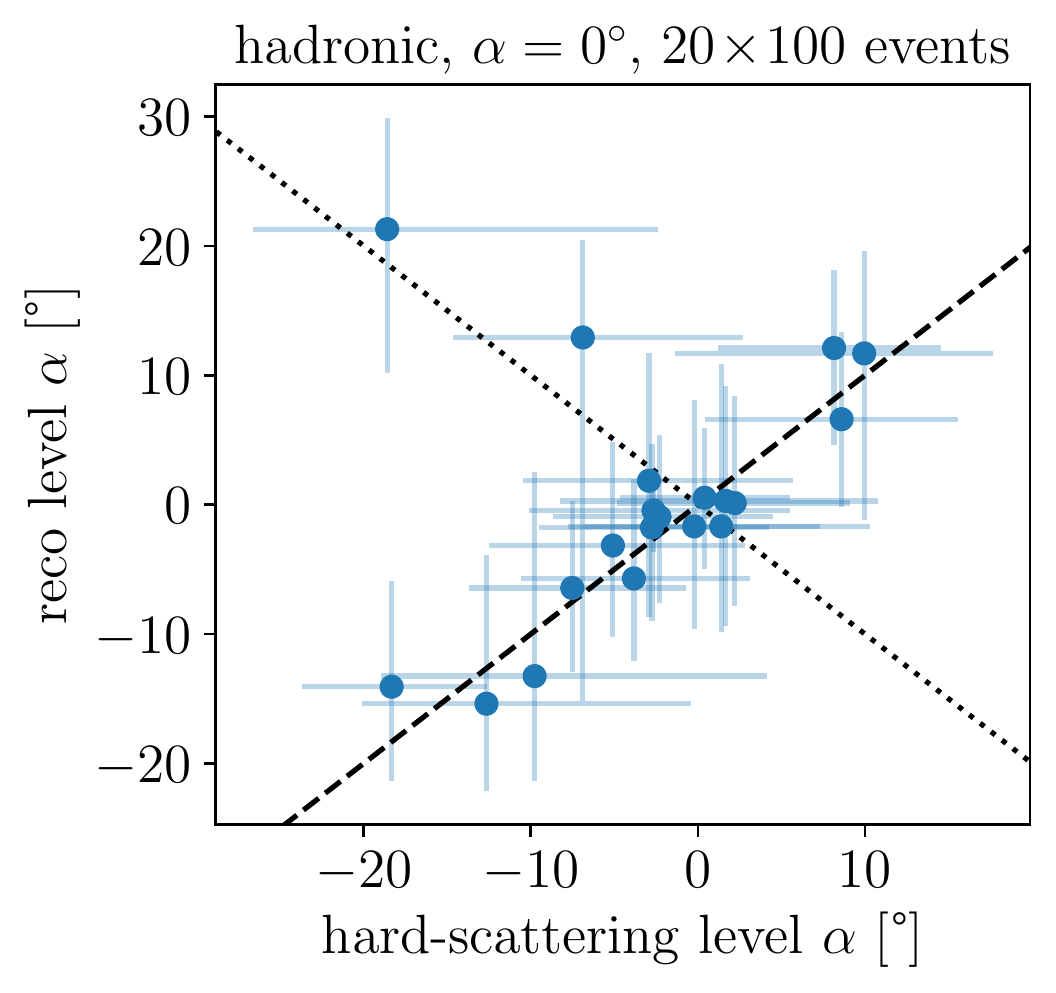}
\includegraphics[width=0.32\textwidth]{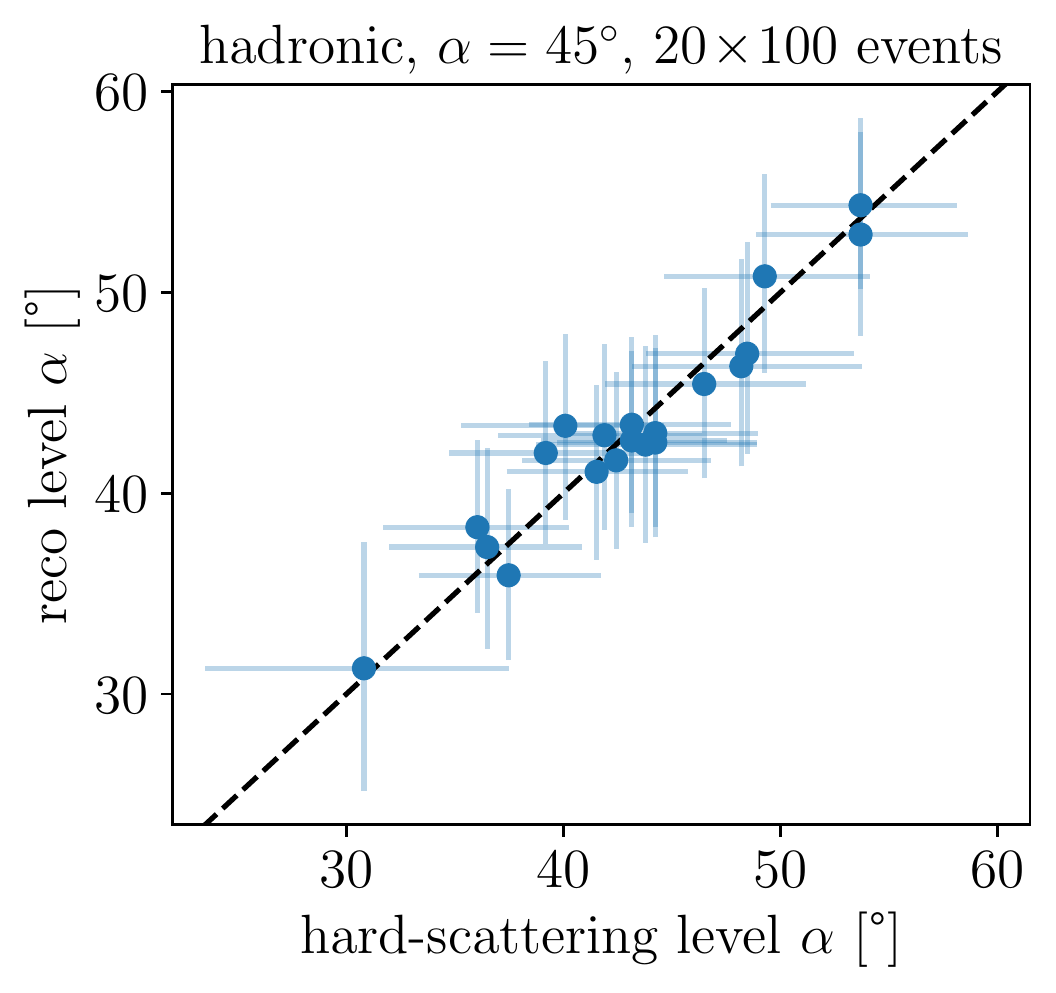}
\includegraphics[width=0.32\textwidth]{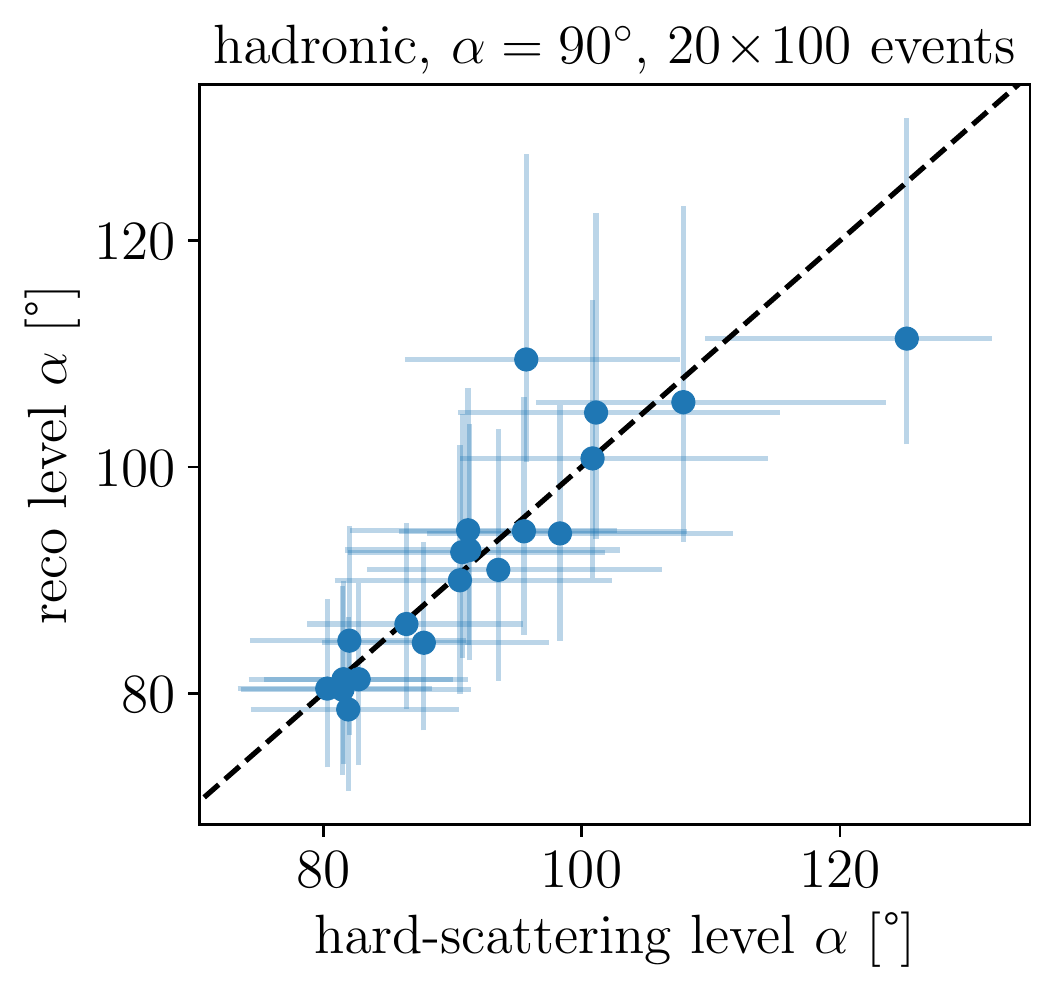}
\caption{Calibration of the $\alpha$-measurement from hadronic top
  decays, in terms of mean values and 68\% confidence intervals
  extracted from 20 sets of 100 events at hard-scattering level and measured.}
\label{fig:par_vs_int_noisr}
\end{figure}

The only change between the leptonic top decay study and the hadronic
top decays is that the reco-level phase space now covers two on-shell
photons, one $b$-jet and three light-flavor jets, leading to $6 \cdot
4 - 2$ dimensions. In Fig.~\ref{fig:likeli400_noisr} we show the
extracted likelihood distributions for 400 events, to be compared with
Fig.~\ref{fig:likeli400_leptonic} for the leptonic case.  We see that
the results are completely comparable, which means that the additional
complication of having to separate $W$-decay jets from the forward jet
is not a problem for the networks. In Fig.~\ref{fig:par_vs_int_noisr}
we see a new feature, as compared to
Fig.~\ref{fig:par_vs_int_leptonic}, where for the SM-value $\alpha =
0^\circ$ the networks sometimes chooses a mismatch of the sign of the
angle between the hard-scattering level and the reconstruction level. This reflects
the approximate symmetry from Eq.\eqref{eq:fiteq} and does not affect
the likelihood extraction in a significant way.

\subsubsection*{With ISR}

\begin{figure}[b!]
\includegraphics[width=0.49\textwidth,page=18]{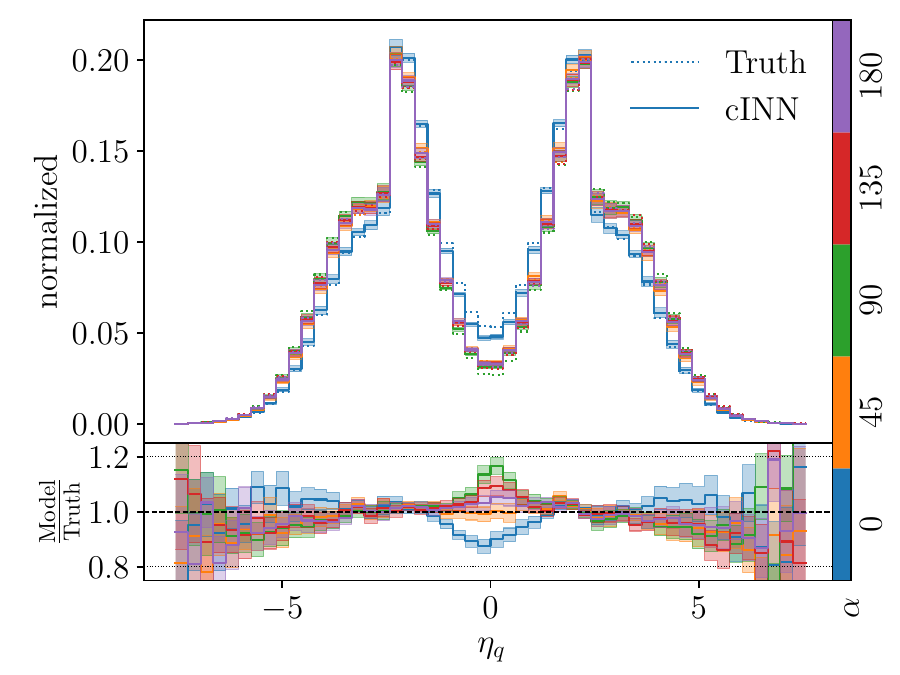}
\includegraphics[width=0.49\textwidth,page=27]{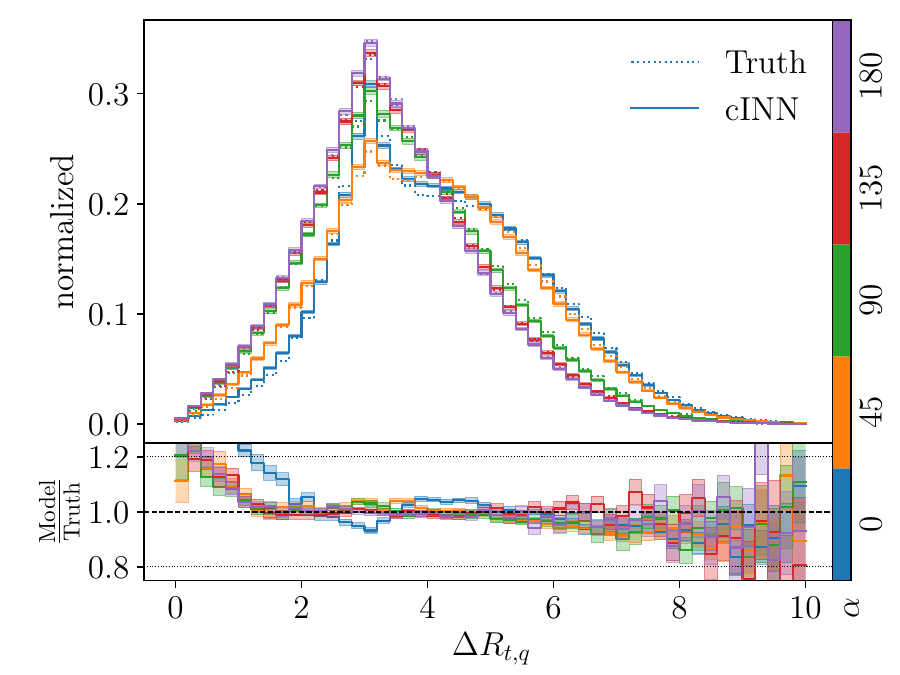}
\includegraphics[width=0.49\textwidth,page=36]{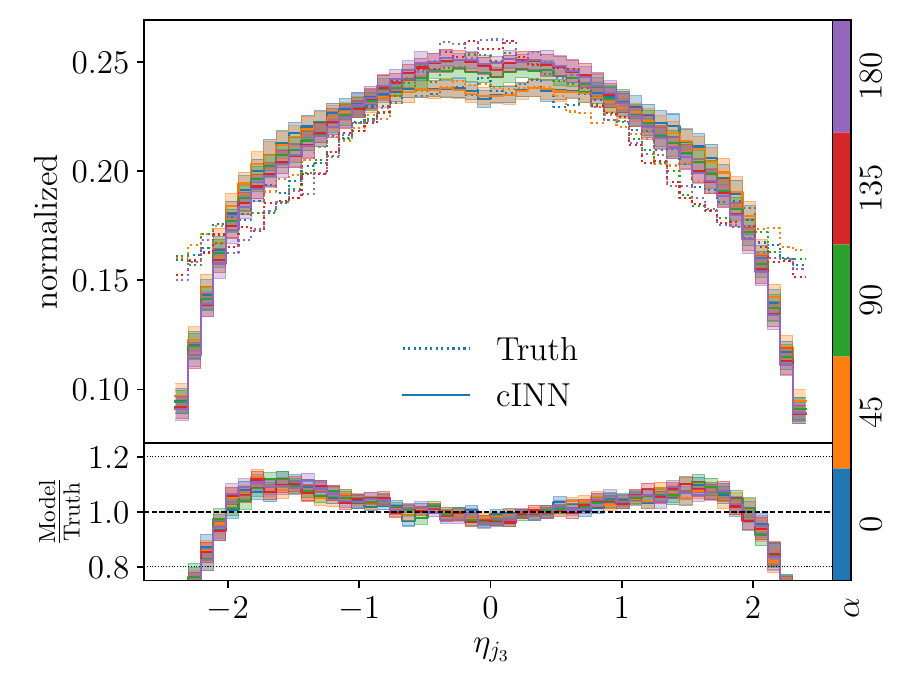}
\includegraphics[width=0.49\textwidth,page=51]{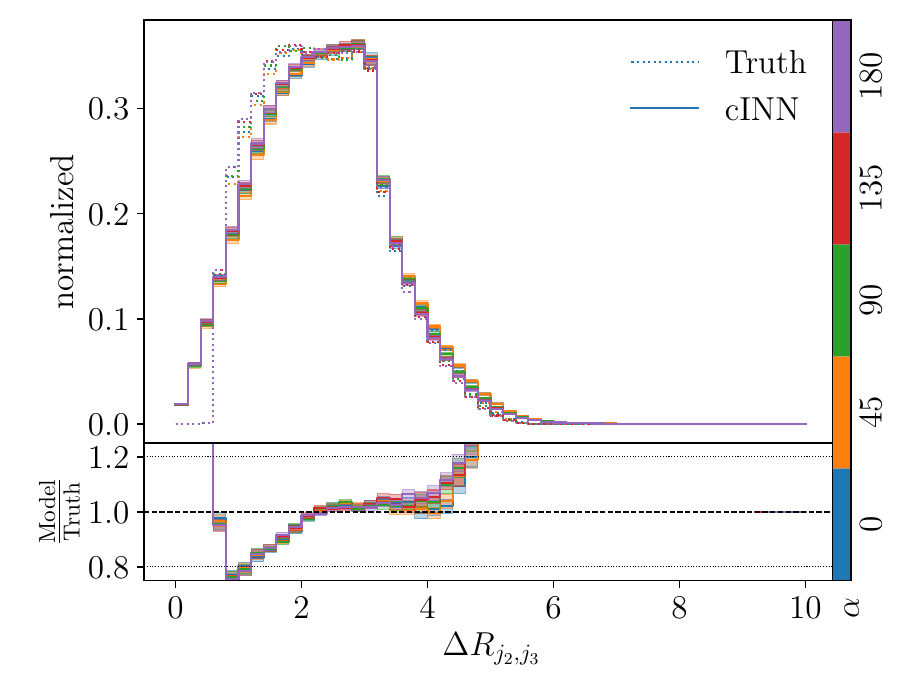}
\caption{Top: unfolded kinematic distributions of the forward quark
  for hadronic top decay with ISR, assuming five different CP-angles
  and including uncertainties from the Bayesian cINN. These
  distributions test the Unfolding-cINN. Bottom: forward-simulated
  kinematic distributions for the hadronic top decay with ISR,
  assuming five different CP-angles and including uncertainties
  from the Bayesian cINN. These distributions test the Transfer-cINN.}
\label{fig:histograms_inverse_fullhad}
\end{figure}

The situation changes when we allow for initial state radiation
(ISR). In the absence of jets radiated from the initial state the
network only has to distinguish decay jets from the forward jet in the
hard process.  The only difference between the two analyses without
ISR and this one is that we now use a larger training dataset with
3.4M events, so the network can learn the more complex kinematic
patterns.  From Fig.~\ref{fig:distris} we know that the kinematic
distribution of this hard forward jet, relative to the top and Higgs,
is intimately tied to the CP-angle $\alpha$.  Final state radiation
can lead to a third decay jet or a splitting of the hard forward jet,
in both cases not affecting the event topology much. This is different
for ISR, because the additional jets can look similar to the hard
forward jet, but they are really not part of the hard process and
therefore only indirectly sensitive to the CP-angle. This makes it
much more complicated to evaluate the hard-scattering likelihood. In
general, the hadronic top decay combined with ISR breaks the
one-to-one correspondence of hard-scattering partons and jets, which
we have confirmed in detail, for instance using geometric
correlations.

In the upper panels of Fig.~\ref{fig:histograms_inverse_fullhad} we
show the unfolded kinematic distributions for the top, Higgs, and
forward jet from the hard process. The Unfolding-cINN generally does
well in reconstructing the hard process, which means the phase space
integration as part of the MEM remains efficient after we include
ISR.

In the lower panels of Fig.~\ref{fig:histograms_inverse_fullhad} we test the
Transfer-cINN. We omit the kinematic distributions related to the top
and $W$-kinematics, where the network does essentially as well as
without ISR, and only show the critical distributions related to the
jets. Here we can see that the performance of the Transfer-cINN is
significantly worse, with typical deviations of up to 20\% on the
underlying phase space density.  At this level the Bayesian
uncertainty does not cover the difference between the truth and the
cINN-generated distributions, and the size of the deviations is going
to affect the extraction of the CP-angle. These results can
immediately be generalized to the forward simulation of QCD jet
radiation and detector effects.

\begin{figure}[t]
\includegraphics[width=0.32\textwidth]{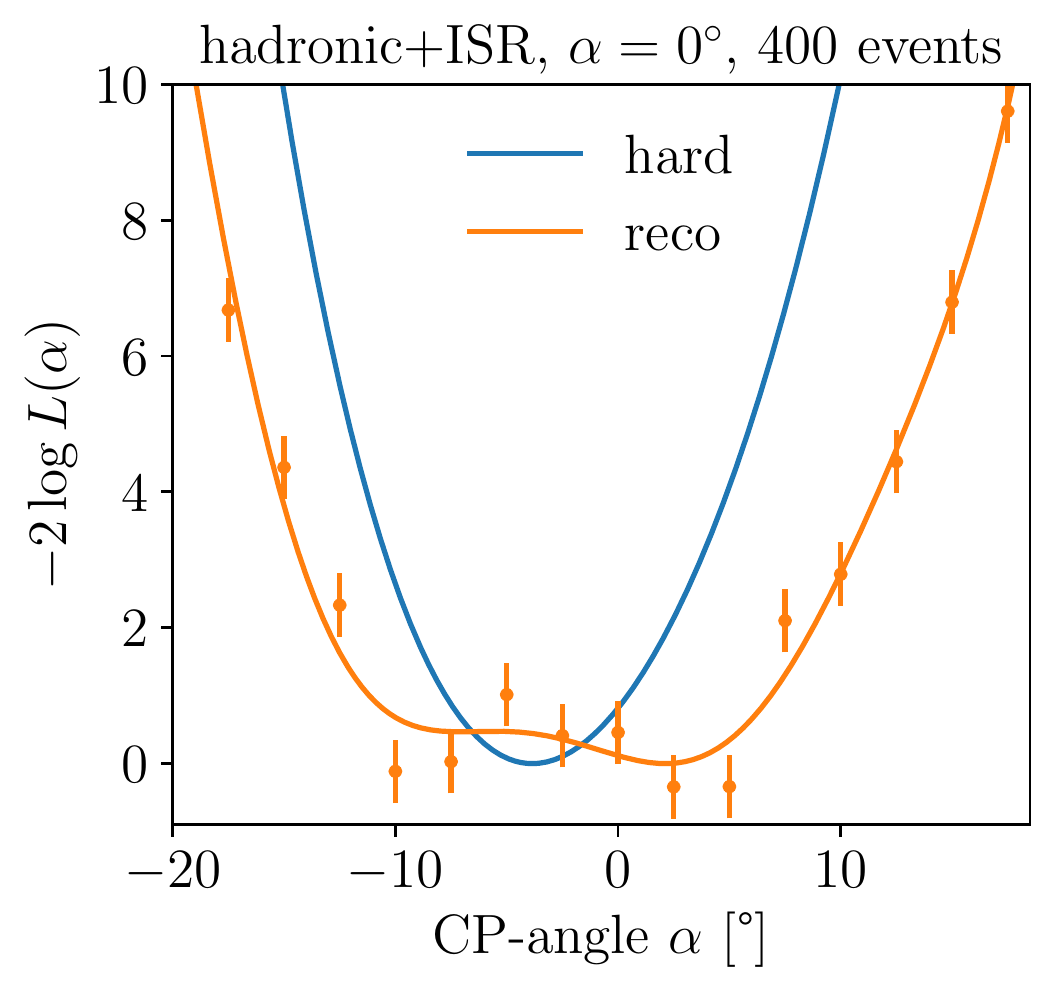}
\includegraphics[width=0.32\textwidth]{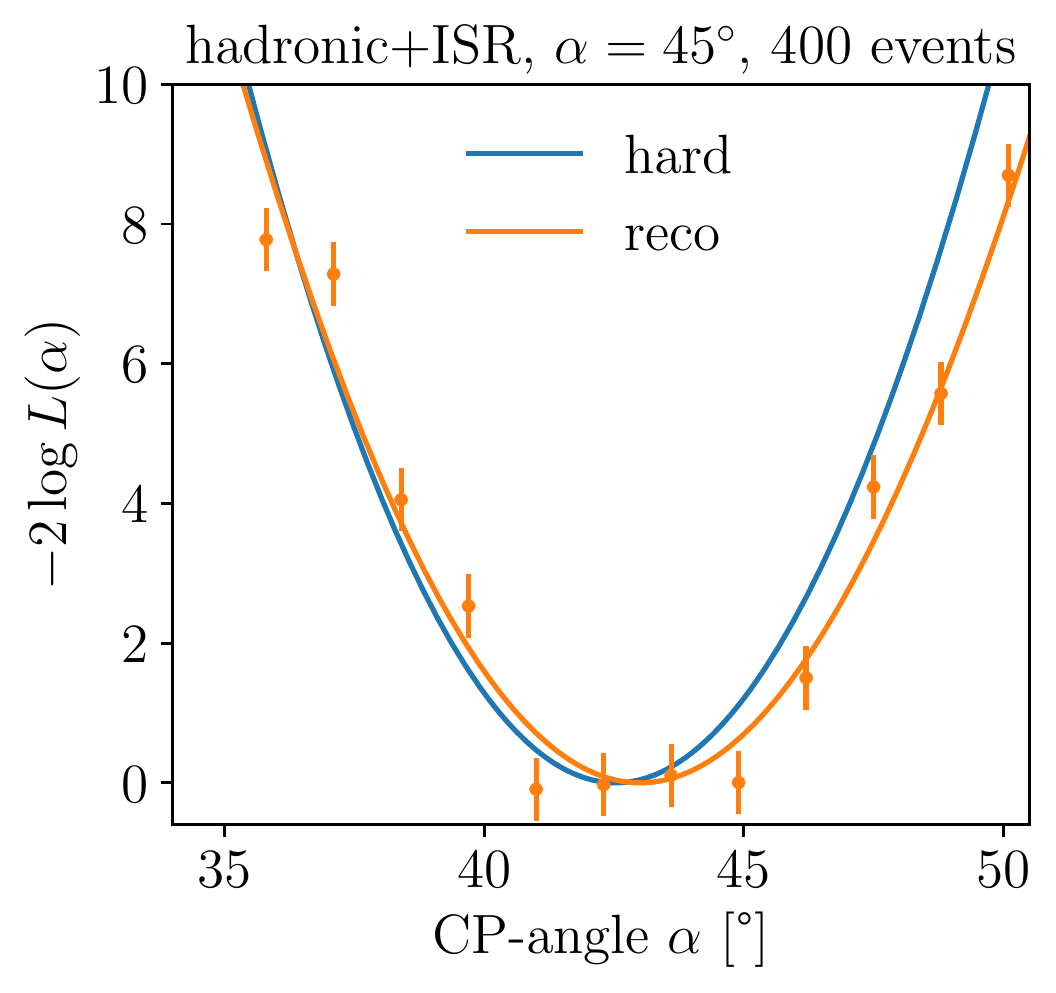}
\includegraphics[width=0.32\textwidth]{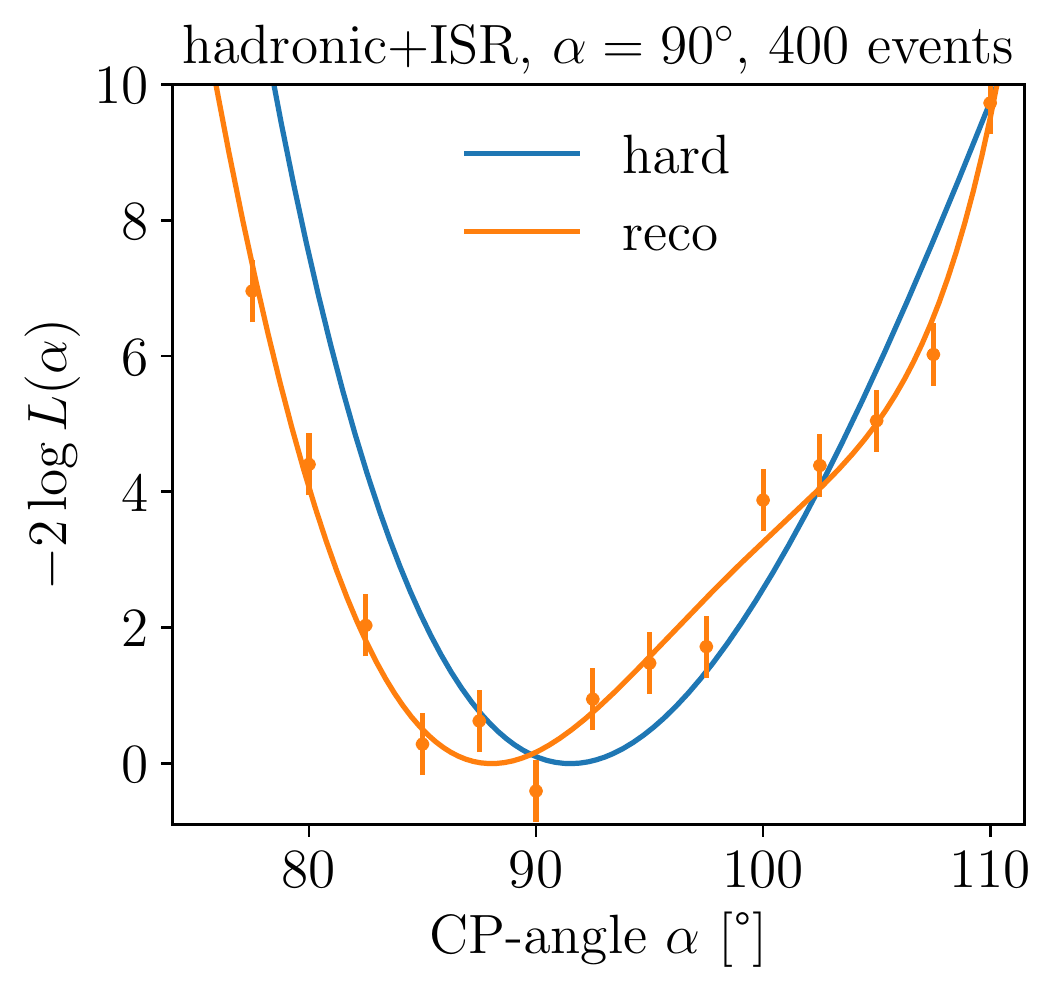} \\
\includegraphics[width=0.32\textwidth]{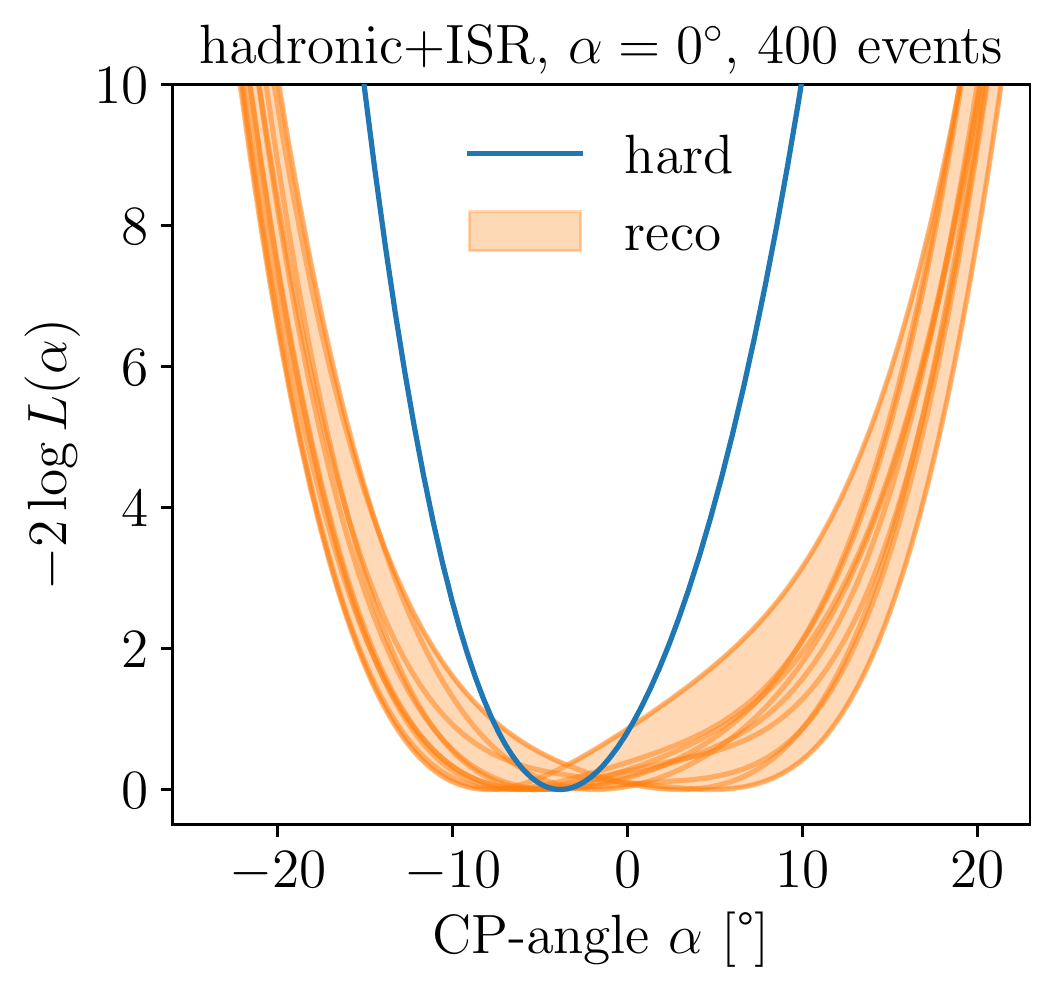}
\includegraphics[width=0.32\textwidth]{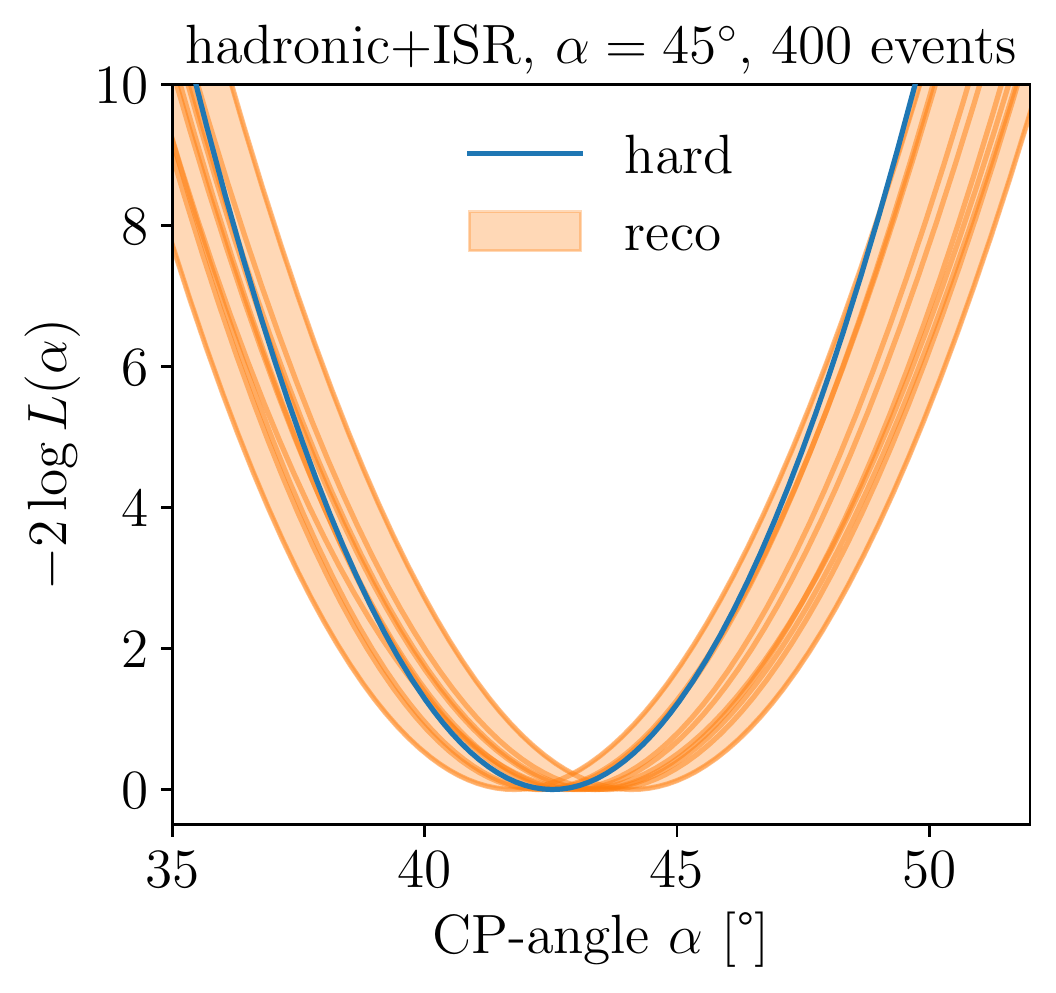}
\includegraphics[width=0.32\textwidth]{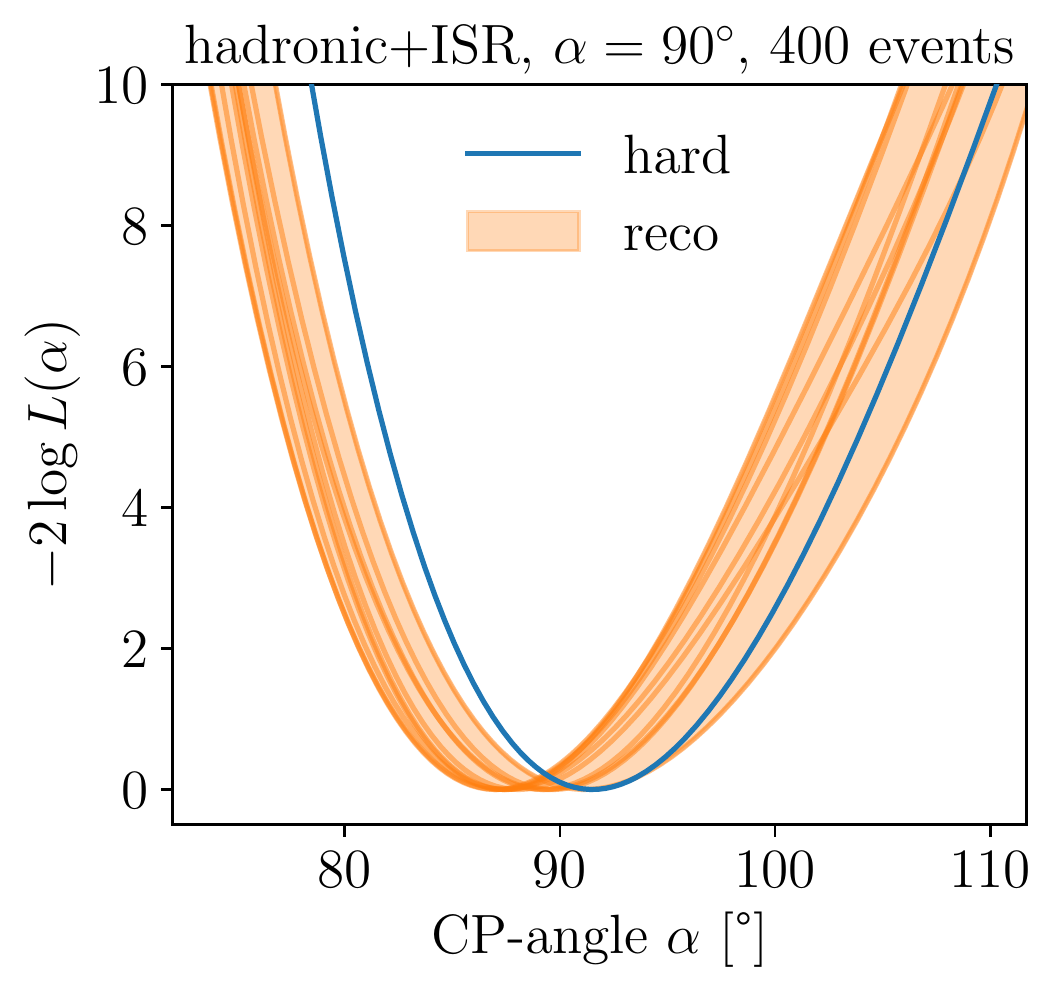}
\caption{Likelihoods for the hadronic top decay, including ISR, as a
  function of the CP-angle $\alpha$, extracted from 400 events for
  three assumed truth angles. For the Bayesian uncertainties we show
  the integrated likelihoods from 10 sampled networks.}
\label{fig:likeli400_fullhad}
\end{figure}

\begin{figure}[t]
\includegraphics[width=0.32\textwidth]{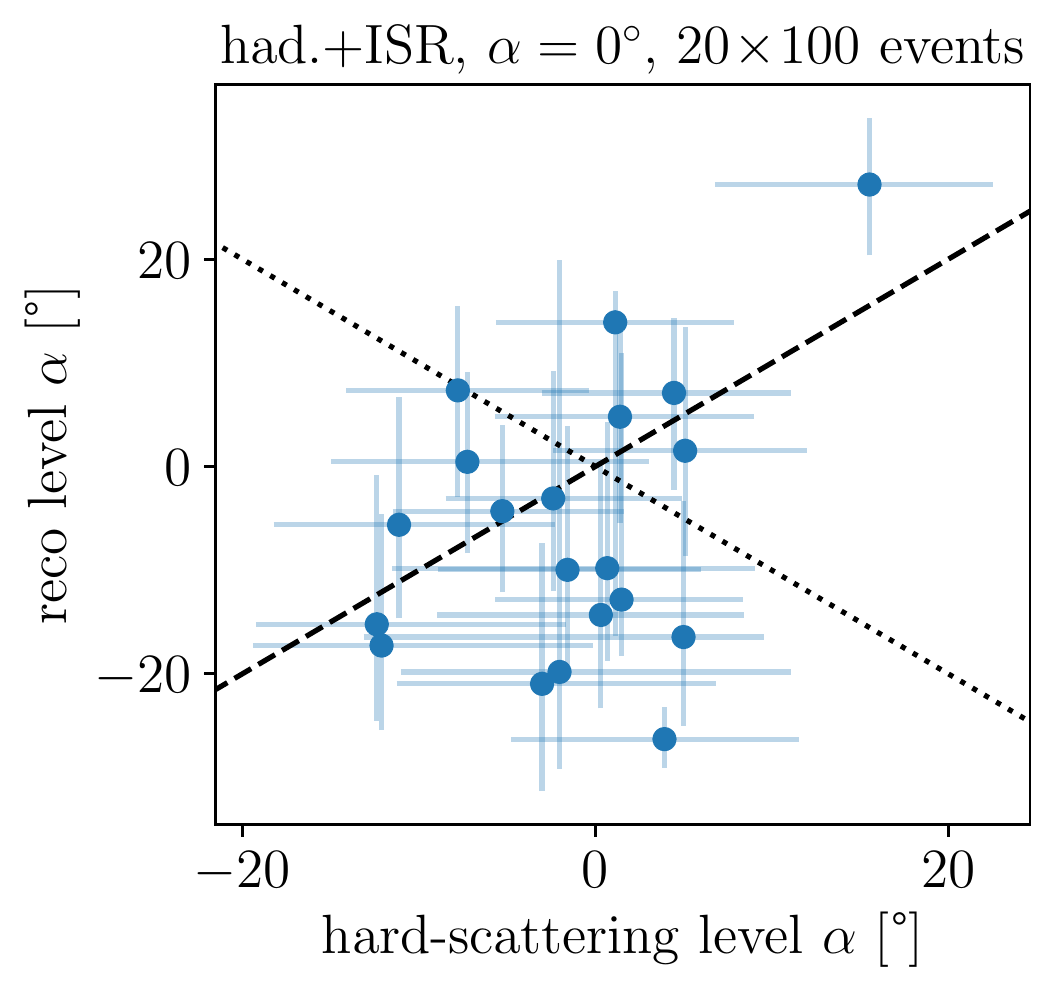}
\includegraphics[width=0.32\textwidth]{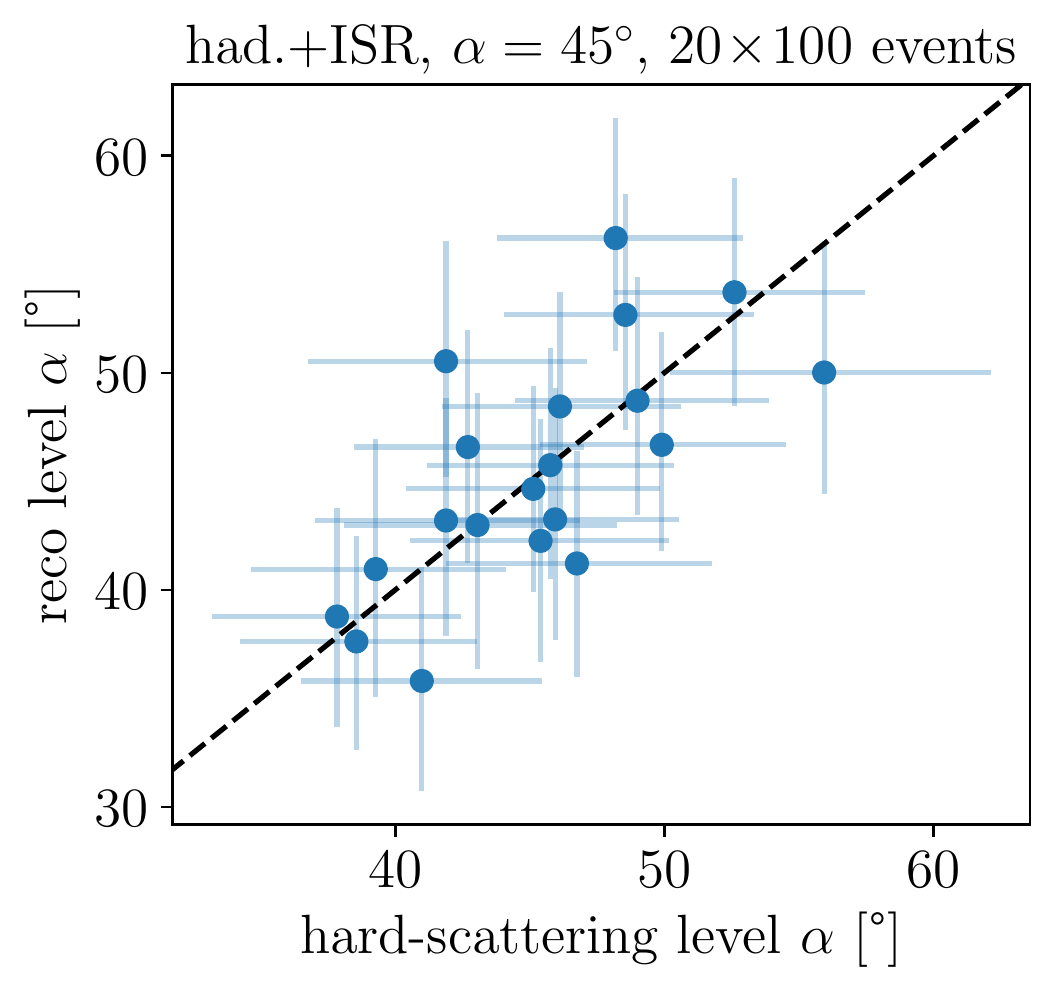}
\includegraphics[width=0.32\textwidth]{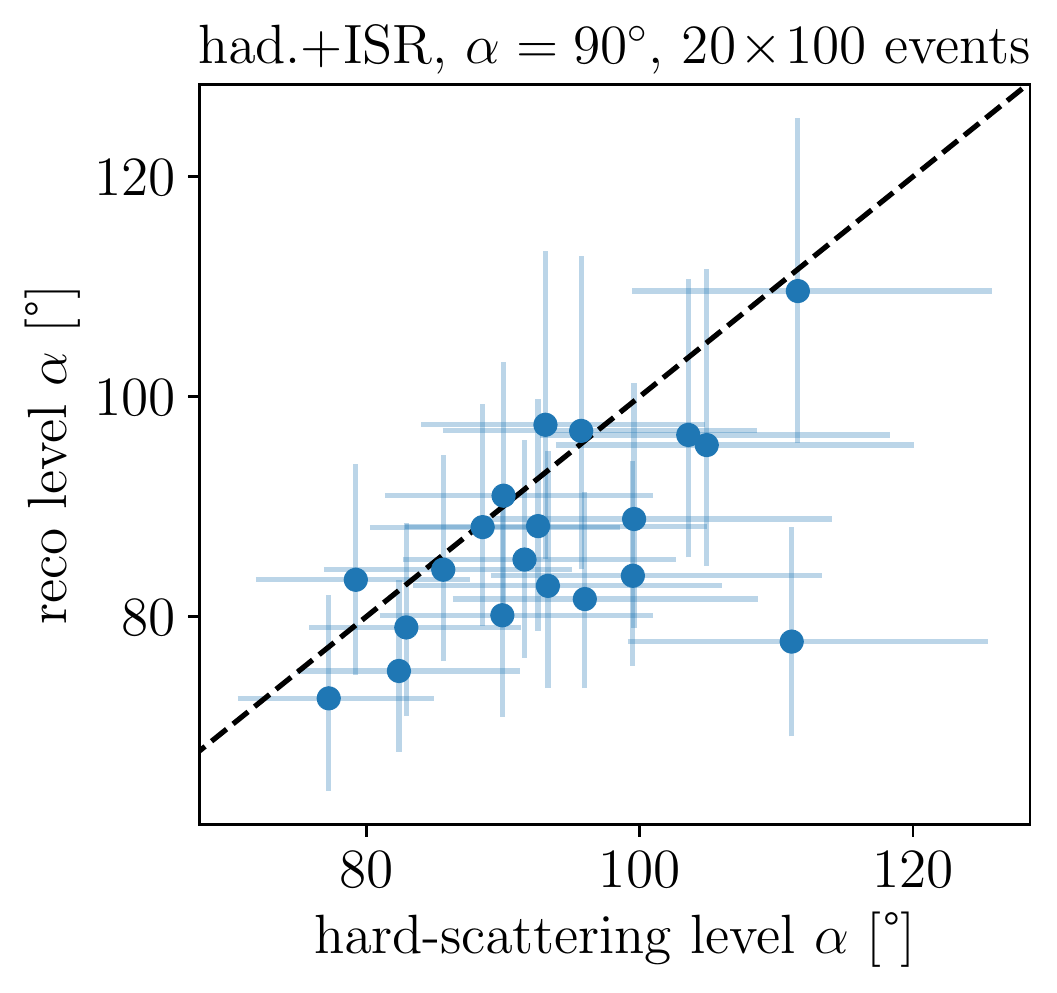}
\caption{Calibration of the $\alpha$-measurement from hadronic top
  decays with ISR, in terms of mean values and 68\% confidence
  intervals extracted from 20 sets of 100 events at hard-scattering level and
  measured.}
\label{fig:par_vs_int_fullhad}
\end{figure}

As before, we show the extracted likelihood as a function of the
CP-angle $\alpha$ in Fig.~\ref{fig:likeli400_fullhad}. For the
deterministic Transfer-cINN we find that the likelihoods extracted
from sets of 400 events reproduce the idealized hard-scattering level results
fairly well. Problems occur around the SM-value, where we know that
the effects of $\alpha$ only grow slowly. We find that for the shown
event sample values $\alpha \lesssim 10^\circ$ cannot really be
distinguished from the Standard Model. This impression is confirmed by
the uncertainties from the Bayesian network, indicating that there is
a significant loss in sensitivity compared to the idealized hard-scattering
level. On the positive side, the insensitive range of $\alpha \lesssim
10^\circ$ is to be compared to the insensitive range of $\alpha
\lesssim 45^\circ$ for the total rate, as seen in Fig.~\ref{fig:xsec}.
For the largest assumed angle $\alpha = 90^\circ$ there is also a bias
towards an underestimation of $\alpha$, which we have confirmed to be
a general feature.

We find the same issues in the calibration curves shown in
Fig.~\ref{fig:par_vs_int_fullhad}. Around $|\alpha| \lesssim 10^\circ$
the network finds hardly any sensitivity to the mixing angle. The
situation improves for the most sensitive region around $\alpha =
45^\circ$, and for $\alpha = 90^\circ$ the measurement indicates a
bias which can, however, be removed through a standard calibration of
the $\alpha$-measurement.

\section{Outlook}
\label{sec:outlook}

The matrix element method is the method of choice to measure
fundamental Lagrangian parameters from a small number of events at
colliders. At hadron colliders, this method is computationally
challenging. We have presented a way to compute the likelihoods at the
hard-scattering level for reconstruction-level events with the help of
two conditional generative neural networks, specifically two
cINNs. First, a Transfer-cINN encodes the effects of QCD jet radiation
and detectors in a forward simulation. This network is nothing but a
fast detector simulation, conditioned on the hard process. Second, the
established Unfolding-cINN allows us to efficiently compute the
integration over the hard-scattering phase space, just like a learned
phase space mapping. In combination, the two conditional generative
networks allow us to compute event-wise likelihood ratios efficently
and without any assumptions on the form of any transfer function.

We illustrate our method using the extraction of the CP-angle in the
top Yukawa coupling, accessible at the LHC through the associated
production of a Higgs with a single top quark. Around the SM-value
$\alpha = 0^\circ$ the total rate of this process shows essentially no
dependence on the CP-angle. We show how this changes once we include
the full kinematic information through the MEM. For a leptonic top
decay and for the hadronic top decay without initial state radiation
our method shows a sensitivity close to the truth at the
hard-scattering level. Once we include ISR, the combinatorics become
more challenging, because the correspondence of jets and partons is
broken and additional jets are hard to distinguish from the forward
jets of the hard process. Nevertheless, even the results with ISR are
promising and indicate that the MEM will allow us to extract maximal
information for LHC processes with a small number of expected events.

\subsection*{Acknowledgments} 

We would like to thank Peter Uwer for his continuing support and
advice.  TP and AB would like to thank the
Baden-W\"urttemberg-Stiftung for support through the program
\textit{Internationale Spitzenforschung}, project
\textsl{Uncertainties --- Teaching AI its Limits} (BWST\_IF2020-010).
AB and TP are supported by the DFG under grant 396021762 – \textsl{TRR
  257 Particle Physics Phenomenology after the Higgs Discovery}.  TH
is supported by the DFG Research Training Group GK-1940,
\textsl{Particle Physics Beyond the Standard Model}.  The authors
acknowledge support by the state of Baden-Württemberg through bwHPC
and the German Research Foundation (DFG) through grant no INST
39/963-1 FUGG (bwForCluster NEMO). This work was supported by the DFG
under Germany's Excellence Strategy EXC 2181/1 - 390900948 \textsl{The
  Heidelberg STRUCTURES Excellence Cluster}.

\end{fmffile}

\bibliography{literature}

\providecommand{\href}[2]{#2}\begingroup\raggedright\begin{thebibliography}{10}

\bibitem{Kondo:1988yd}
K.~Kondo, {\it {Dynamical Likelihood Method for Reconstruction of Events With
  Missing Momentum. 1: Method and Toy Models}},
\href{http://dx.doi.org/10.1143/JPSJ.57.4126}{J. Phys. Soc. Jap. {\bfseries 57}
  (1988)  4126}.

\bibitem{Kondo:1991dw}
K.~Kondo, {\it {Dynamical likelihood method for reconstruction of events with
  missing momentum. 2: Mass spectra for 2 ---\ensuremath{>} 2 processes}},
  \href{http://dx.doi.org/10.1143/JPSJ.60.836}{J. Phys. Soc. Jap. {\bfseries
  60} (1991)  836}.

\bibitem{D0:1998eiz}
D0, B.~Abbott {\em et al.}, {\it {Measurement of the Top Quark Mass in the
  Dilepton Channel}},
  \href{http://dx.doi.org/10.1103/PhysRevD.60.052001}{Phys. Rev. D {\bfseries
  60} (1999)  052001},
  \href{http://arxiv.org/abs/hep-ex/9808029}{{arXiv:hep-ex/9808029}}.

\bibitem{Abazov:2004cs}
D0, V.~M. Abazov {\em et al.}, {\it {A precision measurement of the mass of the
  top quark}},  \href{http://dx.doi.org/10.1038/nature02589}{Nature {\bfseries
  429} (2004)  638},
\href{http://arxiv.org/abs/hep-ex/0406031}{{arXiv:hep-ex/0406031 [hep-ex]}}.

\bibitem{CDF:2006nne}
CDF, A.~Abulencia {\em et al.}, {\it {Top quark mass measurement from dilepton
  events at CDF II with the matrix-element method}},
  \href{http://dx.doi.org/10.1103/PhysRevD.74.032009}{Phys. Rev. D {\bfseries
  74} (2006)  032009},
  \href{http://arxiv.org/abs/hep-ex/0605118}{{arXiv:hep-ex/0605118}}.

\bibitem{Fiedler:2010sg}
F.~Fiedler, A.~Grohsjean, P.~Haefner, and P.~Schieferdecker, {\it {The Matrix
  Element Method and its Application in Measurements of the Top Quark Mass}},
  \href{http://dx.doi.org/10.1016/j.nima.2010.09.024}{Nucl. Instrum. Meth. A
  {\bfseries 624} (2010)  203},
  \href{http://arxiv.org/abs/1003.1316}{{arXiv:1003.1316 [hep-ex]}}.

\bibitem{Giammanco:2017xyn}
A.~Giammanco and R.~Schwienhorst, {\it {Single top-quark production at the
  Tevatron and the LHC}},
  \href{http://dx.doi.org/10.1103/RevModPhys.90.035001}{Rev. Mod. Phys.
  {\bfseries 90} (2018) 3, 035001},
  \href{http://arxiv.org/abs/1710.10699}{{arXiv:1710.10699 [hep-ex]}}.

\bibitem{Andersen:2012kn}
J.~R. Andersen, C.~Englert, and M.~Spannowsky, {\it {Extracting precise Higgs
  couplings by using the matrix element method}},
  \href{http://dx.doi.org/10.1103/PhysRevD.87.015019}{Phys. Rev. {\bfseries
  D87} (2013) 1, 015019},
\href{http://arxiv.org/abs/1211.3011}{{arXiv:1211.3011 [hep-ph]}}.

\bibitem{Artoisenet:2013vfa}
P.~Artoisenet, P.~de~Aquino, F.~Maltoni, and O.~Mattelaer, {\it {Unravelling
  $t\overline{t}h$ via the Matrix Element Method}},
  \href{http://dx.doi.org/10.1103/PhysRevLett.111.091802}{Phys. Rev. Lett.
  {\bfseries 111} (2013) 9, 091802},
\href{http://arxiv.org/abs/1304.6414}{{arXiv:1304.6414 [hep-ph]}}.

\bibitem{Englert:2015dlp}
C.~Englert, O.~Mattelaer, and M.~Spannowsky, {\it {Measuring the Higgs-bottom
  coupling in weak boson fusion}},
  \href{http://dx.doi.org/10.1016/j.physletb.2016.02.074}{Phys. Lett. B
  {\bfseries 756} (2016)  103},
  \href{http://arxiv.org/abs/1512.03429}{{arXiv:1512.03429 [hep-ph]}}.

\bibitem{FerreiradeLima:2017iwx}
D.~E. Ferreira~de Lima, O.~Mattelaer, and M.~Spannowsky, {\it {Searching for
  processes with invisible particles using a matrix element-based method}},
  \href{http://dx.doi.org/10.1016/j.physletb.2018.10.044}{Phys. Lett.
  {\bfseries B787} (2018)  100},
\href{http://arxiv.org/abs/1712.03266}{{arXiv:1712.03266 [hep-ph]}}.

\bibitem{CMS:2015enw}
CMS, V.~Khachatryan {\em et al.}, {\it {Search for a Standard Model Higgs Boson
  Produced in Association with a Top-Quark Pair and Decaying to Bottom Quarks
  Using a Matrix Element Method}},
  \href{http://dx.doi.org/10.1140/epjc/s10052-015-3454-1}{Eur. Phys. J. C
  {\bfseries 75} (2015) 6, 251},
  \href{http://arxiv.org/abs/1502.02485}{{arXiv:1502.02485 [hep-ex]}}.

\bibitem{ATLAS:2015jmq}
ATLAS, G.~Aad {\em et al.}, {\it {Evidence for single top-quark production in
  the $s$-channel in proton-proton collisions at $\sqrt{s}=$8 TeV with the
  ATLAS detector using the Matrix Element Method}},
  \href{http://dx.doi.org/10.1016/j.physletb.2016.03.017}{Phys. Lett. B
  {\bfseries 756} (2016)  228},
  \href{http://arxiv.org/abs/1511.05980}{{arXiv:1511.05980 [hep-ex]}}.

\bibitem{CMS:2015cal}
CMS, V.~Khachatryan {\em et al.}, {\it {Measurement of Spin Correlations in
  $t\bar{t}$ Production using the Matrix Element Method in the Muon+Jets Final
  State in $pp$ Collisions at $\sqrt{s} =$ 8 TeV}},
  \href{http://dx.doi.org/10.1016/j.physletb.2016.05.005}{Phys. Lett. B
  {\bfseries 758} (2016)  321},
  \href{http://arxiv.org/abs/1511.06170}{{arXiv:1511.06170 [hep-ex]}}.

\bibitem{Gritsan:2016hjl}
A.~V. Gritsan, R.~Röntsch, M.~Schulze, and M.~Xiao, {\it {Constraining
  anomalous Higgs boson couplings to the heavy flavor fermions using matrix
  element techniques}},
  \href{http://dx.doi.org/10.1103/PhysRevD.94.055023}{Phys. Rev. {\bfseries
  D94} (2016) 5, 055023},
\href{http://arxiv.org/abs/1606.03107}{{arXiv:1606.03107 [hep-ph]}}.

\bibitem{Artoisenet:2010cn}
P.~Artoisenet, V.~Lemaitre, F.~Maltoni, and O.~Mattelaer, {\it {Automation of
  the matrix element reweighting method}},
  \href{http://dx.doi.org/10.1007/JHEP12(2010)068}{JHEP {\bfseries 12} (2010)
  068},
\href{http://arxiv.org/abs/1007.3300}{{arXiv:1007.3300 [hep-ph]}}.

\bibitem{Alwall:2010cq}
J.~Alwall, A.~Freitas, and O.~Mattelaer, {\it {The Matrix Element Method and
  QCD Radiation}},  \href{http://dx.doi.org/10.1103/PhysRevD.83.074010}{Phys.
  Rev. {\bfseries D83} (2011)  074010},
\href{http://arxiv.org/abs/1010.2263}{{arXiv:1010.2263 [hep-ph]}}.

\bibitem{Campbell:2012cz}
J.~M. Campbell, W.~T. Giele, and C.~Williams, {\it {The Matrix Element Method
  at Next-to-Leading Order}},
  \href{http://dx.doi.org/10.1007/JHEP11(2012)043}{JHEP {\bfseries 11} (2012)
  043},
\href{http://arxiv.org/abs/1204.4424}{{arXiv:1204.4424 [hep-ph]}}.

\bibitem{Campbell:2013hz}
J.~M. Campbell, R.~K. Ellis, W.~T. Giele, and C.~Williams, {\it {Finding the
  Higgs boson in decays to $Z \gamma$ using the matrix element method at
  Next-to-Leading Order}},
  \href{http://dx.doi.org/10.1103/PhysRevD.87.073005}{Phys. Rev. {\bfseries
  D87} (2013) 7, 073005},
\href{http://arxiv.org/abs/1301.7086}{{arXiv:1301.7086 [hep-ph]}}.

\bibitem{Williams:2013kfb}
C.~Williams, J.~M. Campbell, and W.~T. Giele, {\it {Event-by-Event Weighting at
  Next-to-Leading Order}},  \href{http://dx.doi.org/10.22323/1.197.0037}{PoS
  {\bfseries RADCOR2013} (2013)  037},
  \href{http://arxiv.org/abs/1311.5811}{{arXiv:1311.5811 [hep-ph]}}.

\bibitem{Martini:2015fsa}
T.~Martini and P.~Uwer, {\it {Extending the Matrix Element Method beyond the
  Born approximation: Calculating event weights at next-to-leading order
  accuracy}},  \href{http://dx.doi.org/10.1007/JHEP09(2015)083}{JHEP {\bfseries
  09} (2015)  083}, \href{http://arxiv.org/abs/1506.08798}{{arXiv:1506.08798
  [hep-ph]}}.

\bibitem{Martini:2017ydu}
T.~Martini and P.~Uwer, {\it {The Matrix Element Method at next-to-leading
  order QCD for hadronic collisions: Single top-quark production at the LHC as
  an example application}},
  \href{http://dx.doi.org/10.1007/JHEP05(2018)141}{JHEP {\bfseries 05} (2018)
  141}, \href{http://arxiv.org/abs/1712.04527}{{arXiv:1712.04527 [hep-ph]}}.

\bibitem{Kraus:2019qoq}
M.~Kraus, T.~Martini, and P.~Uwer, {\it {Matrix Element Method at NLO for
  (anti-)$k_T$-jet algorithms}},
  \href{http://dx.doi.org/10.1103/PhysRevD.100.076010}{Phys. Rev. D {\bfseries
  100} (2019) 7, 076010},
  \href{http://arxiv.org/abs/1901.08008}{{arXiv:1901.08008 [hep-ph]}}.

\bibitem{Baumeister:2016maz}
R.~Baumeister and S.~Weinzierl, {\it {Matrix element method at next-to-leading
  order for arbitrary jet algorithms}},
  \href{http://dx.doi.org/10.1103/PhysRevD.95.036019}{Phys. Rev. D {\bfseries
  95} (2017) 3, 036019},
  \href{http://arxiv.org/abs/1612.07252}{{arXiv:1612.07252 [hep-ph]}}.

\bibitem{Butter:2020tvl}
A.~Butter and T.~Plehn, {\it {Generative Networks for LHC events}},
  \href{http://arxiv.org/abs/2008.08558}{{arXiv:2008.08558 [hep-ph]}}.

\bibitem{Butter:2022rso}
A.~Butter, T.~Plehn, S.~Schumann, {\em et al.}, {\it {Machine Learning and LHC
  Event Generation}},  \href{http://arxiv.org/abs/2203.07460}{{arXiv:2203.07460
  [hep-ph]}}.

\bibitem{ml_notes}
T.~Plehn, A.~Butter, B.~Dillon, and C.~Krause, {\it {Modern Machine Learning
  for LHC Physicists}},
  \href{http://arxiv.org/abs/2211.01421}{{arXiv:2211.01421 [hep-ph]}}.

\bibitem{Butter:2020qhk}
A.~Butter, S.~Diefenbacher, G.~Kasieczka, B.~Nachman, and T.~Plehn, {\it
  {GANplifying event samples}},
  \href{http://dx.doi.org/10.21468/SciPostPhys.10.6.139}{SciPost Phys.
  {\bfseries 10} (2021) 6, 139},
  \href{http://arxiv.org/abs/2008.06545}{{arXiv:2008.06545 [hep-ph]}}.

\bibitem{inn}
L.~Ardizzone, J.~Kruse, S.~Wirkert, D.~Rahner, E.~W. Pellegrini, R.~S. Klessen,
  L.~Maier-Hein, C.~Rother, and U.~Köthe, {\it Analyzing inverse problems with
  invertible neural networks},
  \href{http://arxiv.org/abs/1808.04730}{{arXiv:1808.04730 [cs.LG]}}.

\bibitem{Winterhalder:2021ngy}
R.~Winterhalder, V.~Magerya, E.~Villa, S.~P. Jones, M.~Kerner, A.~Butter,
  G.~Heinrich, and T.~Plehn, {\it {Targeting multi-loop integrals with neural
  networks}},  \href{http://dx.doi.org/10.21468/SciPostPhys.12.4.129}{SciPost
  Phys. {\bfseries 12} (2022) 4, 129},
  \href{http://arxiv.org/abs/2112.09145}{{arXiv:2112.09145 [hep-ph]}}.

\bibitem{maxim}
M.~D. Klimek and M.~Perelstein, {\it {Neural Network-Based Approach to Phase
  Space Integration}},
\href{http://arxiv.org/abs/1810.11509}{{arXiv:1810.11509 [hep-ph]}}.

\bibitem{Chen:2020nfb}
I.-K. Chen, M.~D. Klimek, and M.~Perelstein, {\it {Improved Neural Network
  Monte Carlo Simulation}},
  \href{http://dx.doi.org/10.21468/SciPostPhys.10.1.023}{SciPost Phys.
  {\bfseries 10} (2021)  023},
  \href{http://arxiv.org/abs/2009.07819}{{arXiv:2009.07819 [hep-ph]}}.

\bibitem{Bothmann:2020ywa}
E.~Bothmann, T.~Jan{\ss}en, M.~Knobbe, T.~Schmale, and S.~Schumann, {\it
  {Exploring phase space with Neural Importance Sampling}},
  \href{http://dx.doi.org/10.21468/SciPostPhys.8.4.069}{SciPost Phys.
  {\bfseries 8} (2020) 4, 069},
  \href{http://arxiv.org/abs/2001.05478}{{arXiv:2001.05478 [hep-ph]}}.

\bibitem{Gao:2020vdv}
C.~Gao, J.~Isaacson, and C.~Krause, {\it {i-flow: High-dimensional Integration
  and Sampling with Normalizing Flows}},
  \href{http://dx.doi.org/10.1088/2632-2153/abab62}{Mach. Learn. Sci. Tech.
  {\bfseries 1} (2020) 4, 045023},
  \href{http://arxiv.org/abs/2001.05486}{{arXiv:2001.05486 [physics.comp-ph]}}.

\bibitem{Gao:2020zvv}
C.~Gao, S.~Höche, J.~Isaacson, C.~Krause, and H.~Schulz, {\it {Event
  Generation with Normalizing Flows}},
  \href{http://dx.doi.org/10.1103/PhysRevD.101.076002}{Phys. Rev. D {\bfseries
  101} (2020) 7, 076002},
  \href{http://arxiv.org/abs/2001.10028}{{arXiv:2001.10028 [hep-ph]}}.

\bibitem{Danziger:2021eeg}
K.~Danziger, T.~Jan\ss{}en, S.~Schumann, and F.~Siegert, {\it {Accelerating
  Monte Carlo event generation -- rejection sampling using neural network
  event-weight estimates}},
  \href{http://arxiv.org/abs/2109.11964}{{arXiv:2109.11964 [hep-ph]}}.

\bibitem{Butter:2019eyo}
A.~Butter, T.~Plehn, and R.~Winterhalder, {\it {How to GAN Event Subtraction}},
   \href{http://dx.doi.org/10.21468/SciPostPhysCore.3.2.009}{SciPost Phys. Core
  {\bfseries 3} (2020)  009},
  \href{http://arxiv.org/abs/1912.08824}{{arXiv:1912.08824 [hep-ph]}}.

\bibitem{Verheyen:2020bjw}
B.~Stienen and R.~Verheyen, {\it {Phase Space Sampling and Inference from
  Weighted Events with Autoregressive Flows}},
  \href{http://dx.doi.org/10.21468/SciPostPhys.10.2.038}{SciPost Phys.
  {\bfseries 10} (2021)  038},
  \href{http://arxiv.org/abs/2011.13445}{{arXiv:2011.13445 [hep-ph]}}.

\bibitem{Backes:2020vka}
M.~Backes, A.~Butter, T.~Plehn, and R.~Winterhalder, {\it {How to GAN Event
  Unweighting}},
  \href{http://dx.doi.org/10.21468/SciPostPhys.10.4.089}{SciPost Phys.
  {\bfseries 10} (2021) 4, 089},
  \href{http://arxiv.org/abs/2012.07873}{{arXiv:2012.07873 [hep-ph]}}.

\bibitem{shower}
E.~Bothmann and L.~Debbio, {\it {Reweighting a parton shower using a neural
  network: the final-state case}},
  \href{http://dx.doi.org/10.1007/JHEP01(2019)033}{JHEP {\bfseries 01} (2019)
  033},
\href{http://arxiv.org/abs/1808.07802}{{arXiv:1808.07802 [hep-ph]}}.

\bibitem{locationGAN}
L.~de~Oliveira, M.~Paganini, and B.~Nachman, {\it {Learning Particle Physics by
  Example: Location-Aware Generative Adversarial Networks for Physics
  Synthesis}},  \href{http://dx.doi.org/10.1007/s41781-017-0004-6}{Comput.
  Softw. Big Sci. {\bfseries 1} (2017) 1, 4},
\href{http://arxiv.org/abs/1701.05927}{{arXiv:1701.05927 [stat.ML]}}.

\bibitem{juniprshower}
A.~Andreassen, I.~Feige, C.~Frye, and M.~D. Schwartz, {\it {JUNIPR: a Framework
  for Unsupervised Machine Learning in Particle Physics}},
  \href{http://dx.doi.org/10.1140/epjc/s10052-019-6607-9}{Eur. Phys. J.
  {\bfseries C79} (2019) 2, 102},
\href{http://arxiv.org/abs/1804.09720}{{arXiv:1804.09720 [hep-ph]}}.

\bibitem{Dohi:2020eda}
K.~Dohi, {\it {Variational Autoencoders for Jet Simulation}},
  \href{http://arxiv.org/abs/2009.04842}{{arXiv:2009.04842 [hep-ph]}}.

\bibitem{DiBello:2020bas}
F.~A. Di~Bello, S.~Ganguly, E.~Gross, M.~Kado, M.~Pitt, L.~Santi, and
  J.~Shlomi, {\it {Towards a Computer Vision Particle Flow}},
  \href{http://dx.doi.org/10.1140/epjc/s10052-021-08897-0}{Eur. Phys. J. C
  {\bfseries 81} (2021) 2, 107},
  \href{http://arxiv.org/abs/2003.08863}{{arXiv:2003.08863 [physics.data-an]}}.

\bibitem{Baldi:2020hjm}
P.~Baldi, L.~Blecher, A.~Butter, J.~Collado, J.~N. Howard, F.~Keilbach,
  T.~Plehn, G.~Kasieczka, and D.~Whiteson, {\it {How to GAN Higher Jet
  Resolution}},  \href{http://arxiv.org/abs/2012.11944}{{arXiv:2012.11944
  [hep-ph]}}.

\bibitem{Paganini:2017dwg}
M.~Paganini, L.~de~Oliveira, and B.~Nachman, {\it {CaloGAN : Simulating 3D high
  energy particle showers in multilayer electromagnetic calorimeters with
  generative adversarial networks}},
  \href{http://dx.doi.org/10.1103/PhysRevD.97.014021}{Phys. Rev. D {\bfseries
  97} (2018) 1, 014021},
  \href{http://arxiv.org/abs/1712.10321}{{arXiv:1712.10321 [hep-ex]}}.

\bibitem{Erdmann:2018jxd}
M.~Erdmann, J.~Glombitza, and T.~Quast, {\it {Precise simulation of
  electromagnetic calorimeter showers using a Wasserstein Generative
  Adversarial Network}},
  \href{http://dx.doi.org/10.1007/s41781-018-0019-7}{Comput. Softw. Big Sci.
  {\bfseries 3} (2019) 1, 4},
  \href{http://arxiv.org/abs/1807.01954}{{arXiv:1807.01954 [physics.ins-det]}}.

\bibitem{Buhmann:2020pmy}
E.~Buhmann, S.~Diefenbacher, E.~Eren, F.~Gaede, G.~Kasieczka, A.~Korol, and
  K.~Kr\"uger, {\it {Getting High: High Fidelity Simulation of High Granularity
  Calorimeters with High Speed}},
  \href{http://dx.doi.org/10.1007/s41781-021-00056-0}{Comput. Softw. Big Sci.
  {\bfseries 5} (2021) 1, 13},
  \href{http://arxiv.org/abs/2005.05334}{{arXiv:2005.05334 [physics.ins-det]}}.

\bibitem{Krause:2021ilc}
C.~Krause and D.~Shih, {\it {CaloFlow: Fast and Accurate Generation of
  Calorimeter Showers with Normalizing Flows}},
  \href{http://arxiv.org/abs/2106.05285}{{arXiv:2106.05285 [physics.ins-det]}}.

\bibitem{Krause:2021wez}
C.~Krause and D.~Shih, {\it {CaloFlow II: Even Faster and Still Accurate
  Generation of Calorimeter Showers with Normalizing Flows}},
  \href{http://arxiv.org/abs/2110.11377}{{arXiv:2110.11377 [physics.ins-det]}}.

\bibitem{dutch}
S.~Otten, S.~Caron, W.~de~Swart, M.~van Beekveld, L.~Hendriks, C.~van Leeuwen,
  D.~Podareanu, R.~Ruiz~de Austri, and R.~Verheyen, {\it {Event Generation and
  Statistical Sampling for Physics with Deep Generative Models and a Density
  Information Buffer}},
  \href{http://dx.doi.org/10.1038/s41467-021-22616-z}{Nature Commun. {\bfseries
  12} (2021) 1, 2985}, \href{http://arxiv.org/abs/1901.00875}{{arXiv:1901.00875
  [hep-ph]}}.

\bibitem{gan_datasets}
B.~Hashemi, N.~Amin, K.~Datta, D.~Olivito, and M.~Pierini, {\it {LHC
  analysis-specific datasets with Generative Adversarial Networks}},
  \href{http://arxiv.org/abs/1901.05282}{{arXiv:1901.05282 [hep-ex]}}.

\bibitem{DijetGAN2}
R.~Di~Sipio, M.~Faucci~Giannelli, S.~Ketabchi~Haghighat, and S.~Palazzo, {\it
  {DijetGAN: A Generative-Adversarial Network Approach for the Simulation of
  QCD Dijet Events at the LHC}},
  \href{http://dx.doi.org/10.1007/JHEP08(2019)110}{JHEP {\bfseries 08} (2020)
  110},
\href{http://arxiv.org/abs/1903.02433}{{arXiv:1903.02433 [hep-ex]}}.

\bibitem{Butter:2019cae}
A.~Butter, T.~Plehn, and R.~Winterhalder, {\it {How to GAN LHC Events}},
  \href{http://dx.doi.org/10.21468/SciPostPhys.7.6.075}{SciPost Phys.
  {\bfseries 7} (2019) 6, 075},
  \href{http://arxiv.org/abs/1907.03764}{{arXiv:1907.03764 [hep-ph]}}.

\bibitem{Alanazi:2020klf}
Y.~Alanazi, N.~Sato, T.~Liu, W.~Melnitchouk, M.~P. Kuchera, E.~Pritchard,
  M.~Robertson, R.~Strauss, L.~Velasco, and Y.~Li, {\it {Simulation of
  electron-proton scattering events by a Feature-Augmented and Transformed
  Generative Adversarial Network (FAT-GAN)}},
  \href{http://arxiv.org/abs/2001.11103}{{arXiv:2001.11103 [hep-ph]}}.

\bibitem{Butter:2021csz}
A.~Butter, T.~Heimel, S.~Hummerich, T.~Krebs, T.~Plehn, A.~Rousselot, and
  S.~Vent, {\it {Generative Networks for Precision Enthusiasts}},
  \href{http://arxiv.org/abs/2110.13632}{{arXiv:2110.13632 [hep-ph]}}.

\bibitem{Datta:2018mwd}
K.~Datta, D.~Kar, and D.~Roy, {\it {Unfolding with Generative Adversarial
  Networks}},
\href{http://arxiv.org/abs/1806.00433}{{arXiv:1806.00433 [physics.data-an]}}.

\bibitem{fcgan}
M.~Bellagente, A.~Butter, G.~Kasieczka, T.~Plehn, and R.~Winterhalder, {\it
  {How to GAN away Detector Effects}},
  \href{http://dx.doi.org/10.21468/SciPostPhys.8.4.070}{SciPost Phys.
  {\bfseries 8} (2020) 4, 070},
  \href{http://arxiv.org/abs/1912.00477}{{arXiv:1912.00477 [hep-ph]}}.

\bibitem{Andreassen:2019cjw}
A.~Andreassen, P.~T. Komiske, E.~M. Metodiev, B.~Nachman, and J.~Thaler, {\it
  {OmniFold: A Method to Simultaneously Unfold All Observables}},
  \href{http://dx.doi.org/10.1103/PhysRevLett.124.182001}{Phys. Rev. Lett.
  {\bfseries 124} (2020) 18, 182001},
  \href{http://arxiv.org/abs/1911.09107}{{arXiv:1911.09107 [hep-ph]}}.

\bibitem{cinn}
L.~Ardizzone, C.~Lüth, J.~Kruse, C.~Rother, and U.~Köthe, {\it Guided image
  generation with conditional invertible neural networks},
  \href{http://arxiv.org/abs/1907.02392}{{arXiv:1907.02392 [cs.CV]}}.

\bibitem{Bellagente:2020piv}
M.~Bellagente, A.~Butter, G.~Kasieczka, T.~Plehn, A.~Rousselot,
  R.~Winterhalder, L.~Ardizzone, and U.~K\"othe, {\it {Invertible Networks or
  Partons to Detector and Back Again}},
  \href{http://dx.doi.org/10.21468/SciPostPhys.9.5.074}{SciPost Phys.
  {\bfseries 9} (2020)  074},
  \href{http://arxiv.org/abs/2006.06685}{{arXiv:2006.06685 [hep-ph]}}.

\bibitem{Bieringer:2020tnw}
S.~Bieringer, A.~Butter, T.~Heimel, S.~H\"oche, U.~K\"othe, T.~Plehn, and S.~T.
  Radev, {\it {Measuring QCD Splittings with Invertible Networks}},
  \href{http://dx.doi.org/10.21468/SciPostPhys.10.6.126}{SciPost Phys.
  {\bfseries 10} (2021) 6, 126},
  \href{http://arxiv.org/abs/2012.09873}{{arXiv:2012.09873 [hep-ph]}}.

\bibitem{Bister:2021arb}
T.~Bister, M.~Erdmann, U.~K\"othe, and J.~Schulte, {\it {Inference of
  cosmic-ray source properties by conditional invertible neural networks}},
  \href{http://dx.doi.org/10.1140/epjc/s10052-022-10138-x}{Eur. Phys. J. C
  {\bfseries 82} (2022) 2, 171},
  \href{http://arxiv.org/abs/2110.09493}{{arXiv:2110.09493 [astro-ph.IM]}}.

\bibitem{Leigh:2022lpn}
M.~Leigh, J.~A. Raine, and T.~Golling, {\it {$\nu$-Flows: conditional neutrino
  regression}},  \href{http://arxiv.org/abs/2207.00664}{{arXiv:2207.00664
  [hep-ph]}}.

\bibitem{Bellagente:2021yyh}
M.~Bellagente, M.~Hau\ss{}mann, M.~Luchmann, and T.~Plehn, {\it {Understanding
  Event-Generation Networks via Uncertainties}},
  \href{http://dx.doi.org/10.21468/SciPostPhys.13.1.003}{SciPost Phys.
  {\bfseries 13} (2022)  003},
  \href{http://arxiv.org/abs/2104.04543}{{arXiv:2104.04543 [hep-ph]}}.

\bibitem{Brehmer:2019xox}
J.~Brehmer, F.~Kling, I.~Espejo, and K.~Cranmer, {\it {MadMiner: Machine
  learning-based inference for particle physics}},
  \href{http://dx.doi.org/10.1007/s41781-020-0035-2}{Comput. Softw. Big Sci.
  {\bfseries 4} (2020) 1, 3},
  \href{http://arxiv.org/abs/1907.10621}{{arXiv:1907.10621 [hep-ph]}}.

\bibitem{HEPSoftwareFoundation:2017ggl}
HEP Software Foundation, J.~Albrecht {\em et al.}, {\it {A Roadmap for HEP
  Software and Computing R\&D for the 2020s}},
  \href{http://dx.doi.org/10.1007/s41781-018-0018-8}{Comput. Softw. Big Sci.
  {\bfseries 3} (2019) 1, 7},
  \href{http://arxiv.org/abs/1712.06982}{{arXiv:1712.06982 [physics.comp-ph]}}.

\bibitem{Bury:2020ewi}
F.~Bury and C.~Delaere, {\it {Matrix element regression with deep neural
  networks \textemdash{} Breaking the CPU barrier}},
  \href{http://dx.doi.org/10.1007/JHEP04(2021)020}{JHEP {\bfseries 04} (2021)
  020}, \href{http://arxiv.org/abs/2008.10949}{{arXiv:2008.10949 [hep-ex]}}.

\bibitem{Buckley:2015vsa}
M.~R. Buckley and D.~Goncalves, {\it {Boosting the Direct CP Measurement of the
  Higgs-Top Coupling}},
  \href{http://dx.doi.org/10.1103/PhysRevLett.116.091801}{Phys. Rev. Lett.
  {\bfseries 116} (2016) 9, 091801},
  \href{http://arxiv.org/abs/1507.07926}{{arXiv:1507.07926 [hep-ph]}}.

\bibitem{Ren:2019xhp}
J.~Ren, L.~Wu, and J.~M. Yang, {\it {Unveiling CP property of top-Higgs
  coupling with graph neural networks at the LHC}},
  \href{http://dx.doi.org/10.1016/j.physletb.2020.135198}{Phys. Lett. B
  {\bfseries 802} (2020)  135198},
  \href{http://arxiv.org/abs/1901.05627}{{arXiv:1901.05627 [hep-ph]}}.

\bibitem{Bortolato:2020zcg}
B.~Bortolato, J.~F. Kamenik, N.~Ko\v{s}nik, and A.~Smolkovi\v{c}, {\it
  {Optimized probes of $CP$ -odd effects in the $t \bar t h$ process at hadron
  colliders}},  \href{http://dx.doi.org/10.1016/j.nuclphysb.2021.115328}{Nucl.
  Phys. B {\bfseries 964} (2021)  115328},
  \href{http://arxiv.org/abs/2006.13110}{{arXiv:2006.13110 [hep-ph]}}.

\bibitem{Bahl:2020wee}
H.~Bahl, P.~Bechtle, S.~Heinemeyer, J.~Katzy, T.~Klingl, K.~Peters,
  M.~Saimpert, T.~Stefaniak, and G.~Weiglein, {\it {Indirect $\mathcal{CP}$
  probes of the Higgs-top-quark interaction: current LHC constraints and future
  opportunities}},  \href{http://dx.doi.org/10.1007/JHEP11(2020)127}{JHEP
  {\bfseries 11} (2020)  127},
  \href{http://arxiv.org/abs/2007.08542}{{arXiv:2007.08542 [hep-ph]}}.

\bibitem{Martini:2021uey}
T.~Martini, R.-Q. Pan, M.~Schulze, and M.~Xiao, {\it {Probing the CP structure
  of the top quark Yukawa coupling: Loop sensitivity versus on-shell
  sensitivity}},  \href{http://dx.doi.org/10.1103/PhysRevD.104.055045}{Phys.
  Rev. D {\bfseries 104} (2021) 5, 055045},
  \href{http://arxiv.org/abs/2104.04277}{{arXiv:2104.04277 [hep-ph]}}.

\bibitem{Goncalves:2021dcu}
D.~Gon\c{c}alves, J.~H. Kim, K.~Kong, and Y.~Wu, {\it {Direct Higgs-top
  CP-phase measurement with $ t\overline{t}h $ at the 14 TeV LHC and 100 TeV
  FCC}},  \href{http://dx.doi.org/10.1007/JHEP01(2022)158}{JHEP {\bfseries 01}
  (2022)  158}, \href{http://arxiv.org/abs/2108.01083}{{arXiv:2108.01083
  [hep-ph]}}.

\bibitem{Barman:2021yfh}
R.~K. Barman, D.~Gon\c{c}alves, and F.~Kling, {\it {Machine learning the Higgs
  boson-top quark CP phase}},
  \href{http://dx.doi.org/10.1103/PhysRevD.105.035023}{Phys. Rev. D {\bfseries
  105} (2022) 3, 035023},
  \href{http://arxiv.org/abs/2110.07635}{{arXiv:2110.07635 [hep-ph]}}.

\bibitem{Bahl:2021dnc}
H.~Bahl and S.~Brass, {\it {Constraining $ \mathcal{CP} $-violation in the
  Higgs-top-quark interaction using machine-learning-based inference}},
  \href{http://dx.doi.org/10.1007/JHEP03(2022)017}{JHEP {\bfseries 03} (2022)
  017}, \href{http://arxiv.org/abs/2110.10177}{{arXiv:2110.10177 [hep-ph]}}.

\bibitem{Kraus:2019myc}
M.~Kraus, T.~Martini, S.~Peitzsch, and P.~Uwer, {\it {Exploring BSM Higgs
  couplings in single top-quark production}},
  \href{http://arxiv.org/abs/1908.09100}{{arXiv:1908.09100 [hep-ph]}}.

\bibitem{DelDebbio:2004xtd}
NNPDF, L.~Del~Debbio, S.~Forte, J.~I. Latorre, A.~Piccione, and J.~Rojo, {\it
  {Unbiased determination of the proton structure function F(2)**p with
  faithful uncertainty estimation}},
  \href{http://dx.doi.org/10.1088/1126-6708/2005/03/080}{JHEP {\bfseries 03}
  (2005)  080},
  \href{http://arxiv.org/abs/hep-ph/0501067}{{arXiv:hep-ph/0501067}}.

\bibitem{pythia8}
T.~Sjöstrand, S.~Ask, J.~R. Christiansen, R.~Corke, N.~Desai, P.~Ilten,
  S.~Mrenna, S.~Prestel, C.~O. Rasmussen, and P.~Z. Skands, {\it {An
  Introduction to PYTHIA 8.2}},
  \href{http://dx.doi.org/10.1016/j.cpc.2015.01.024}{Comput. Phys. Commun.
  {\bfseries 191} (2015)  159},
  \href{http://arxiv.org/abs/1410.3012}{{arXiv:1410.3012 [hep-ph]}}.

\bibitem{delphes}
DELPHES 3, J.~de~Favereau, C.~Delaere, P.~Demin, A.~Giammanco, V.~Lemaître,
  A.~Mertens, and M.~Selvaggi, {\it {DELPHES 3, A modular framework for fast
  simulation of a generic collider experiment}},
  \href{http://dx.doi.org/10.1007/JHEP02(2014)057}{JHEP {\bfseries 02} (2014)
  057},
\href{http://arxiv.org/abs/1307.6346}{{arXiv:1307.6346 [hep-ex]}}.

\bibitem{fastjet}
M.~Cacciari, G.~P. Salam, and G.~Soyez, {\it {FastJet User Manual}},
  \href{http://dx.doi.org/10.1140/epjc/s10052-012-1896-2}{Eur. Phys. J. C
  {\bfseries 72} (2012)  1896},
  \href{http://arxiv.org/abs/1111.6097}{{arXiv:1111.6097 [hep-ph]}}.

\bibitem{anti-kt}
M.~Cacciari, G.~P. Salam, and G.~Soyez, {\it {The anti-$k_t$ jet clustering
  algorithm}},  \href{http://dx.doi.org/10.1088/1126-6708/2008/04/063}{JHEP
  {\bfseries 04} (2008)  063},
  \href{http://arxiv.org/abs/0802.1189}{{arXiv:0802.1189 [hep-ph]}}.

\bibitem{Artoisenet:2013puc}
P.~Artoisenet {\em et al.}, {\it {A framework for Higgs characterisation}},
  \href{http://dx.doi.org/10.1007/JHEP11(2013)043}{JHEP {\bfseries 11} (2013)
  043}, \href{http://arxiv.org/abs/1306.6464}{{arXiv:1306.6464 [hep-ph]}}.

\bibitem{Demartin:2015uha}
F.~Demartin, F.~Maltoni, K.~Mawatari, and M.~Zaro, {\it {Higgs production in
  association with a single top quark at the LHC}},
  \href{http://dx.doi.org/10.1140/epjc/s10052-015-3475-9}{Eur. Phys. J. C
  {\bfseries 75} (2015) 6, 267},
  \href{http://arxiv.org/abs/1504.00611}{{arXiv:1504.00611 [hep-ph]}}.

\bibitem{durkan2019neural}
C.~Durkan, A.~Bekasov, I.~Murray, and G.~Papamakarios, {\it Neural spline
  flows},  Advances in Neural Information Processing Systems {\bfseries 32}
  (2019)  7511.

\bibitem{pytorch}
A.~Paszke, S.~Gross, F.~Massa, A.~Lerer, J.~Bradbury, G.~Chanan, T.~Killeen,
  Z.~Lin, N.~Gimelshein, L.~Antiga, A.~Desmaison, A.~Kopf, E.~Yang, Z.~DeVito,
  M.~Raison, A.~Tejani, S.~Chilamkurthy, B.~Steiner, L.~Fang, J.~Bai, and
  S.~Chintala, {\it Pytorch: An imperative style, high-performance deep
  learning library},  in {\em Advances in Neural Information Processing Systems
  32}, H.~Wallach, H.~Larochelle, A.~Beygelzimer, F.~d'Alch\'{e} Buc, E.~Fox,
  and R.~Garnett, eds., pp.~8024--8035.
\newblock Curran Associates, Inc., 2019.
\newblock
\newblock
  \href{http://papers.neurips.cc/paper/9015-pytorch-an-imperative-style-high-performance-deep-learning-library.pdf}{\href{http://arxiv.org/abs/1912.01703}{{arXiv:1912.01703
  [cs.LG]}}}.

\bibitem{adam}
D.~P. {Kingma} and J.~{Ba}, {\it {Adam: A Method for Stochastic Optimization}},
   \href{http://arxiv.org/abs/1412.6980}{{arXiv:1412.6980 [cs.LG]}}.

\bibitem{one-cycle-lr}
L.~N. Smith and N.~Topin, {\it Super-convergence: Very fast training of neural
  networks using large learning rates},  in {\em Artificial Intelligence and
  Machine Learning for Multi-Domain Operations Applications}, International
  Society for Optics and Photonics.
\newblock 2019.

\bibitem{bnn_early}
D.~MacKay, {\it {Probable Networks and Plausible Predictions – A Review of
  Practical Bayesian Methods for Supervised Neural Networks}},
  \href{http://www.inference.org.uk/mackay/network.pdf}{Comp. in Neural Systems
  {\bfseries 6} (1995)  4679}.

\bibitem{bnn_early2}
R.~M. Neal, {\em {Bayesian learning for neural networks}}.
\newblock PhD thesis, Toronto,
\newblock \href{ftp://www.cs.toronto.edu/dist/radford/thesis.pdf}{1995}.

\bibitem{bnn_early3}
Y.~Gal, {\em {Uncertainty in Deep Learning}}.
\newblock PhD thesis, Cambridge,
\newblock \href{http://mlg.eng.cam.ac.uk/yarin/thesis/thesis.pdf}{2016}.

\bibitem{deep_errors}
A.~Kendall and Y.~Gal, {\it {What Uncertainties Do We Need in Bayesian Deep
  Learning for Computer Vision?}},  Proc. NIPS (2017)  ,
  \href{http://arxiv.org/abs/1703.04977}{{arXiv:1703.04977 [cs.CV]}}.

\bibitem{Bollweg:2019skg}
S.~Bollweg, M.~Hau\ss{}mann, G.~Kasieczka, M.~Luchmann, T.~Plehn, and
  J.~Thompson, {\it {Deep-Learning Jets with Uncertainties and More}},
  \href{http://dx.doi.org/10.21468/SciPostPhys.8.1.006}{SciPost Phys.
  {\bfseries 8} (2020) 1, 006},
  \href{http://arxiv.org/abs/1904.10004}{{arXiv:1904.10004 [hep-ph]}}.

\bibitem{Kasieczka:2020vlh}
G.~Kasieczka, M.~Luchmann, F.~Otterpohl, and T.~Plehn, {\it {Per-Object
  Systematics using Deep-Learned Calibration}},
  \href{http://dx.doi.org/10.21468/SciPostPhys.9.6.089}{SciPost Phys.
  {\bfseries 9} (2020)  089},
  \href{http://arxiv.org/abs/2003.11099}{{arXiv:2003.11099 [hep-ph]}}.

\end{thebibliography}\endgroup

\end{document}